\documentclass[pdftex,twocolumn,epjc3_preprint,runningheads]{svjour3}

\usepackage[T1]{fontenc}
\usepackage{lmodern}
\usepackage{calc}
\usepackage{graphicx}
\usepackage{booktabs}
\usepackage{textcomp}
\usepackage{xspace}
\usepackage{relsize}
\usepackage{amssymb}
\usepackage{amsmath}
\usepackage{listings}
\usepackage{microtype}
\usepackage{multirow}
\usepackage{tabularx}
\usepackage{array}
\usepackage{placeins}
\usepackage{cuted}
\usepackage{soul} % only for \st; delete if this causes you problems.
\usepackage{fixltx2e}
\usepackage[numbers,sort&compress]{natbib}
\usepackage[labelfont=bf,font=small]{caption}
\usepackage[skip=-2pt]{subcaption}
\usepackage[colorlinks,citecolor=blue,urlcolor=blue,linkcolor=blue,breaklinks=breakall]{hyperref}
\usepackage{breakurl}
\usepackage[dvipsnames]{xcolor}
\usepackage[clockwise,figuresright]{rotating}
\usepackage{siunitx}
\usepackage{tikz}

\usepackage{etoolbox}
\AfterEndEnvironment{strip}{\leavevmode}

\allowdisplaybreaks

\newcolumntype{L}{>{\raggedright\let\newline\\\arraybackslash\hspace{0pt}}X}
\newcolumntype{R}{>{\raggedleft\let\newline\\\arraybackslash\hspace{0pt}}X}
\newcolumntype{C}{>{\centering\let\newline\\\arraybackslash\hspace{0pt}}X}

\newcommand{\imperial}{Department of Physics, Imperial College London, Blackett Laboratory, Prince Consort Road, London SW7 2AZ, UK}
\newcommand{\nordita}{NORDITA, Roslagstullsbacken 23, SE-10691 Stockholm, Sweden}
\newcommand{\oslo}{Department of Physics, University of Oslo, N-0316 Oslo, Norway}
\newcommand{\adelaide}{Department of Physics, University of Adelaide, Adelaide, SA 5005, Australia}

\newcommand{\coepp}{Australian Research Council Centre of Excellence for Particle Physics at the Tera-scale}
\newcommand{\okc}{Oskar Klein Centre for Cosmoparticle Physics, AlbaNova University Centre, SE-10691 Stockholm, Sweden}
\newcommand{\su}{Department of Physics, Stockholm University, SE-10691 Stockholm, Sweden}
\newcommand{\mcgill}{Department of Physics, McGill University, 3600 rue University, Montr\'eal, Qu\'ebec H3A 2T8, Canada}

\newcommand{\annecy}{LAPTh, Universit\'e de Savoie, CNRS, 9 chemin de Bellevue B.P.110, F-74941 Annecy-le-Vieux, France}

\newcommand{\grappa}{GRAPPA, Institute of Physics, University of Amsterdam, Science Park 904, 1098 XH Amsterdam, Netherlands}

\newcommand{\desy}{DESY, Notkestra\ss e 85, D-22607 Hamburg, Germany}

\newcommand{\gambitacknosplus}{We warmly thank the Casa Matem\'aticas Oaxaca, affiliated with the Banff International Research Station, for hospitality whilst part of this work was completed, and the staff at Cyfronet, for their always helpful supercomputing support.  \GB has been supported by STFC (UK; ST/K00414X/1, ST/P000762/1), the Royal Society (UK; UF110191), Glasgow University (UK; Leadership Fellowship), the Research Council of Norway (FRIPRO 230546/F20), NOTUR (Norway; NN9284K), the Knut and Alice Wallenberg Foundation (Sweden; Wallenberg Academy Fellowship), the Swedish Research Council (621-2014-5772), the Australian Research Council (CE110001004, FT130100018, FT140100244, FT160100274), The University of Sydney (Australia; IRCA-G162448), PLGrid Infrastructure (Poland), Polish National Science Center (Sonata UMO-2015/17/D/ST2/03532), the Swiss National Science Foundation (PP00P2-144674), European Commission Horizon 2020 (Marie Sk\l{}odowska-Curie actions H2020-MSCA-RISE-2015-691164, European Research Council Starting Grant ERC-2014-STG-638528), the ERA-CAN+ Twinning Program (EU \& Canada), the Netherlands Organisation for Scientific Research (NWO-Vidi 016.149.331), the National Science Foundation (USA; DGE-1339067), the FRQNT (Qu\'ebec) and NSERC/The Canadian Tri-Agencies Research Councils (BPDF-424460-2012).}

\makeatletter

\newcommand{\preprintnumber}[1]{\gdef\@preprintnumber{\begin{flushright}{#1}\end{flushright}}}

\g@addto@macro\bfseries{\boldmath}
\makeatother

\newcommand{\subparagraph}{} %< workaround for svjour not defining subparagraph
\usepackage{titlesec}
\titleformat*{\paragraph}{\bfseries}
\journalname{Eur. Phys. J. C}
\smartqed
\let\underscore\_
\renewcommand{\_}{\discretionary{\underscore}{}{\underscore}}

\makeatletter
\let\orgdescriptionlabel\descriptionlabel
\renewcommand*{\descriptionlabel}[1]{%
  \let\orglabel\label
  \let\label\@gobble
  \phantomsection
  \protected@edef\@currentlabel{#1}%
  \let\label\orglabel
  \orgdescriptionlabel{#1}%
}
\makeatother

\newcommand\postnewlinemarker{\hbox{\ensuremath{\hookrightarrow}}}
\newcommand\cpp[1]{{\lstinline!#1!}}  % Apparently curly braces are only "experimental"
\newcommand\cpppragma[1]{{\CPPcommentstyle#1}}
\newcommand\yaml[1]{{\lstset{style=yaml}\lstinline!#1!\lstset{style=cpp}}}
\newcommand\yamlvalue[1]{{\YAMLvaluestyle\ttfamily#1}}
\newcommand\term[1]{{\lstset{style=terminal}\lstinline!#1!\lstset{style=cpp}}}
\newcommand\fortran[1]{{\lstset{style=fortran}\lstinline!#1!\lstset{style=cpp}}}
\newcommand\py[1]{{\lstset{style=python}\lstinline!#1!\lstset{style=cpp}}}
\newcommand\customtilde{{\raisebox{0.2ex}{\scalebox{0.6}{\boldmath$\sim$}}}}
\newcommand\mathematica[1]{{\lstset{style=Mathematica}\lstinline!#1!\lstset{style=cpp}}}

\lstnewenvironment{lstlistingyaml}{\lstset{style=yaml}}{\lstset{style=cpp}}
\lstnewenvironment{lstlistingterm}{\lstset{style=terminal}}{\lstset{style=cpp}}
\lstnewenvironment{lstlistingfortran}{\lstset{style=fortran}}{\lstset{style=cpp}}
\lstnewenvironment{lstcpp}{\lstset{style=cpp}}{\lstset{style=cpp}}
\lstnewenvironment{lstcppalt}{\lstset{style=cppalt}}{\lstset{style=cpp}}
\lstnewenvironment{lstcppnum}{\lstset{style=cppnum}}{\lstset{style=cpp}}
\lstnewenvironment{lstyaml}{\lstset{style=yaml}}{\lstset{style=cpp}}
\lstnewenvironment{lstterm}{\lstset{style=terminal}}{\lstset{style=cpp}}
\lstnewenvironment{lsttermalt}{\lstset{style=terminalalt}}{\lstset{style=cpp}}
\lstnewenvironment{lsttext}{\lstset{style=text}}{\lstset{style=cpp}}
\lstnewenvironment{lstfortran}{\lstset{style=fortran}}{\lstset{style=cpp}}
\lstnewenvironment{lstpy}{\lstset{style=python}}{\lstset{style=cpp}}
\lstnewenvironment{lstmathematica}{\lstset{style=mathematica}}{\lstset{style=cpp}}

\newcommand{\tmpname}{}
\newcommand{\tmplistingname}{}
\makeatletter
\newif\ifATOlabelname
\lst@Key{labelname}{Listing}{\def\ATOlabelname{#1}\global\ATOlabelnametrue}
\makeatother
\lstnewenvironment{lstcpplabel}[1][]{
  \lstset{style=cpp,#1} % #1 allows to add new options with [] same as for normal lstlistings environment
  \ifATOlabelname
    \renewcommand{\tmpname}{\lstlistingname}
    \renewcommand{\tmplistingname}{\lstlistlistingname}
    \renewcommand{\lstlistingname}{\ATOlabelname}% Listing -> labelname
    \renewcommand{\lstlistlistingname}{List of \lstlistingname s}% List of Listings -> List of labelname
  \fi
}{
  \renewcommand{\lstlistingname}{\tmpname}
  \renewcommand{\lstlistlistingname}{\tmplistingname}
  \lstset{style=cpp}
}
\definecolor{solarized@base03}{HTML}{002B36}
\definecolor{solarized@base02}{HTML}{073642}
\definecolor{solarized@base01}{HTML}{586e75}
\definecolor{solarized@base00}{HTML}{657b83}
\definecolor{solarized@base0}{HTML}{839496}
\definecolor{solarized@base1}{HTML}{93a1a1}
\definecolor{solarized@base2}{HTML}{EEE8D5}
\definecolor{solarized@base3}{HTML}{FDF6E3}
\definecolor{solarized@yellow}{HTML}{B58900}
\definecolor{solarized@orange}{HTML}{CB4B16}
\definecolor{solarized@red}{HTML}{DC322F}
\definecolor{solarized@magenta}{HTML}{D33682}
\definecolor{solarized@violet}{HTML}{6C71C4}
\definecolor{solarized@blue}{HTML}{268BD2}
\definecolor{solarized@cyan}{HTML}{2AA198}
\definecolor{solarized@green}{HTML}{859900}
\definecolor{darkred}{HTML}{550003}
\definecolor{darkgreen}{HTML}{00AA00}

\newcommand\YAMLstringstyle{\footnotesize\color{solarized@green}\mdseries}
\newcommand\YAMLkeystyle{\footnotesize\color{solarized@blue}\ttfamily}
\newcommand\YAMLvaluestyle{\footnotesize\color{blue}\mdseries}
\newcommand\ProcessThreeDashes{\llap{\color{cyan}\mdseries-{-}-}}
\newcommand\CPPidentifierstyle{\color{solarized@blue}\footnotesize\ttfamily}
\newcommand\CPPcommentstyle{\color{solarized@violet}\footnotesize\ttfamily}
\newcommand\CPPdirectivestyle{\color{solarized@magenta}\footnotesize\ttfamily}
\newcommand\termplainstyle{\footnotesize\ttfamily}

\newcommand\processLongMacroDelimiter
{%
\CPPdirectivestyle%
\#define%
}

\lstdefinestyle{cpp}
{
  language=C++,
  basicstyle=\footnotesize\ttfamily,
  basewidth={0.53em,0.44em}, %Ben: experimenting a bit with the fixed-width width (first argument); feels a bit more readable to me with the slightly smaller width (was 0.6em by default)
  numbers=none,
  tabsize=2,
  breaklines=true,
  escapeinside={@}{@},
  showstringspaces=false,
  numberstyle=\tiny\color{solarized@base01},
  keywordstyle=\color{solarized@orange},
  stringstyle=\color{solarized@red}\ttfamily,
  identifierstyle=\color{solarized@blue},
  commentstyle=\CPPcommentstyle,
  directivestyle=\CPPdirectivestyle,
  emphstyle=\color{solarized@green},
  frame=single,
  rulecolor=\color{solarized@base2},
  rulesepcolor=\color{solarized@base2},
  literate={~} {\customtilde}1,
  moredelim=*[directive]\ \ \#,
  moredelim=*[directive]\ \ \ \ \#
}

\lstdefinestyle{cppalt}
{
  language=C++,
  basicstyle=\footnotesize\ttfamily,
  basewidth={0.53em,0.44em}, %Ben: experimenting a bit with the fixed-width width (first argument); feels a bit more readable to me with the slightly smaller width (was 0.6em by default)
  numbers=none,
  tabsize=2,
  breaklines=true,
  escapeinside={*@}{@*},
  showstringspaces=false,
  numberstyle=\tiny\color{solarized@base01},
  keywordstyle=\color{solarized@orange},
  stringstyle=\color{solarized@red}\ttfamily,
  identifierstyle=\color{solarized@blue},
  commentstyle=\CPPcommentstyle,
  directivestyle=\CPPdirectivestyle,
  emphstyle=\color{solarized@green},
  frame=single,
  rulecolor=\color{solarized@base2},
  rulesepcolor=\color{solarized@base2},
  literate={~}{\customtilde}1,
  moredelim=**[is][\processLongMacroDelimiter]{BeginLongMacro}{EndLongMacro} %special delimiter for long macros that go over several lines
}

\lstdefinestyle{cppnum}
{
  language=C++,
  basicstyle=\footnotesize\ttfamily,
  basewidth={0.53em,0.44em}, %Ben: experimenting a bit with the fixed-width width (first argument); feels a bit more readable to me with the slightly smaller width (was 0.6em by default)
  numbers=none,
  tabsize=2,
  breaklines=true,
  escapeinside={@}{@},
  numberstyle=\tiny\color{solarized@base01},
  showstringspaces=false,
  numberstyle=\tiny\color{solarized@base01},
  keywordstyle=\color{solarized@orange},
  stringstyle=\color{solarized@red}\ttfamily,
  identifierstyle=\color{solarized@blue},
  commentstyle=\CPPcommentstyle,
  directivestyle=\CPPdirectivestyle,
  emphstyle=\color{solarized@green},
  frame=single,
  rulecolor=\color{solarized@base2},
  rulesepcolor=\color{solarized@base2},
  literate={~} {\customtilde}1,
  moredelim=*[directive]\ \ \#,
  moredelim=*[directive]\ \ \ \ \#
}

\lstdefinestyle{python}
{
  language=Python,
  basicstyle=\footnotesize\ttfamily,
  basewidth={0.53em,0.44em},
  numbers=none,
  tabsize=2,
  breaklines=true,
  escapeinside={@}{@},
  showstringspaces=false,
  numberstyle=\tiny\color{solarized@base01},
  keywordstyle=\color{blue},
  stringstyle=\color{orange}\ttfamily,
  identifierstyle=\color{darkred},
  commentstyle=\color{purple},
  emphstyle=\color{green},
  frame=single,
  rulecolor=\color{solarized@base2},
  rulesepcolor=\color{solarized@base2},
  literate = {~}{\customtilde}1
             {\ as\ }{{\color{blue}\ as\ \color{black}}}3
}

\lstdefinestyle{fortran}
{
  language=Fortran,
  basicstyle=\footnotesize\ttfamily,
  basewidth={0.53em,0.44em},
  numbers=none,
  tabsize=2,
  breaklines=true,
  escapeinside={@}{@},
  showstringspaces=false,
  numberstyle=\tiny\color{solarized@base01},
  keywordstyle=\color{blue},
  stringstyle=\color{orange}\ttfamily,
  identifierstyle=\color{Periwinkle},
  commentstyle=\color{purple},
  emphstyle=\color{green},
  morekeywords={and, or, true, false},
  frame=single,
  rulecolor=\color{solarized@base2},
  rulesepcolor=\color{solarized@base2},
  literate={~}{\customtilde}1
}

\lstdefinestyle{terminal}
{
  language=bash,
  basicstyle=\termplainstyle,
  numbers=none,
  tabsize=2,
  breaklines=true,
  escapeinside={@}{@},
  frame=single,
  showstringspaces=false,
  numberstyle=\tiny\color{solarized@base01},
  keywordstyle=\color{solarized@orange},
  stringstyle=\color{solarized@red}\ttfamily,
  identifierstyle=\color{black},
  commentstyle=\color{solarized@violet},
  emphstyle=\color{solarized@green},
  frame=single,
  rulecolor=\color{solarized@base2},
  rulesepcolor=\color{solarized@base2},
  morekeywords={gambit, cmake, make, mkdir},
  deletekeywords={test},
  literate = {\ gambit}{{\ }{\color{black}}gambit}7
             {/gambit}{{/}{\color{black}}gambit}6
             {gambit/}{{\color{black}}gambit{/}}6
             {/include}{{/}{\color{black}}include}8
             {cmake/}{{\color{black}}cmake/}6
             {.cmake}{{.}{\color{black}}cmake}6
             {~}{\customtilde}1
}

\lstdefinestyle{terminalalt}
{
  language=bash,
  basicstyle=\footnotesize\ttfamily,
  numbers=none,
  tabsize=2,
  breaklines=true,
  escapeinside={*@}{@*},
  frame=single,
  showstringspaces=false,
  numberstyle=\tiny\color{solarized@base01},
  keywordstyle=\color{solarized@orange},
  stringstyle=\color{solarized@red}\ttfamily,
  identifierstyle=\color{black},
  commentstyle=\color{solarized@violet},
  emphstyle=\color{solarized@green},
  frame=single,
  rulecolor=\color{solarized@base2},
  rulesepcolor=\color{solarized@base2},
  morekeywords={gambit, cmake, make, mkdir},
  deletekeywords={test},
  literate = {\ gambit}{{\ }{\color{black}}gambit}7
             {/gambit}{{/}{\color{black}}gambit}6
             {gambit/}{{\color{black}}gambit{/}}6
             {/include}{{/}{\color{black}}include}8
             {cmake/}{{\color{black}}cmake/}6
             {.cmake}{{.}{\color{black}}cmake}6
             {~}{\customtilde}1
}

\lstdefinestyle{text}
{
  language={},
  basicstyle=\footnotesize\ttfamily,
  identifierstyle=\color{black},
  numbers=none,
  tabsize=2,
  breaklines=true,
  escapeinside={*@}{@*},
  showstringspaces=false,
  frame=single,
  rulecolor=\color{solarized@base2},
  rulesepcolor=\color{solarized@base2},
  literate={~}{\customtilde}1
}

\lstdefinestyle{yaml}
{
  language=bash,
  escapeinside={@}{@},
  keywords={true,false,null},
  otherkeywords={},
  keywordstyle=\color{solarized@base0}\bfseries,
  basicstyle=\footnotesize\color{black}\ttfamily,
  identifierstyle=\YAMLkeystyle,
  sensitive=false,
  commentstyle=\color{solarized@orange}\ttfamily,
  morecomment=[l]{\#},
  morecomment=[s]{/*}{*/},
  stringstyle=\YAMLstringstyle\ttfamily,
  moredelim=**[s][\YAMLkeystyle]{,}{:},   % switch to value style at : but back to key style at,
  moredelim=**[l][\YAMLvaluestyle]{:},    % switch to value style at :
  morestring=[b]',
  morestring=[b]",
  literate =    {---}{{\ProcessThreeDashes}}3
                {>}{{\textcolor{solarized@red}\textgreater}}1
                {|}{{\textcolor{solarized@red}\textbar}}1
                {\ -\ }{{\mdseries\color{black}\ -\ \negmedspace}}3
                {\}}{{{\color{black} \}}}}1
                {\{}{{{\color{black} \{}}}1
                {[}{{{\color{black} [}}}1
                {]}{{{\color{black} ]}}}1
                {~}{\customtilde}1,
  breakindent=0pt,
  breakatwhitespace,
  columns=fullflexible
}

\lstdefinestyle{mathematica}
{
  language={Mathematica},
  basicstyle=\footnotesize\ttfamily,
  basewidth={0.53em,0.44em},
  numbers=none,
  tabsize=2,
  breaklines=true,
  escapeinside={@}{@},
  numberstyle=\tiny\color{black},
  showstringspaces=false,
  numberstyle=\tiny\color{solarized@base01},
  keywordstyle=\color{solarized@orange},
  stringstyle=\color{solarized@red}\ttfamily,
  identifierstyle=\color{solarized@orange}\ttfamily,
  commentstyle=\color{solarized@gray}\ttfamily,
  directivestyle=\color{solarized@orange}\ttfamily,
  emphstyle=\color{solarized@green},
  frame=single,
  rulecolor=\color{solarized@base2},
  rulesepcolor=\color{solarized@base2},
  literate={~} {\customtilde}1,
  moredelim=*[directive]\ \ \#,
  moredelim=*[directive]\ \ \ \ \#,
  mathescape=true
}

\newcommand{\cross}[1]{\ref{#1}}
\newcommand{\doublecross}[2]{\hyperref[#2]{\textbf{#1}}}
\newcommand{\doublecrosssf}[2]{\hyperref[#2]{\textbf{\textsf{#1}}}}
\newcommand{\gitem}[1]{\item[\textbf{#1}\label{#1}]}

\newcommand{\startglossary}{\section{Glossary}\label{glossary}Here we explain some terms that have specific technical definitions in \GB.\begin{description}}
\newcommand{\finishglossary}{\end{description}}

\newcommand{\bcode}{\begin{lstlisting}}
\newcommand{\ecode}{\end{lstlisting}}
\newcommand{\metavarf}[1]{\textit{\color{darkgreen}\footnotesize\textrm{#1}}}
\newcommand{\metavars}[1]{\textit{\color{darkgreen}\scriptsize\textrm{#1}}}
\newcommand{\metavar}{\metavarf}

\DeclareMathOperator\erf{erf}

\newcommand{\eV}{\ensuremath{\text{e}\mspace{-0.8mu}\text{V}}\xspace}

\newcommand{\GeV}{\text{G\eV}\xspace}

\newcommand{\gambit}{\textsf{GAMBIT}\xspace}

\newcommand{\darkbit}{\textsf{DarkBit}\xspace}
\newcommand{\colliderbit}{\textsf{ColliderBit}\xspace}

\newcommand{\specbit}{\textsf{SpecBit}\xspace}

\newcommand{\GB}{\gambit}
\newcommand{\DB}{\darkbit}
\newcommand{\omp}{\textsf{OpenMP}\xspace}

\newcommand{\pythia}{\textsf{Pythia}\xspace}

\newcommand{\ds}{\textsf{DarkSUSY}\xspace}
\newcommand{\darksusy}{\ds}

\newcommand{\threebit}{\textsf{3-BIT-HIT}\xspace}
\newcommand{\pppc}{\textsf{PPPC4DMID}\xspace}
\newcommand{\mo}{\micromegas}
\newcommand{\micromegas}{\textsf{micrOMEGAs}\xspace}

\newcommand{\feynhiggs}{\textsf{FeynHiggs}\xspace}

\newcommand\flexiblesusy{\FlexibleSUSY}
\newcommand\FlexibleSUSY{\textsf{FlexibleSUSY}\xspace}

\newcommand\SOFTSUSY{\textsf{SOFTSUSY}\xspace}

\newcommand\SUSYHIT{\textsf{SUSY-HIT}\xspace}

\newcommand\nulike{\textsf{nulike}\xspace}
\newcommand\gamLike{\textsf{gamLike}\xspace}
\newcommand\gamlike{\gamLike}
\newcommand\daFunk{\textsf{daFunk}\xspace}

\newcommand\ddcalc{\textsf{DDCalc}\xspace}
\newcommand\tpcmc{\textsf{TPCMC}\xspace}
\newcommand\nest{\textsf{NEST}\xspace}
\newcommand\luxcalc{\textsf{LUXCalc}\xspace}
\newcommand\xx{\raisebox{0.2ex}{\smaller ++}\xspace}
\newcommand\Cpp{\textsf{C\xx}\xspace}

\newcommand\plainC{\textsf{C}\xspace}

\newcommand\Fortran{\textsf{Fortran}\xspace}
\newcommand\YAML{\textsf{YAML}\xspace}

\newcommand\beq{\begin{equation}}
\newcommand\eeq{\end{equation}}

\renewcommand{\url}[1]{\href{#1}{#1}}

\newcommand{\Enr}{\ensuremath{E}}
\newcommand{\dRdE}{\ensuremath{\frac{dR}{dE}}}
\newcommand{\dRdEnr}{\ensuremath{\frac{dR}{dE}}}

\newcommand{\mchi}{\ensuremath{m_{\chi}}}
\newcommand{\rhochi}{\ensuremath{\rho_{0}}}
\newcommand{\vmin}{\ensuremath{v_\mathrm{min}}}
\newcommand{\vmp}{\ensuremath{v_0}}
\newcommand{\vrot}{\ensuremath{v_\mathrm{rot}}}
\newcommand{\vobs}{\ensuremath{v_\mathrm{obs}}}

\newcommand{\vesc}{\ensuremath{v_\mathrm{esc}}}

\newcommand{\bv}{\ensuremath{\mathbf{v}}}  % bold v (velocity vector)
\newcommand{\bvsunpec}{\ensuremath{\mathbf{v}_{\odot,\mathrm{pec}}}}
\newcommand{\bvLSR}{\ensuremath{\mathbf{v}_{\mathrm{LSR}}}}

\newcommand{\qmax}{\ensuremath{q_{\mathrm{max}}}}  % max recoil momentum

\newcommand{\fpSI}{\ensuremath{f_{\mathrm{p}}}}
\newcommand{\fnSI}{\ensuremath{f_{\mathrm{n}}}}

\newcommand{\apSD}{\ensuremath{a_{\mathrm{p}}}}
\newcommand{\anSD}{\ensuremath{a_{\mathrm{n}}}}

\newcommand{\GNSI}{\ensuremath{G_{\scriptscriptstyle\mathrm{SI}}^{N}}}
\newcommand{\GpSI}{\ensuremath{G_{\scriptscriptstyle\mathrm{SI}}^{\mathrm{p}}}}
\newcommand{\GnSI}{\ensuremath{G_{\scriptscriptstyle\mathrm{SI}}^{\mathrm{n}}}}

\newcommand{\GNSD}{\ensuremath{G_{\scriptscriptstyle\mathrm{SD}}^{N}}}
\newcommand{\GpSD}{\ensuremath{G_{\scriptscriptstyle\mathrm{SD}}^{\mathrm{p}}}}
\newcommand{\GnSD}{\ensuremath{G_{\scriptscriptstyle\mathrm{SD}}^{\mathrm{n}}}}

\newcommand{\sigmaSI}{\ensuremath{\sigma_{\mathrm{SI}}}}
\newcommand{\sigmaSD}{\ensuremath{\sigma_{\mathrm{SD}}}}

\newcommand{\sigmapSI}{\ensuremath{\sigma_{\mathrm{SI,p}}}}
\newcommand{\sigmanSI}{\ensuremath{\sigma_{\mathrm{SI,n}}}}

\newcommand{\sigmapSD}{\ensuremath{\sigma_{\mathrm{SD,p}}}}
\newcommand{\sigmanSD}{\ensuremath{\sigma_{\mathrm{SD,n}}}}

\newcommand{\mup}{\ensuremath{\mu_{\mathrm{p}}}}

\newcommand{\Sp}{\ensuremath{\langle S_{\mathrm{p}} \rangle}}
\newcommand{\Sn}{\ensuremath{\langle S_{\mathrm{n}} \rangle}}
\newcommand{\Spp}{\ensuremath{S_{\mathrm{pp}}}}
\newcommand{\Spn}{\ensuremath{S_{\mathrm{pn}}}}
\newcommand{\Snn}{\ensuremath{S_{\mathrm{nn}}}}

\newcommand*\justify{%
  \fontdimen2\font=0.4em% interword space
  \fontdimen3\font=0.2em% interword stretch
  \fontdimen4\font=0.1em% interword shrink
  \fontdimen7\font=0.1em% extra space
  \hyphenchar\font=`\-% allowing hyphenation
}

\begin{document}

\preprintnumber{DESY-17-235, NORDITA-2017-076}

\title{DarkBit: A GAMBIT module for computing dark matter observables and likelihoods}

\titlerunning{DarkBit}

\author{The GAMBIT Dark Matter Workgroup:
Torsten Bringmann\thanksref{inst:a,e1} \and
Jan Conrad\thanksref{inst:b,inst:c} \and
Jonathan M.~Cornell\thanksref{inst:d,e2} \and
Lars A.~Dal\thanksref{inst:a} \and
Joakim Edsj\"o\thanksref{inst:b,inst:c} \and
Ben Farmer\thanksref{inst:b,inst:c} \and
Felix Kahlhoefer\thanksref{inst:f} \and
Anders Kvellestad\thanksref{inst:g} \and
Antje Putze\thanksref{inst:h} \and
Christopher Savage\thanksref{inst:g} \and
Pat Scott\thanksref{inst:i,e3} \and
Christoph Weniger\thanksref{inst:j,e4} \and
Martin White\thanksref{inst:k,inst:l} \and
Sebastian Wild\thanksref{inst:f}
}

\institute{%
  \oslo\label{inst:a} \and
  \okc\label{inst:b} \and
  \su\label{inst:c} \and
  \mcgill\label{inst:d} \and
  \desy\label{inst:f} \and
  \nordita\label{inst:g} \and
  \annecy\label{inst:h} \and
  \imperial\label{inst:i} \and
  \grappa\label{inst:j} \and
  \adelaide\label{inst:k} \and
  \coepp\label{inst:l}
}

\thankstext{e1}{torsten.bringmann@fys.uio.no}
\thankstext{e2}{cornellj@physics.mcgill.ca}
\thankstext{e3}{p.scott@imperial.ac.uk}
\thankstext{e4}{c.weniger@uva.nl}

\titlerunning{DarkBit}
\authorrunning{GAMBIT Dark Matter Workgroup}

\date{Received: date / Accepted: date}

\maketitle

\begin{abstract}
We introduce \DB, an advanced software code for computing dark matter constraints on various
extensions to the Standard Model of particle physics, comprising both new native code and
interfaces to external packages. This release includes a dedicated signal yield
calculator for gamma-ray observations, which significantly extends current tools by
implementing a cascade decay Monte Carlo, as well as a dedicated likelihood calculator for current and future experiments (\gamlike).
This provides a general solution for studying complex particle physics
models that predict dark matter annihilation to a multitude of final states. We also supply
a direct detection package that models a large range of direct detection
experiments (\ddcalc), and provides the corresponding likelihoods for arbitrary combinations of
spin-independent and spin-dependent scattering processes. Finally, we provide custom
relic density routines along with interfaces to \ds, \micromegas, and the neutrino telescope
likelihood package \nulike. \DB is written in the framework of the Global And Modular
Beyond the Standard Model Inference Tool (\GB), providing seamless integration into a
comprehensive statistical fitting framework that allows users to explore new models with both
particle and astrophysics constraints, and a consistent treatment of systematic
uncertainties. In this paper we describe its main functionality,
provide a guide to getting started quickly, and show illustrative examples for results obtained
with \DB (both as a standalone tool and as a \GB module). This includes a quantitative
comparison between two of the main dark matter codes (\ds and \micromegas), and
application of \darkbit's advanced direct and indirect detection routines to a
simple effective dark matter model.
\end{abstract}

\tableofcontents

%%%%%%%%%%%%%%%%%%%%%%%%%%%%%%%%%%%%%%%%%%%%%%%%%%%%%%%
%%%%%%%%%%%%%%%%%%%%%%%%%%%%%%%%%%%%%%%%%%%%%%%%%%%%%%%
\section{Introduction}
\label{intro}

The identity of dark matter (DM) remains one of the most vexing puzzles of fundamental physics.
After decades of intense effort, its cosmological abundance has been determined at a precision of better than one percent
\cite{Planck15cosmo}, but so far no experiment has reported any clear evidence of
its non-gravitational interactions. Despite these null searches, the leading hypothesis remains that DM
consists of a new type of
elementary particle \cite{Bertone:2010zza}. Out of the many possibilities
\cite{Steffen:2008qp,Feng:2010gw,Baer:2014eja}, weakly interacting massive particles
(WIMPs) are often argued to be particularly appealing candidates, both because they almost
inevitably appear in well-motivated extensions of the Standard Model -- like supersymmetry
\cite{Jungman:1995df} or universal extra dimensions \cite{Hooper:2007qk} -- but also because their
thermal production in the early Universe naturally results in a relic abundance in broad agreement with the observed DM density today.

Traditionally, the particle identity of DM has been tested with three different strategies: {\it i)} by
trying to directly produce it in {\it accelerator searches}, {\it ii)} by performing {\it direct searches} for recoiling nuclei caused by collisions with passing DM particles in large underground detectors, or {\it iii)} by {\it indirect
searches} for the debris from DM annihilation or decay in the Sun or outer space.
Although these approaches are particularly suitable for WIMPs, several other candidates
can be probed by some of these methods as well. More recently, another approach has emerged
that is particularly relevant for DM scenarios beyond the standard WIMP case,
e.g.~for self-interacting DM,
namely to {\it iv)} use {\it astrophysical probes} related to the distribution of matter on galactic and cosmological
scales \cite{Tulin:2013teo,Cyr-Racine:2015ihg}.
For each of these methods, an immense amount of experimental data is expected during the next decade(s).
In order to extract the maximal amount of information and narrow down the properties of a given DM
candidate, or exclude it, it is mandatory to combine these measurements in a statistically rigorous way.

With this article we introduce \DB, a new numerical tool for tackling this task.  \DB calculates DM
observables and likelihoods in a comprehensive and flexible way, making them available for both phenomenological DM studies and broader Beyond-the-Standard Model (BSM) global fits.   In particular, the first release of \DB
contains up-to-date limits and likelihoods for indirect DM searches with gamma rays and neutrinos, for the spin-dependent and spin-independent cross-sections relevant to direct detection, and for the
relic density. In order to increase the efficiency of observable and likelihood calculations by reusing as much code as possible,
\DB relies on highly flexible data structures that can easily accommodate the specific needs of most particle
models. Examples include the Process Catalogue (Sec.\ \ref{process_catalog}), which contains all relevant particles and
interaction rates, a general halo model, and a fully model-independent framework to calculate the relic
density.

\begin{figure*}[t]
  \centering
  \includegraphics[width=0.60\linewidth]{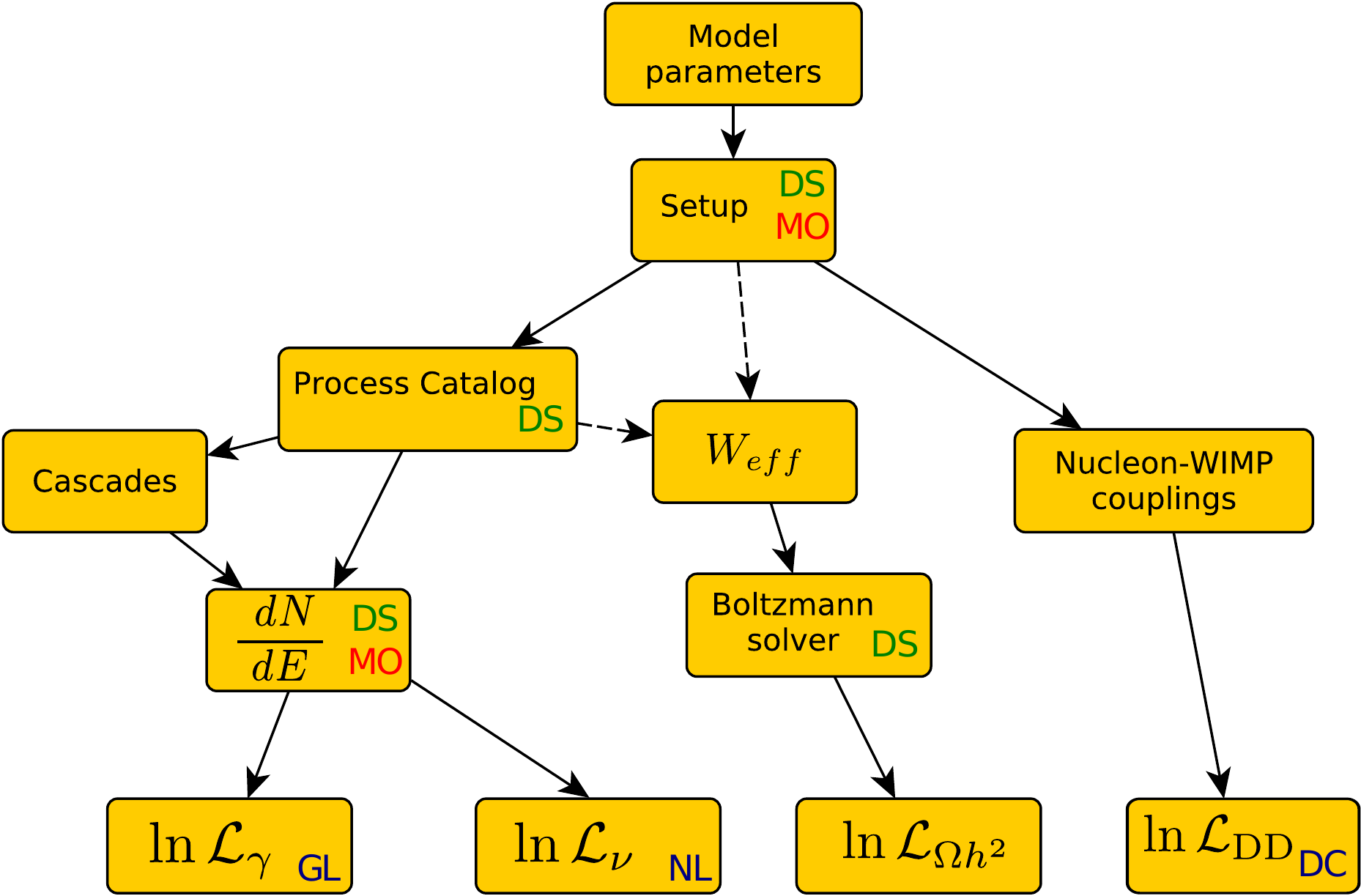}
  \caption{Schematic overview over different \DB components.  Based on
    the parameters of the scanned model(s), a Process Catalogue is initialised.  This
    contains the relevant particle physics processes to infer dark matter
    annihilation spectra.  The effective annihilation rate relevant for relic
    density calculations can (in simple cases) be inferred from the process
    catalogue, or it can be set directly.  WIMP-nucleon couplings are also set
    depending on the model.  All information is channelled into various
    likelihood routines.  The two-letter insets indicate what backend codes can
    be used:  \ds (DS), \micromegas (MO), \gamLike (GL), \nulike (NL) and
    \ddcalc (DC). Currently the neutrino spectra can calculated in \DB using only \ds,
    while the gamma-ray spectra can be taken from either \ds or \micromegas.
    }
  \label{fig:flow}
\end{figure*}

\DB is designed as a \cross{module} within the \GB framework \cite{gambit,ColliderBit,FlavBit,SDPBit,ScannerBit}.  Where we introduce key terms with specific meanings in the context of \GB, we \doublecross{highlight}{glossary} and link them to the \doublecross{glossary}{glossary} of standard \GB terms at the end of this paper.  \GB defines a series of \doublecross{physics modules}{physics module}, each consisting of a collection of \doublecross{module functions}{module function}.  Each module function is able to compute an observable, a likelihood or some intermediate quantity required in the calculation of other observables or likelihoods.  At runtime, the user informs \GB of the observables they want to compute, the theoretical model and parameter ranges over which those observables should be calculated, and how they would like \GB to sample the parameter space.  \GB then identifies the module functions necessary for delivering the requested observables and arranges them into a \cross{dependency tree}, describing which module functions must be run, and in what order.  It then chooses parameter combinations, passes them to the module functions, and outputs the resulting samples.  Module functions can call on additional functions from external \cross{backend} codes, which \GB also sorts into the dependency tree.  Rules that dictate how different module functions, models and backends may rely on each other can be defined in either the source code or input file, allowing the user to force the resulting dependency tree to obey arbitrarily detailed physical conditions and relations.

One of the main design features of \DB, as compared to other DM codes, is its extremely
modular structure. This allows users to interface essentially any external code in a straightforward way, allowing them to
extract any given functionality for use within \DB. Examples of such \doublecross{backends}{backend} used in the first release
include \ds \cite{darksusy} and \micromegas \cite{micromegas_nu} (for various direct and indirect rates, as
well as Boltzmann solvers),
\ddcalc
(introduced in Sec.\ \ref{sec:ddcalc}; for direct detection rates and likelihoods), \gamlike (introduced in Sec.\ \ref{sec:gamlike}; for gamma-ray likelihoods) and \nulike \cite{IC79_SUSY} (for
neutrino likelihoods). In a situation where several backends provide the same functionality, the user
can easily switch between them -- or instead use powerful \DB internal routines, like an on-the-fly cascade decay spectrum generator (which we have implemented from scratch).
On the technical side, \DB makes use of dynamic \Cpp function objects to
facilitate the manipulation and exchange of real-valued functions between
different backends and \GB; these are implemented in the \daFunk library
(Appendix~\ref{sec:dafunk}).

In standalone mode, \DB can be used for directly computing observables and likelihoods, for any combination of parameter values in some
underlying particle model.
When employed as a \GB module, it can be used to do this over an entire parameter space of a
chosen BSM model, providing various independent likelihoods for automatic combination with those from other
\GB modules in a statistically consistent way.
This usage mode makes it possible to not only simultaneously include all possible
constraints from different observation channels, but also to incorporate the full uncertainties
arising from poorly-constrained astrophysical or nuclear parameters in the scan, by treating them as nuisance parameters.  Typical examples include nuclear form factors and halo model uncertainties.

This article is organised as follows. In Section \ref{sec:overview} we give an
overview of the physics, observables and likelihoods contained in \DB, along
with the basic module structure.  Section \ref{sec:halo} details the DM halo
models that we employ in \DB.  In Sections \ref{sec:rd}, \ref{sec:dd} and
\ref{sec:id} we go through \DB's abilities in relic density calculations,
direct and indirect detection, respectively.  In particular, Section
\ref{sec:ddcalc} introduces the new direct detection pheno-likelihood code
\ddcalc and Section \ref{sec:gamlike} introduces the new gamma-ray indirect
detection likelihood code \gamlike. We show validation tests and illustrative examples of typical \DB usage in Section \ref{examples}. We continue with an outlook on planned code expansions in Section \ref{sec:out}, and conclude in Section \ref{conc}. In Appendix \ref{code_init} we provide a quick-start guide to installing \DB and running a simple test example.
Appendix \ref{sec:dafunk} introduces the new dynamic functions library \daFunk.  Appendix \ref{glossary} provides a glossary containing the \GB terms used in this paper.

The source code for \darkbit is available from \href{http://gambit.hepforge.org}{gambit.hepforge.org}, and is released under the terms of the standard 3-clause BSD license.\footnote{\href{http://opensource.org/licenses/BSD-3-Clause}{http://opensource.org/licenses/BSD-3-Clause}.  Note that \textsf{fjcore} \cite{Cacciari:2011ma} and some outputs of \flexiblesusy \cite{Athron:2014yba} (incorporating routines from \SOFTSUSY \cite{Allanach:2001kg}) are also shipped with \GB \textsf{1.0} and \textsf{1.1}.  These code snippets are distributed under the GNU General Public License (GPL; \href{http://opensource.org/licenses/GPL-3.0}{http://opensource.org/licenses/GPL-3.0}), with the special exception, granted to \GB by the authors, that they do not require the rest of \GB to inherit the GPL.}

%%%%%%%%%%%%%%%%%%%%%%%%%%%%%%%%%%%%%%%%%%%%%%%%%%%%%%%
%%%%%%%%%%%%%%%%%%%%%%%%%%%%%%%%%%%%%%%%%%%%%%%%%%%%%%%
\section{Module overview}
\label{sec:overview}

\GB is built around the ideas of modularity and re-usability.  \textit{All}
calculations required to
get from experimental data and model parameters to likelihood values and
observables are
performed in separate \GB~\doublecross{module functions}{module function}.  Each module function is able to calculate
exactly one quantity, its \cross{capability}.  The \cross{type} of this
capability can be just about anything, from a simple integer to any complex \Cpp
structure required to carry the result of the calculation.  Examples
of capabilities are: model parameters, particle spectra,
experimental data, and the values of likelihood functions.  Most module
functions will also have \doublecross{dependencies}{dependency} on capabilities that were
calculated by other module functions.  These dependencies will be automatically resolved at
run time, based on choices of the user about what particle physics model to
analyse, what observables to include etc.
%In general, the dependency tree of
%module functions used for a specific scan will form a
%directed a-cyclic graph.
For details, we refer the reader to the main \GB\
paper~\cite{gambit}.

\DB is one of the central \GB modules, and essentially a collection of module
functions that compute dark matter observables and likelihoods.
Some of the basic elements of \DB and their relations are sketched in
Fig.~\ref{fig:flow} (the full dependency tree with all module functions is
far more complex than this sketch).  Based on the model parameters of a particular point in
the scan, \DB sets up several central computational structures (like the Process
Catalogue, the effective annihilaton rate, and WIMP-nucleon couplings), which are used
for relic density calculation (see Sec.~\ref{sec:rd}),
the calculation of
direct detection constraints (Sec.~\ref{sec:dd}),
and the calculation of
gamma-ray and neutrino yields (Sec.~\ref{sec:id}).
%In many of the steps,
%\DB relies on the functionality of various backends, which are also indicated
%in the figure.

In many cases, and as indicated in Fig.~\ref{fig:flow}, the calculations
performed by \DB module functions build on functionality of external independent
codes like \ds or \mo.  These codes provide a
lot of functionality that can often be used beyond their original scope
(examples are Boltzmann solvers, tabulated particle yields, routines to
calculate J-values, etc).  From the perspective of \GB, these codes are
\doublecross{backends}{backend}.  Technically, they are coupled to \GB by compiling them as
shared libraries, which are loaded by \GB at runtime.
The interface to these backends is provided by convenient
\doublecross{frontends}{frontend}, which specify the form and subset of functionality of the
backend that is accessible by \GB.  For details we again refer to
Ref.~\cite{gambit}.

Although the main purpose of \DB is to provide dark matter-related
functionality for global scans with \GB, it can also be used as a
standalone code.  In fact, most of the functionality of \DB could also
be used outside of scans, e.g.~implemented in a command line tool, if so
desired.  We will show a few examples for this below in
Sec.~\ref{sec:simpleWIMP}.

%%%%%%%%%%%%%%%%%%%%%%%%%%%%%%%%%%%%%%%%%%%%%%%%%%%%%%%
%%%%%%%%%%%%%%%%%%%%%%%%%%%%%%%%%%%%%%%%%%%%%%%%%%%%%%%
\section{Halo Modeling}
\label{sec:halo}

\subsection{Background}
All of the direct and indirect detection observables that can be calculated in \DB are strongly dependent on
the spatial distribution of DM particles, and often velocities as well.  Predicted event rates in direct detection experiments
and the rate of DM annihilation in the Sun depend on the local
density of dark matter in the Milky Way. Indirect detection signals from
annihilations to gamma rays and neutrinos (and charged cosmic rays) depend on
the spatial DM distribution in the source being observed. In order to assure
consistency in the calculations of these observables, \GB contains halo models
that describe the density and velocity of DM in the Milky Way and other
astronomical objects.

\subsubsection{Density profiles}
Multiple forms of halo density profiles exist in the literature. Early analytic calculations of infalling dark matter onto collapsed density perturbations showed that the dark matter density should approximately scale like $r^{-2}$ \cite{Bertschinger:1985pd}, the same behaviour that one would expect for a halo with a constant velocity dispersion (or equivalently, constant temperature). This led to the modelling of the dark matter distribution by the modified isothermal profile
\begin{equation}
\rho(r) = \frac{2 \rho_s}{1 + (r/r_s)^2} \, ,
\label{eqn:isothermal}
\end{equation}
where $r_s$ is a scale radius, and $\rho_s$ is the density at $r=r_s$.

Subsequent $N$-body simulations of the gravitational interactions of dark matter lead Navarro, Frenk, and White (NFW) to conclude that the structure of dark matter halos could be described by a cusped profile of the form \cite{Navarro:1995iw}
\begin{equation}
\rho(r) = \frac{4 \rho_s}{(r/r_s) \left[1 + (r/r_s) \right]^2} \, .
\label{Eq:NFW}
\end{equation}
They showed that the dark matter density profile appears to be roughly universal, meaning that for halos of a range of sizes from dwarf galaxies to galaxy clusters, the form of the profile is the same. Other groups \cite{Kravtsov:1997dp,Moore:1999gc} found better fits to simulation data could be obtained by slightly modifying the NFW profile. To take these modifications into account, it is common to write the halo profile in the following form:
\begin{equation}
\rho(r) = \frac{2^{(\beta - \gamma)/\alpha} \rho_s}{(r/r_s)^\gamma \left[1 + (r/r_s)^\alpha \right]^{(\beta - \gamma) / \alpha}} \, .
\label{eqn:gNFW}
\end{equation}
Here, $\gamma$ describes the inner slope of the profile, $\beta$ the outer slope, and
$\alpha$ the shape in the transition region around $r\sim r_s$.

More recently, it has been pointed out that a better fit to $N$-body simulations can be given by what is known as an Einasto profile \cite{Graham:2005xx}, named after Einasto's use of the profile to model the mass distribution of galaxies \cite{Haud:1986yj}. In this model the logarithm of the slope varies continuously with radius, leading to a density profile of the form
\begin{equation}
\rho(r) = \rho_s \exp \left\{-\frac{2}{\alpha} \left[ \left(\frac{r}{r_s}
\right)^\alpha -1 \right] \right\} \, .
\label{eqn:Einasto}
\end{equation}

\subsubsection{Velocity distribution}

The velocities $\mathbf{v}$ of dark matter particles in a halo are usually taken to follow the Maxwell-Boltzmann distribution for an ideal gas at constant temperature. This distribution, truncated to reflect the fact that any particle with a speed beyond the escape velocity $v_{\rm esc}$ will leave the halo, takes the form
\begin{equation}
\label{eq:MB}
\tilde{f}(\mathbf{v}) = \frac{1}{N_{\rm esc}} (\pi v_0^2)^{-3/2} e^{-\mathbf{v}^2/v_0^2} \, ,
\end{equation}
where $v_0$ is the most probable speed. Note that the above formula is only valid when $|\mathbf{v}| < v_{\rm esc}$; we assume that the probability of a speed higher than $v_{\rm esc}$ is 0. The normalisation that corrects for the truncation, $N_{\rm esc}$, is given by
\begin{equation}
N_{\rm esc} = \erf \left(\frac{v_{\rm esc}}{v_0} \right) - \frac{2 v_{\rm esc}}{\sqrt{\pi} v_0} \exp \left(- \frac{v_{\rm esc}^2}{v_0^2} \right) \, .
\end{equation}

Our Milky Way galaxy rotates in a dark matter halo that is essentially stationary. The velocity of the Earth in the halo is given by the sum
\begin{equation}
\mathbf{v}_{\rm obs} = \mathbf{v}_{\rm LSR} + \mathbf{v}_{\rm \odot, pec} + \mathbf{V}_\oplus(t) \, .
\end{equation}
Here $\mathbf{v}_{\rm LSR} = (0, v_{\rm rot}, 0)$ is the motion of the Local Standard of Rest in Galactic coordinates, $\mathbf{v}_{\rm \odot, pec} = (11, 12, 7)$\,km\,s$^{-1}$ is the well known peculiar velocity of the Sun \cite{Schoenrich:2009bx}, and $\mathbf{V}_\oplus(t)$ is the velocity of the Earth relative to the Sun. The magnitude of $\mathrm{V}_\oplus$ is well measured at 29.78\,km\,s$^{-1}$ \cite{Freese:2012xd}, and its  changing direction is expected to give rise to an annual modulation of scattering rates in direct detection experiments \cite{Drukier:1986tm}. The distribution of velocities $\mathbf{u}$ of dark matter particles in the Earth's frame is given by
\begin{equation}
f(\mathbf{u}, t) = \tilde{f}(\mathbf{v}_{\rm obs}(t) + \mathbf{u}) \, .
\label{eq:velocity_dist_shift}
\end{equation}
Assuming that the density profile of the halo surrounding the Milky Way is smooth and spherical like those discussed above, $v_0$ is approximately the same as the rotation speed $v_{\rm rot}$ of the galactic disk (for an isothermal density profile it is exactly the same, whereas for the NFW profile of Eq.\ \ref{Eq:NFW}, the two values can vary by over 10\% \cite{Serpico:2010ae}).

\subsection{Halo model implementation in \GB}

\subsubsection{Halo models and associated capabilities}

In \GB, the radial distribution $\rho(r)$ of dark matter in the Milky Way, the local density $\rho_0$, the distance from the Sun to the Galactic centre $r_{\rm sun}$, as well as the local velocity distribution $f(\vec{u})$ are simultaneously described by a given halo model. In the first release, we provide two main halo models: \yaml{Halo\_gNFW} and \yaml{Halo\_Einasto}. The former corresponds to the generalised NFW profile with parameters $\rho_s$, $r_s$, $\alpha$, $\beta$ and $\gamma$ as defined in Eq.\ \ref{eqn:gNFW}, together with the Maxwell-Boltzmann velocity distribution given by Eqs.\ \ref{eq:MB}--\ref{eq:velocity_dist_shift}, specified by the model parameters $v_0$, $v_{\rm rot}$ and $v_{\rm esc}$\footnote{The remaining velocity parameters $v_{\rm \odot, pec}$ and $|\mathbf{V}_\oplus(t)|$ are much better known. Hence, instead of being part of the halo models, $v_{\rm \odot, pec}$ simply defaults to the value assumed by the \ddcalc backend, (11, 12, 7)\,km\,s$^{-1}$, and $|\mathbf{V}_\oplus|$ to 29.78\,km\,s$^{-1}$.  The magnitude of $\mathbf{V}_\oplus$ can however be overridden in module functions that use it, by setting the \YAML option \yaml{v_earth}.}. Lastly, the model contains $r_{\rm sun}$ and $\rho_0$ as additional free parameters. Analogously, the \yaml{Halo\_Einasto} model describes the density profile given by Eq.\ \ref{eqn:Einasto}, with free parameters $\rho_s$, $r_s$ and $\alpha$, and otherwise is identical to the \yaml{Halo\_gNFW} model. Note that with appropriate choices of $\alpha$, $\beta$, and $\gamma$, the density profile in the \yaml{Halo\_gNFW} model is equivalent to the isothermal profile of Eq.\ \ref{eqn:isothermal} ($\alpha = 2$, $\beta = 2$, $\gamma = 0$) or the NFW profile of Eq.\ \ref{Eq:NFW} ($\alpha = 1$, $\beta = 3$, $\gamma = 1$).

In these two halo models, the density profile $\rho(r)$ relevant for calculating the gamma-ray flux induced by dark matter annihilations is completely decoupled from the local properties of dark matter (in particular from the local density $\rho_0$), which set the event rate in direct detection experiments and neutrino telescopes. However, it is also possible to directly link the density profile to the local density by enforcing the relation $\rho(r_{\rm sun}) \equiv \rho_0$. For the case of the generalised NFW profile, this is realised by two child models of \yaml{Halo\_gNFW}, denoted \yaml{Halo\_gNFW\_rho0} and \yaml{Halo\_gNFW\_rhos}. When employing the former model, the user need only specify the value of the local density $\rho_0$, which is then internally converted to the corresponding value of the scale density $\rho_s$ using Eq.\ \ref{eqn:gNFW}. Conversely, in the latter model one specifies $\rho_s$, and $\rho_0$ is determined by \GB. A completely analogous choice is possible for the Einasto profile via the halo models \yaml{Halo\_Einasto\_rho0} and \yaml{Halo\_Einasto\_rhos}.

In order to communicate the astrophysical properties of the Galactic dark matter population to the module and backend functions relevant for direct and indirect searches, \GB employs two central capabilities: (1) \cpp{GalacticHalo}, which is of type \cpp{GalacticHaloProperties}, a data structure containing (i) a \cpp{daFunk::Funk} object (Appendix~\ref{sec:dafunk}), which describes the radial density profile as a function of the radius \cpp{"r"}, and (ii) the distance $r_{\rm sun}$ from the Sun to the Galactic centre. This capability is required in particular by the \gamlike backend for the computation of gamma-ray fluxes from dark matter annihilations within the Milky Way. The other capability describing the dark matter halo is (2) \cpp{LocalHalo}, which is of type \cpp{LocalMaxwellianHalo}. This object is simply a container for the parameters relevant to direct detection and capture in the Sun, i.e.~the local density $\rho_0$ as well as the velocity parameters $v_0$, $v_{\rm rot}$ and $v_{\rm esc}$.

Depending on the halo model in use, the capability \cpp{GalacticHalo} can be provided by the module functions \cpp{GalacticHalo\_gNFW} or \cpp{GalacticHalo\_Einasto}; for all halo models, the \cpp{LocalHalo} capability is obtained through the module function \cpp{ExtractLocalMaxwellianHalo} (see also Table\ \ref{tab:halocap}). The rest of the \GB code is designed such that only the module functions providing the capabilities \cpp{GalacticHalo} or \cpp{LocalHalo} explicitly depend on a halo model, while all other module and backend functions requiring access to the astrophysical properties of dark matter instead depend on these capabilities. This setup allows \GB to be straightforwardly extended to incorporate new halo models, in particular to density profiles different from the generalised NFW and Einasto parameterisations.

\subsubsection{Likelihoods}

\GB provides several likelihood functions for the (typically quite large) uncertainties of the parameters included in the halo models.  These are summarised in Tables\ \ref{tab:HaloModelLikelihoods} and\ \ref{tab:halocap}. For the local dark matter density $\rho_0$, we implement the likelihood as a log-normal function:
\begin{equation}
\mathcal{L}_{\rho_0} =  \frac{1}{\sqrt{2\pi} \sigma'_{\rho_0} \rho_0} \exp \left(- \frac{\ln(\rho_0 / \bar\rho_0)^2}{2 {\sigma^{\prime 2}_{\rho_0}}} \right) \, ,
\end{equation}
where $\sigma'_{\rho_0} = \ln(1 + \sigma_{\rho_0}/\rho_0)$.

The methods that have been used to determine $\rho_0$ can be roughly categorised into two approaches (see \cite{Read:2014qva} for a review). Local measures (e.g \cite{Bovy:2012tw}) use the kinematics of nearby stars to find $\rho_0$, whereas global measures (e.g \cite{Caldwell:1981rj, Catena:2009mf, Salucci:2010qr, Pato:2015dua}) extrapolate the local density from the galactic rotation curve. The latter method often leads to results with smaller errors than the former, but they are very much dependent on assumptions about the shape of the halo \cite{Pato:2010yq}. There appears to be a growing consensus that $\rho_0 \approx 0.4 \ \mathrm{GeV}/\mathrm{cm}^3$, so by default we set $\bar\rho_0$ to this value. We take $\sigma_{\rho_0}$ to be $ 0.15 \ \mathrm{GeV}/\mathrm{cm}^3$, to represent the range of determinations of the local density in the literature (see e.g. \cite{Akrami:2010dn}).

\begin{table*}
  \centering
  \begin{tabular}{llcrrr}
    \toprule
    Parameter & Units & Likelihood Form & Central Value & Uncertainty & Function
    \\\midrule
    Local dark matter density ($\rho_0$) & GeV/${\rm cm}^3$ & log-normal & 0.4 & 0.15 & {\lstinline[style=yaml]!lnL_rho0_lognormal!}\\
    Maxwellian most-probable speed ($v_0$) & km\,s$^{-1}$ & Gaussian & 235 & 20 & {\lstinline[style=yaml]!lnL_v0_gaussian!}\\
    Local disk rotation speed ($v_{\rm rot}$) & km\,s$^{-1}$ & Gauusian & 235 & 20 & {\lstinline[style=yaml]!lnL_vrot_gaussian!}\\
    Local galactic escape speed ($v_{\rm esc}$) & km\,s$^{-1}$ & Gaussian & 550 & 35 & {\lstinline[style=yaml]!lnL_vesc_gaussian!}\\
    \bottomrule
  \end{tabular}
  \caption{Milky Way halo parameters used in \DB, the form of their likelihood
    functions, their central values and 1 $\sigma$ uncertainties, and
    the function that provides each likelihood. The capabilities associated
    with each likelihood are the same as the function name without the
    likelihood form, e.g.\ the capability for the $\rho_0$ likelihood is
    {\lstinline[style=yaml]!lnL_rh0!}.}
  \label{tab:HaloModelLikelihoods}
\end{table*}

We also provide likelihood functions for $v_0$, $v_{\rm rot}$, and $v_{\rm esc}$. For these parameters, we assume that the likelihood follows a standard Gaussian distribution
\begin{equation}
\mathcal{L}_x = \frac{1}{\sqrt{2 \pi} \sigma_x} \exp \left(- \frac{(x - \bar x)^2}{2 {\sigma_x}^2} \right) \, .
\label{eq:Gaussian}
\end{equation}
For the disk rotational velocity, by default the central value and error of the
distribution are set to $235 \pm 20$\,km\,s$^{-1}$, based on measurements of galactic masers \cite{Reid:2009nj, Bovy:2009dr}. To take into account the possible discrepancies between $v_0$ and $v_{\rm rot}$ due to variances in the density profile away from the simple isothermal model, we have implemented an independent Gaussian likelihood for $v_0$ with the same parameters as $v_{\rm rot}$. Finally, for the escape velocity we use $v_{\rm esc} = 550 \pm 35$\,km\,s$^{-1}$ based on measurements of high velocity stars in the RAVE survey \cite{Smith:2006ym}. The likelihood functions discussed in this section are listed in Tab.~\ref{tab:HaloModelLikelihoods}, along with their corresponding capabilities. The central values and errors for these likelihoods can be adjusted by setting the \YAML options \metavar{param}\yaml{_obs} and \metavar{param}\yaml{_obserr} respectively, where \metavar{param} is the name of the parameter (e.g.\ to override the default likelihood for $\rho_0$ one would set \yaml{rho0_obs} and \yaml{rho0_obserr}).

Currently, no specific likelihoods are included to constrain the Galactic halo
profile parameters.  This can, however, easily be done by specifying appropriate
parameter ranges and priors in the \GB initialisation file.  For typical
standard values we refer the reader to Ref.~\cite{pppc}; for recent kinematical
constraints we refer to Ref.~\cite{Pato:2015dua}.

%%%%%%%%%%%%%%%%%%%%%%%%%%%%%%%%%%%%%%%%%%%%%%%%%%%%%%%
%%%%%%%%%%%%%%%%%%%%%%%%%%%%%%%%%%%%%%%%%%%%%%%%%%%%%%%
\section{Relic Density}
\label{sec:rd}

%%%%%%%%%%%%%%%%%%%%%%%%%%%%%%%%%%%%%%%%%%%%%%%%%%%%%%%
\subsection{Background}

To calculate the relic density we need to solve the Boltzmann equation for the
number density $n$ of dark matter particles, which
in general can be written as \cite{Edsjo:1997bg}
\begin{equation} \label{eq:Boltzmann}
  \frac{dn}{dt} =
  -3Hn - \langle \sigma_{\rm{eff}} v \rangle
  \left( n^2 - n_{\rm{eq}}^2 \right)\;,
\end{equation}
where $n_\text{eq}$ denotes the equilibrium density, and $H$ the Hubble
constant.
Furthermore, the thermal average of the effective annihilation cross section is defined as
\begin{equation} \label{eq:sigmavefff}
  \langle \sigma_{\rm{eff}}v \rangle = \frac{\int_0^\infty
  dp_{\rm{eff}} p_{\rm{eff}}^2 W_{\rm{eff}} K_1 \left(
  \frac{\sqrt{s}}{T} \right) } { m_1^4 T \left[ \sum_i \frac{g_i}{g_1}
  \frac{m_i^2}{m_1^2} K_2 \left(\frac{m_i}{T}\right) \right]^2}\;,
\end{equation}
where $p_{\rm eff}=\frac12\sqrt{s-4m_1^2}$ is the effective momentum in the centre-of-momentum frame of the lightest DM species (assumed to be particle 1). $K_1$ and $K_2$ are modified Bessel functions, $g_i$ is the number of internal degrees of freedom of co-annihilating particle $i$, $m_i$ is its mass, $T$ is the temperature, and $s$ is the centre-of-momentum energy squared. Finally, $W_{\rm eff}$ is the effective annihilation rate, given by
\begin{eqnarray} \label{eq:weff}
  W_{\rm{eff}} & = & \sum_{ij}\frac{p_{ij}}{p_{11}}
  \frac{g_ig_j}{g_1^2} W_{ij} \\ \nonumber
  & \!\!\!\!\!\!\!\!= &
  \!\!\!\!\sum_{ij} \sqrt{\frac{[s-(m_{i}-m_{j})^2][s-(m_{i}+m_{j})^2]}
  {s(s-4m_1^2)}} \frac{g_ig_j}{g_1^2} W_{ij}\,.
\end{eqnarray}
Here,
\begin{equation}
  p_{ij}\equiv\sqrt{\frac{(p_i \cdot p_j)^2 - m_i^2 m_j^2}{s}}
\end{equation}
is the effective momentum for $ij$ annihilation,
 with $p_{11}=p_\mathrm{eff}$, and $W_{ij}$ is
related to the annihilation cross section by
\begin{eqnarray} \label{eq:Wijcross}
  W_{ij} &=& 4 p_{ij} \sqrt{s} \sigma_{ij}
  = 4 E_{i} E_{j} \sigma_{ij} v_{ij}\,,
\end{eqnarray}
where $E_i$ is the energy of particle $i$.

The main particle physics-specific
quantity, to be provided by the respective particle model, is thus the invariant rate
$W_\mathrm{eff}(p_\mathrm{eff})$. While its computation can be computationally
expensive for complex models, it is independent of temperature; it is thus usually
advantageous to tabulate this function (as done in e.g.\ \ds).

The integration of the Boltzmann equation,
Eq.~\ref{eq:Boltzmann}, can then proceed in a model-independent way and the
final relic density of the DM particle is given by
\begin{equation}
\label{ochi}
\Omega_\chi=m_\chi n_0/\rho_\mathrm{crit}\,.
\end{equation}
Here, $n_0$ is the asymptotic value of $n(t\to\infty)$ as expected today and $\rho_\mathrm{crit}=3H_0^2/8\pi G$. Note that there are two equivalent ways of dealing with a situation where there is more than one DM particle with the same mass $m_\chi$ -- like for example for Dirac particles where there would be both DM particles and {\it anti}particles. The first option is to treat the DM particles as separate species; the relic density given in Eq.~\ref{ochi} then only refers to the density of {\it one} of the species. Alternatively, all
DM particles can be treated as a single effective species with a correspondingly larger value of $g_1$ in
the definition of $W_\mathrm{eff}$ in Eq.~\ref{eq:Wijcross}; for the case of Dirac DM, e.g., one would have to replace $g_1\to2g_1$. In this case, the expression in Eq.~\ref{ochi} will refer to the {\it total} DM density.

Numerically, the integration of the Boltzmann equation is simplified by changing variables from
$n$ and $t$ to the dimensionless quantities $x\equiv m_\chi/T$ and $Y\equiv n/s$, where $s$
now denotes the entropy density of the heat bath \cite{Gondolo:1990dk}.
In \ds, the thermal average in Eq.~\ref{eq:sigmavefff} is done by using an adaptive
Gaussian method, employing splines to interpolate between the tabulated points of $W_\mathrm{eff}$ and taking
special care around the known locations of thresholds and resonances. The
actual integration of the Boltzmann equation is then performed via an implicit trapezoidal method with
adaptive stepsize; see \cite{darksusy} for further details.

%%%%%%%%%%%%%%%%%%%%%%%%%%%%%%%%%%%%%%%%%%%%%%%%%%%%%%%
\subsection{Interfaces to \ds and \micromegas}

Here we discuss the general features of the \GB interface to \ds and \micromegas, two of the most important backends for \DB.

\ds\footnote{\href{http://www.darksusy.org/}{http://www.darksusy.org}}
\cite{Gondolo:2004sc} has been fully implemented into the modular framework of
\GB, with the calculation of all observables broken down into discrete parts
that can be easily replaced with calculations from other backends. \ds is
used by \GB to obtain multiple theoretical quantities, including $W_{\rm eff}$,
the DM relic density (through a fully numerical solution of the Boltzmann
equation), effective couplings between nucleons and WIMPs, the rate of dark
matter capture in the Sun, and the spectra of gamma rays from DM annihilation.

If \ds is used as a backend for \DB, it is important that it be
correctly initialised at each point in the scan.  The most basic
model-independent initialisation happens in the \ds backend initialisation
function.  However, in order to ensure that MSSM observables are also
calculated correctly, \DB module functions that rely on the
model-dependent capabilities of \ds have an auxiliary
dependency on the capability \cpp{DarkSUSY_PointInit} (see Table~\ref{tab:darkbitmisc}).  It is then the
responsibility of the function that provides this capability to initialise
\ds correctly. A separate capability \cpp{DarkSUSY_PointInit_LocalHalo} is
provided to initialise the DM Halo model in \ds for those backend functions where
it is necessary.  The full set of relic density capabilities, functions and dependencies in \DB can be found in Tables \ref{tab:darkbitrelcap} and \ref{tab:darkbitmainrelcap}.

\textsf{MicrOMEGAs}\footnote{\href{https://lapth.cnrs.fr/micromegas/}{https://lapth.cnrs.fr/micromegas}}
\cite{Belanger:2013oya, Belanger:2010gh, Belanger:2008sj, Belanger:2006is} can
be used by \DB to obtain many of the same quantities as \ds, including the
relic density, WIMP-nucleon effective couplings, and gamma-ray spectra.
However, the interface of \micromegas with the \GB framework is currently more
coarse-grained than the one for \ds. An
illustrative example of this is the calculation of the relic density: in the
case of \ds, \GB calls different \ds functions for each part of the
calculation, including the calculation of $W_{\rm eff}$, and the solution of
the Boltzmann equation, whereas with \micromegas, the entire calculation is done
by calling one function.

As \micromegas is not set up to take information from the \DB Process Catalog, a
dark matter model must be implemented in \micromegas following the normal method for the
code. This consists of writing a compatible set of \textsf{CalcHEP} \cite{Belyaev:2012qa} model files and then
compiling \micromegas with these files. All of the relevant objects, both model
specific and then generic, are then combined into a shared library that is
used by the \micromegas frontend. As there are model-specific functions in
the library, separate libraries are needed for each particle physics model.
\GB comes with two \micromegas frontends: one for the MSSM
\cite{Belanger:2004yn, Belanger:2001fz} and one for the scalar singlet DM
model. The latter is not included by default with \micromegas, so we have
included the \textsf{CalcHEP} files needed to implement it with the \GB
distribution.

The intialisation functions for these frontends, \cpp{MicrOmegas_MSSM_3_6_9_2_init}
and \cpp{MicrOmegas_SingletDM} \cpp{_3_6_9_2_init}, both have the \YAML options
\yaml{VZdecay} and \yaml{VWdecay}. These control how \micromegas treats
annihilations to 3-body final states via virtual $W$ and $Z$ bosons. If these
options are set to 0 these processes are ignored, while a value of 1 causes them be
included in the case of DM self-annihilations, and with a value of 2 they are taken
into account for all coannihilation processes as well. To initialise the backend, \cpp{MicrOmegas_SingletDM_3_6_9_2_init} passes information about couplings, masses, and decays directly to \micromegas, while \cpp{MicrOmegas_MSSM_3_6_9_2_init} writes an SLHA1 file to disk, which is subsequently read by the \cpp{lesHinput} function of \cpp{MicrOmegas} (this function is not compatible with the more general SLHA2 format). The latter function also has the \YAML option
\yaml{internal_decays}, which, when set to \yaml{false}, causes information from the \GB
decay table to be passed via the \term{DECAY} blocks of the SLHA1 file to \micromegas; otherwise \micromegas calculates widths internally. Starting with \GB  \textsf{1.1.0}, this option is set to \yaml{false} by default.

\subsection{Relic density implementation in \DB}
\label{code_rd}

\begin{figure}[t]
  \centering
  \includegraphics[width=0.9\linewidth]{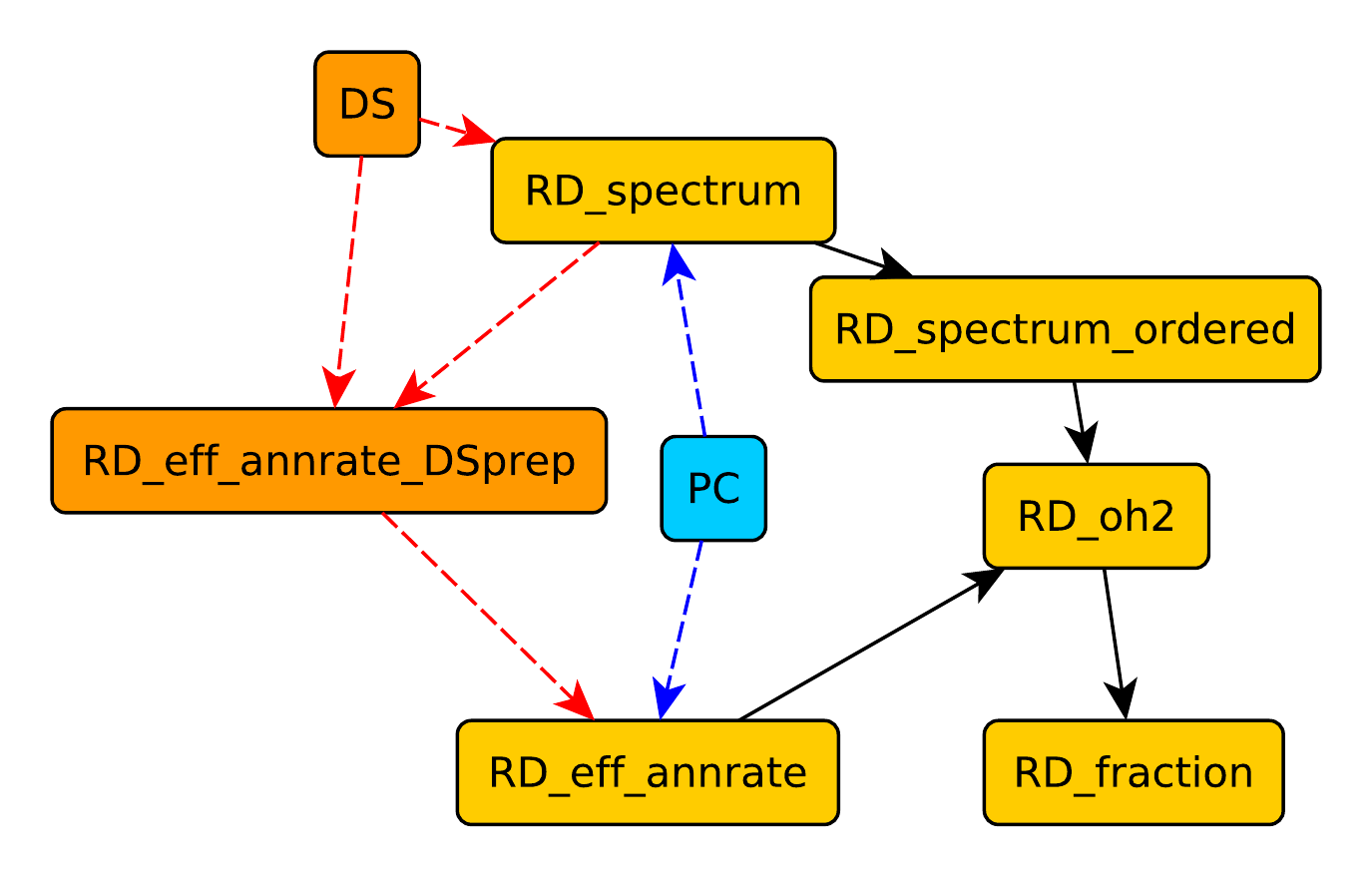}
  \caption{Relic density overview plot.  The capabilities required for relic
    density calculations are the effective annihilation rate $W_\mathrm{eff}$ and the ordered
    spectrum of coannihilating particles.  Currently, these quantities can be set up
    either directly with \ds (DS) or in simpler cases without
    coannihilation via the Process Catalogue (PC).  However, each of these
    functions can be easily replaced with a user-defined function.}
  \label{fig:RD_flow}
\end{figure}

The general structure and main capabilities of relic density calculations in \DB are
summarised in Fig.~\ref{fig:RD_flow}. As explained above, the most important particle physics input needed to calculate the
DM relic density is the effective invariant annihilation rate $W_\mathrm{eff}$.
In \DB, this is represented by the capability \cpp{RD_eff_annrate}.

Currently,
there are two functions that provide this capability (Table~\ref{tab:darkbitrelcap}), though
further user-defined functions can easily be added. The first function,
\yaml{RD_eff_annrate_SUSY}, returns $W_\mathrm{eff}$
for SUSY models using \ds as a backend. It depends on the capability
\yaml{RD_eff_annrate_DSprep}, which ensures  that \ds is
correctly set up and configured to provide $W_\mathrm{eff}$ for neutralino annihilation.
In this case two boolean options can be provided in the \YAML file:
\yaml{CoannCharginosNeutralinos} and \yaml{CoannSfermions}.  These specify whether
chargino, neutralino or sfermion coannihilations are taken into account.  These options default to \yaml{true}, but
the relevant coannihilations can be disabled to speed up the relic density calculation (at the expense of accuracy).\footnote{\ds does not currently include gluino coannihilations.}
Another important option to steer the numerical performance is \yaml{CoannMaxMass}
(default 1.6), which specifies up to what mass, in units of the DM mass, coannihilating particles
are taken into account when calculating $\langle \sigma_\mathrm{eff} v\rangle$.
The second alternative is to determine $W_\mathrm{eff}$ directly from the Process Catalogue (Sec.\ \ref{process_catalog}).
The corresponding function obviously has \yaml{TH_ProcessCatalogue} as a dependency, as
well as \yaml{DarkMatter_ID}, and can
be used for models where coannihilations are not important.
Here, the identity of the dark matter particle, as it exists in
the \GB particle database (see Ref.~\cite{gambit}), is provided by the
capability \cpp{DarkMatter_ID} (Table~\ref{tab:darkbitdirect}).

The main capability of the relic density part of \DB is \yaml{RD_oh2}, which returns
$\Omega_\chi h^2$ for a given model point.
If the user wishes, the result can be expressed as a fraction of the total, measured DM
density. This capability is \cpp{RD_fraction}, and is provided by three different module functions: \yaml{RD_fraction_one}, \yaml{RD_fraction_leq_one} and \yaml{RD_fraction_rescale}. The choice of function dictates whether indirect and direct detection routines should use the observed relic density or the one calculated for the particular parameter point.  \yaml{RD_fraction_one} simply returns 1, and causes the observed relic density to be used in all direct and indirect detection routines.  \yaml{RD_fraction_leq_one} returns the lesser of 1 and the ratio of the computed relic density to the observed value.  \yaml{RD_fraction_rescale} always returns the computed-to-reference ratio, regardless of whether or not it exceeds 1.  The latter two functions accept an option \yaml{"oh2\_obs"}, which defaults to \yamlvalue{0.1188}; this is the observed value of $\Omega h^2$ to use for rescaling.

\DB currently includes three
functions that provide \yaml{RD_oh2} (Table~\ref{tab:darkbitmainrelcap}). Two of those (\yaml{RD_oh2_DarkSUSY} and \yaml{RD_oh2_MicrOmegas}) are
direct calls to the unmodified relic density routines of \ds and \micromegas, with all particle
model parameters initialised as
per default in the respective backend code. \yaml{RD_oh2_DarkSUSY} is only compatible with the MSSM, while \yaml{RD_oh2_MicrOmegas} can be used with any model implemented in \micromegas. This latter function has two options: \yaml{fast}, which determines the numerical accuracy of the calculation (\cpp{int}; \yaml{0}:
 accurate  [default], \yaml{1}: fast), and \yaml{Beps} (\cpp{double}; default $10^{-5}$), which corresponds to a minimum value for the parameter
\begin{equation}
B = \exp \left(-\frac{T}{m_1} \frac{(m_i + m_j - 2 m_1)}{m_1} \right) \, .
\end{equation}
If $B$ is less than the \yaml{Beps} value for a certain process, that process will not be included in the calculation of the effective annihilation rate (this serves the same purpose as \yaml{CoannMaxMass} for \ds).
 The third function (\yaml{RD_oh2_general}) is a
general Boltzmann solver, which again largely relies on various subroutines provided by the \ds
backend, but not on any \ds-specific initialisation of particle parameters (e.g.~the
 setting of sparticle masses and couplings): the option \yaml{fast} (\cpp{int}; \yaml{0}:
 accurate, \yaml{1}: fast [default]) again steers the
 numerical performance of the backend code.
  In order to calculate $\langle\sigma_\mathrm{eff}\rangle$,
 c.f.~Eq.~\ref{eq:sigmavefff}, the Boltzmann solver needs not only the invariant rate, but also
 the internal
 degrees of freedom and masses of all (co)annihilating particles. For a high-precision result
 of this integral, one will in general also need to know the exact location of thresholds and resonances in
 $W_\mathrm{eff}$. \yaml{RD_oh2} therefore depends on the capability \cpp{RD_spectrum_ordered} which
 contains all this information, ordered by increasing $p_\mathrm{eff}$. \yaml{RD_spectrum_ordered} in turn
 depends on the capability  \yaml{RD_spectrum} which contains the same information, except for
 coannihilation thresholds, but not necessarily in an ordered form. Presently, \yaml{RD_spectrum} can be
 provided in two ways, either by
 the Process Catalogue (if coannihilations are not important; see Sec.\ \ref{process_catalog}) or by \ds.

Lastly, the capability of the likelihood function constraining the relic
density is \yaml{lnL_oh2}.  This capability is provided by two module
functions.  First, \yaml{lnL_oh2_Simple} is a Gaussian likelihood that
implements the observational limits from Ref.~\cite{Planck15cosmo} ($\Omega_\chi h^2 = 0.1188 \pm 0.0010$ at $1\sigma$). In addition to the experimental error, we also take into the account possible theoretical errors in the calculation of
the relic density. We assume that the likelihood for the prediction for $\Omega_\chi h^2$ is a Gaussian distribution centred around the calculated value with a standard deviation that is taken, by default, to be 5\% of that calculated value.
This error is conservative for most parameter combinations, and underestimates the $\mathcal{O}(50\%$) corrections that can occur due to loop corrections in a few specific scenarios \cite{Baro:2007em,Baro:2009na,Herrmann:2014kma,arXiv:1510.06295,arXiv:1602.08103}. The choice is thus a pragmatic compromise that represents the best that can be done with a single uncertainty. The user is free to change it, and we emphasise  the importance of the user choosing an error appropriate for the particular model he or she is studying. The exact form of the Gaussian likelihood for the model parameters is determined by the \YAML option \yaml{profile_systematics}, which determines whether the prediction for $\Omega_\chi h^2$ is profiled or marginalised over (by default this option is set to \yaml{false}, corresponding to marginalisation).
The mean value, the experimental error
and the theoretical error can be changed with the \YAML file options
\yaml{oh2_obs}, \yaml{oh2_obserr}, and \yaml{oh2_fractional_theory_err},
respectively. For the exact form of the likelihood function and the details of its derivation, we refer the reader to section 8.3.1 and 8.3.2 of \cite{gambit}.

Alternatively, \yaml{lnL_oh2_upperlimit} implements the observational
constraint as a one-sided limit, leaving open the possibility that a certain DM candidate makes up only a fraction of the total abundance. The likelihood function is roughly a Gaussian similar to the one described in the previous paragraph for predictions greater than the observed relic density and flat for predictions below the observed value. Here the exact form is again dependent on the \YAML parameter \yaml{profile_systematics} (see section 8.3.3 and 8.3.4 of \cite{gambit} for many more details). The other three \YAML parameters for this function are the same as before.

%%%%%%%%%%%%%%%%%%%%%%%%%%%%%%%%%%%%%%%%%%%%%%%%%%%%%%%
%%%%%%%%%%%%%%%%%%%%%%%%%%%%%%%%%%%%%%%%%%%%%%%%%%%%%%%
\section{Direct Detection}
\label{sec:dd}

\subsection{Background}

Given that the solar system sits within a DM halo, DM particles are expected to pass through Earth continuously.  If DM has any ability to interact with regular matter at all, it will occasionally scatter on terrestrial nuclei.  Direct detection experiments \cite{Goodman:1984dc} search for these scattering events, by looking for nuclear recoils in large volumes of inert target material placed in ultra-clean environments deep underground.

In natural units, the differential rate of recoil events in a direct detection experiment is
\begin{equation}\label{eqn:dRdEnr}
  \dRdEnr
    = \frac{2\rhochi}{\mchi}
      \int v f(\bv,t) \frac{d\sigma}{dq^2}(q^2,v) \, d^3v,
\end{equation}
where $\mchi$ is the WIMP mass, $\rhochi$ is the local DM mass density, $f(\bv,t)$ is the three-dimensional,
time-dependent WIMP velocity distribution, $\frac{d\sigma}{dq^2}(q^2,v)$ is the velocity-dependent differential
cross-section, and $q^2 = 2 m_\mathrm{nuc} \Enr$ is the momentum exchanged in the
scattering process (for a nucleon mass $m_\mathrm{nuc}$ and recoil energy $E$).  Numerical values of this differential rate are typically expressed in cpd (counts per day), per kg of target material, per keV recoil energy.
Most direct search detectors contain more than one isotope, in which case the differential rate is given by a sum over
Eq.~\ref{eqn:dRdEnr} for each isotope, weighted according to the mass fraction of the isotope in the detector.

The expected number of signal events in an analysis by a direct search experiment is given by
\begin{equation} \label{eqn:signal}
  N_\mathrm{p} = MT \int_0^{\infty} \phi(E) \dRdE(E) \, dE,
\end{equation}
where $M$ is the detector mass and $T$ is the exposure time. The detector response function $\phi(E)$ describes the fraction of recoil events of energy $E$ that will be observed within some pre-defined analysis region. The precise definition of such an analysis region depends on the experiment under consideration. In the simplest case it would be given by a lower and an upper bound on the reconstructed energy, so that the response function $\phi(E)$ can be calculated in terms of the energy resolution of the detector and the various trigger efficiencies. The energy range over which the experiment is sensitive is then encoded within $\phi(E)$, so that there is no need to impose a finite upper or lower cutoff in the integral in Eq. \ref{eqn:signal}.

Some experiments implement more elaborate analyses, imposing further cuts on observables that depend on the recoil energy in a more complicated way. All of these possibilities can be captured by an appropriate function $\phi(E)$, because the detector response is always independent of the nature of the particle interaction of the WIMP with nuclei, which is contained in the differential event rate. The detector response $\phi(E)$ can therefore be tabulated in advance for different analyses and re-used for any WIMP model. Eq. \ref{eqn:signal} can be generalised to experiments with more than one analysis region (e.g.\ binned event rates) by defining a separate function $\phi_i(E)$ for each analysis region.

For many WIMP candidates (which we refer to as $\chi$), the dominant WIMP-quark interactions arise from a combination of
\begin{itemize}
\item a scalar ($\bar{\chi}\chi\bar{q}q$) or vector ($\bar{\chi}\gamma_{\mu}\chi\bar{q} \gamma^{\mu} q$) coupling, which give rise to a spin-independent (SI) cross-section, and
\item an axial-vector coupling ($\bar{\chi}\gamma_{\mu}\gamma_5 \chi\bar{q} \gamma^{\mu} \gamma_5 q$), which gives rise to a spin-dependent (SD) cross-section\end{itemize}
(for a complete list of possible operators see Ref.~\cite{Kumar:2013iva}).
In both of these cases, the matrix element involved has no intrinsic dependence upon either the momentum exchanged in the collision, nor on the relative velocity of the DM and the nucleus.  In such cases, it is convenient to write the scattering cross-section as a simple product of a cross-section $\sigma_0$ defined at zero-momentum-transfer, and a form factor $F^2(q)$ that accounts for the
finite size of the nucleus.  For such velocity and momentum-independent interactions, the differential cross-section becomes
\begin{equation}\label{eqn:dsigmadq}
  \frac{d\sigma}{dq^2}(q^2,v) = \frac{\sigma_{0}}{4 \mu^2 v^2}
                              F^2(q) \, \Theta(\qmax-q)\;,
\end{equation}
where $\Theta$ is the Heaviside step function, $\mu$ is the WIMP-nucleon reduced mass, $\qmax = 2 \mu v$ is the maximum momentum transfer in a collision at a
relative velocity $v$, and the velocity dependence is entirely due to
kinematics rather than the interaction.  The requirement that $q \le \qmax$ for an interaction to be kinematically possible translates into a lower limit $v\ge\vmin =\sqrt{m_\mathrm{nuc} \Enr/2\mu^2}$ in the integral over the WIMP velocity distribution (Eq.\ \ref{eqn:dRdEnr}).  The total WIMP-nucleus differential cross-section is the sum over the SI and SD contributions, each with its own form factor.

The zero-momentum cross-section $\sigma_0$ for SI WIMP-nucleus interactions is
\begin{eqnarray} \label{eqn:sigmaSI}
  \sigmaSI
     \ =\ \frac{\mu^2}{\pi} \, \Big[ Z \GpSI + (A-Z) \GnSI \Big]^{2} \\
     \ =\ \frac{4\mu^2}{\pi} \, \Big[ Z \fpSI + (A-Z) \fnSI \Big]^{2} \; ,
  \nonumber
\end{eqnarray}
where $Z$ is the atomic number and $A$ is the atomic mass number.  $Z$ and $A-Z$ are respectively the number of protons and neutrons in the
nucleus, and $\fpSI$ and $\fnSI$ are the effective couplings to them.  The
latter depend on both the precise nature of the interaction between WIMPs and
quarks (and/or gluons), and on the contents of the proton and neutron. We note
that the alternative normalisation involving $\GpSI=2\fpSI$ and $\GnSI=2\fnSI$
is often found in the literature, where $\GNSI$ with $N=n,p$ are the $G_F$-like effective four-fermion coupling constants in the case of scalar interactions. The \micromegas manual, meanwhile, uses $\lambda_N=\frac{1}{2}\GNSI$. The nucleon contents are described by the nuclear hadronic matrix elements, discussed in detail in Sec.\ \ref{matrix_elements} below.

For most DM candidates with scalar couplings, the proton and neutron SI cross-sections are roughly the same, so $\fnSI \simeq \fpSI$. For identical couplings ($\fnSI = \fpSI$), the SI
cross-section reduces to
\begin{equation} \label{eqn:sigmaSI2}
  \sigmaSI = \frac{\mu^2}{\mup^2} A^2 \, \sigmapSI \, ,
\end{equation}
where $\mup$ is the WIMP-proton reduced mass. Direct detection experiments are often designed to use heavy nuclei, as the SI cross-section grows rapidly with $A$.

The SI form factor is essentially a Fourier transform of the mass
distribution of the nucleus, and it is reasonably approximated by the Helm form factor \cite{Helm:1956zz,Lewin:1995rx},
\begin{equation} \label{eqn:SIFF}
  F(q) = 3 e^{-q^2 s^2/2} \; \frac{\sin(qr_n)- qr_n\cos(qr_n)}{(qr_n)^3} \, ,
\end{equation}
where $s\simeq 0.9$~fm and $r_n^2 = c^2 + \frac{7}{3} \pi^2 a^2 - 5
s^2$ is an effective nuclear radius with $a \simeq 0.52$~fm and $c
\simeq 1.23 A^{1/3} - 0.60$~fm.  Further details on SI form factors
can be found in Refs.~\cite{Lewin:1995rx,Duda:2006uk}.

SD scattering is only present for detectors that contain isotopes with net nuclear spin.  This generally requires the nucleus to possess an unpaired proton and/or neutron in its shell structure. The relevant WIMP-nucleon cross-section is
\begin{align} \label{eqn:sigmaSD}
  \sigmaSD
    \ &=\ \frac{4 \mu^2}{\pi} \, \frac{(J+1)}{J} \,
          \Big[ \GpSD \Sp + \GnSD \Sn \Big]^2 \nonumber\\
    \ &=\ \frac{32 \mu^2 \, G_F^2}{\pi} \, \frac{(J+1)}{J} \,
          \Big[ \apSD \Sp + \anSD \Sn \Big]^2 \, ,
\end{align}
where $G_F$ is the Fermi constant, $J$ is the spin of the nucleus,
$\Sp$ and $\Sn$ are the average spin contributions from the
proton and neutron groups respectively, and $\apSD$ and $\anSD$ are the
effective couplings to the proton and the neutron in units of $2\sqrt{2}G_F$.  Similarly to $\fpSI$ and $\fpSI$, $\apSD$ and $\anSD$ depend on both the WIMP-quark interaction and on the relative contributions of different quark flavours to the nucleon spin; the latter is discussed further in Sec.\ \ref{matrix_elements}. As in the spin-independent case, alternative normalisations can be found in the literature.  The \ds manual, for example, refers to $\GNSD=2\sqrt{2}G_F a_N$, while the \micromegas manual uses $\xi_N=\frac{1}{2}\GNSD$. In addition, whilst we here use $a_N$ and $\GNSD$ to distinguish the two notations, $a_N$ is frequently used within the literature for \textit{both} cases.

Unlike the SI case, the two SD couplings $\apSD$ and $\anSD$ differ substantially in many theories. Individual experiments typically only strongly constrain either $\apSD$ or $\anSD$, as most detector materials do not contain isotopes with both unpaired neutrons and protons; experimental results on the SD cross-section are therefore generally presented in terms of $\sigmapSD = \sigmaSD(\anSD=0)$ or $\sigmanSD = \sigmaSD(\apSD=0)$.

The SD form factor is given in terms of the structure function $S(q)$ normalised so that $F^2(0)=1$,
\begin{equation}
  F^2(q)=S(q)/S(0),
\end{equation}
with
\begin{equation}
  S(q) = \apSD^2\,\Spp(q) + \anSD^2\,\Snn(q) + \apSD\;\!\anSD\;\!\Spn(q) \;.
\end{equation}
In the limit $q\rightarrow 0$, the functions $S_{nn}(0)$ and $S_{pp}(0)$ are proportional to the expectation values of spins of the the proton and neutron subsystems \cite{Bednyakov:2004xq,Bednyakov:2006ux},
\begin{align}
  \Spp(0) &= \frac{(J+1)(2J+1)}{\pi J}\Sp^2,\nonumber \\
  \Snn(0) &= \frac{(J+1)(2J+1)}{\pi J}\Sn^2.
\end{align}

\subsection{\ddcalc}
\label{sec:ddcalc}

The traditional presentation of results from direct searches for dark matter is an exclusion curve for the SI or SD WIMP-nucleon scattering cross-section, as a function of the WIMP mass.  This invariably comes with some rather specific restrictions:\begin{enumerate}
\item the exclusion is given at only a single confidence level (CL; traditionally 90\%),
\item $\fpSI=\fnSI$ (often somewhat loosely referred to as `isospin conservation') is assumed,
\item either $\apSD=\anSD=0$ (pure SI coupling), or $\fpSI=\fnSI=(\apSD\ \mathrm{or}\ \anSD) = 0$ (pure SD$_p$ or SD$_n$ coupling) is assumed,
\item it is assumed that the local density and velocities of DM follow the Standard Halo Model (cf.\ Sec.\ \ref{sec:halo}),
\item a specific set of nuclear form factors $F^2(q)$ is adopted, and
\item the nuclear parameters are fixed to assumed values when calculating $\fpSI$, $\fnSI$, $\apSD$ and $\anSD$ for any comparison against theory predictions.
\end{enumerate}
These restrictions are all problematic when trying to recast direct search results, as one must go from the idealised effective WIMP frameworks in which they are presented to real constraints on actual theories.  Ultimately, we are interested in the overall degree to which a model with some arbitrary combination of couplings $\fpSI$, $\fnSI$, $\apSD$ and $\anSD$ agrees or disagrees with data, not merely which side of a 90\% CL curve it lies on under different limiting approximations about $\fpSI$, $\fnSI$, $\apSD$ and $\anSD$.  Ideally, this would also include the impacts of systematic errors on that exclusion, due to uncertainties from the halo, nuclear and form-factor models that one must assume in order to obtain that result.

For these reasons, here we present \ddcalc: a new, general solution for recasting direct search limits.  \ddcalc is released and maintained as a standalone backend code by the \GB Dark Matter Workgroup.  It can be obtained from \url{http://ddcalc.hepforge.org} under an academic use license.  A development version of the code including only the first run of the LUX experiment was previously released as \luxcalc \cite{Savage:2015xta}, and has been used in a number of analyses (e.g.\ \cite{Berlin,Beniwal,Liem}).

Yet another issue is that the limits presented by experimental collaborations
almost always assume Eq.\ \ref{eqn:dsigmadq}, i.e.\ that the scattering matrix
element has no explicit velocity or momentum dependence.  Many DM models
involve non-trivial momentum- or velocity-dependences in their cross-sections.
This means that extra velocity factors must be incorporated into the integral
over the WIMP velocity distribution (Eq.\ \ref{eqn:dRdEnr}), and extra momenta
must be included in the final integral over the differential rate when
calculating the total event yield (Eq.\ \ref{eqn:signal}). Furthermore, these models often probe properties of the target nuclei not captured by the standard SI and SD nuclear form factors. Although the first
release of \ddcalc does not include such generalised couplings out of the box,
its structure is designed to easily accommodate them (and they will be
explicitly included in a future release).

\subsubsection{Methods}

\ddcalc calculates predicted signal rates and likelihoods for various experiments, given input WIMP and halo models.

A \ddcalc WIMP model consists of the DM mass and the four couplings $\fpSI$, $\fnSI$, $\apSD$ and $\anSD$, specifiable also directly as SI/SD proton/neutron cross-sections. \ddcalc does not deal directly with nuclear uncertainties; users (or \GB as the case may be) are expected to vary these externally in the calculation of the couplings.  Momentum-dependent couplings can be implemented by adding the requisite additional power of $q$ to the integrand of Eq.\ \ref{eqn:signal}, as implemented in the source file \term{DDRates.f90}.  Similarly, velocity-dependent cross-sections can be implemented in the integrand of Eq.\ \ref{eqn:dRdEnr}, found in \term{DDHalo.f90}.  Some more explicit tips for the brave are provided in the \ddcalc~\term{README}.  SI form factors in the first release default to Helm (Eq.\ \ref{eqn:SIFF}).  SD form factors are included for $^{19}$F, $^{23}$Na, $^{27}$Al, $^{29}$Si, $^{73}$Ge, $^{127}$I, $^{129}$Xe and $^{131}$Xe from Ref.\ \cite{Klos}.  Alternative form factors can be encoded in \term{DDNuclear.f90}.

The local halo model consists of a constant local density and a truncated
Maxwell-Boltzmann velocity distribution (Eq.\ \ref{eq:MB}).  The most-probable speed $\vmp$, truncation speed $\vesc$, local speed relative to the halo $\vobs$ and local density $\rho_\chi$ are all individually configurable; $\vobs$ can also be constructed automatically from the local standard of rest $\bvLSR$ or disk rotation speed $v_\mathrm{loc}$, along with the Sun's peculiar velocity $\bvsunpec$ relative to it.

A particular set of WIMP and halo model parameters will produce a set of predicted yields in various direct search experiments. Having calculated the expected number of signal events $N_{\mathrm{p},i}$ for the $i$th experimental analysis region (Eq.\ \ref{eqn:signal}), \ddcalc calculates a Poisson likelihood for the model as
\begin{equation} \label{eqn:Poisson}
  \mathcal{L}_i(N_{\mathrm{p},i}|N_{\mathrm{o},i})
  = \frac{(b_i+N_{\mathrm{p},i})^{N_{\mathrm{o},i}} \, e^{-(b_i+N_{\mathrm{p},i})}}{N_{\mathrm{o},i}!},
\end{equation}
where $N_{\mathrm{o},i}$ is the number of observed events in the analysis region, and $b_i$ is the expected number of background events in that region. Note that a single experiment may comprise more than one analysis region (for example bins in reconstructed energy). Unbinned analyses can in principle also be implemented, provided sufficient information from the experiment is available. The likelihoods for each experiment and analysis region, $\{\mathcal{L}_i\}$ can then be combined to form a composite direct search likelihood, which can itself be used in combination with other likelihood terms from other experiments.

Some direct detection experiments do not provide explicit background estimates or prefer not to perform a background subtraction in order to test the WIMP hypothesis. In this case, $b_i$ should be set to the value that maximises the likelihood:
\begin{equation}
 b_i =
\begin{cases}
N_{\mathrm{o},i} - N_{\mathrm{p},i} \, , & N_{\mathrm{o},i} > N_{\mathrm{p},i}   \\
0  \, , & N_{\mathrm{o},i} \leq N_{\mathrm{p},i} \; .
\end{cases}
\end{equation}
This leads to a one-sided likelihood, i.e.\ a non-zero WIMP signal can only be disfavoured but not preferred relative to the background-only hypothesis.

The likelihood functions can be used directly to calculate constraints in the $\sigma$-$m_{\chi}$ plane. A parameter point is considered to be excluded at 90\% confidence level if
\begin{equation}
2  \log \mathcal{L}(\sigma = 0) - 2 \log  \mathcal{L}(\sigma, m_\chi) > 1.64 \; ,
\end{equation}
where $\mathcal{L}(\sigma = 0)$ denotes the likelihood of the background-only hypothesis. Alternatively, one can also use \ddcalc to obtain constraints in the $\sigma$, $m_{\chi}$ plane using one of two $p$-value methods:

\paragraph{Feldman-Cousins}
\ddcalc implements the standard Feldman-Cousins method~\cite{Feldman:1997qc} for generation of one- or two-sided confidence intervals, based on the Poisson likelihood in Eq. \ref{eqn:Poisson}. Note that the likelihood in this case uses the total expected signal and background yields across the entire analysis region, and hence does not incorporate spectral information that might lead to a stronger result.

\paragraph{Maximum gap}

Yellin's maximum gap method~\cite{Yellin:2002xd} was proposed as a way of handling spectral information in the case that the magnitude and shape of the background are unknown and a background subtraction is therefore not possible. The method assumes that all of the observed events could in principle be signal events, leading to a conservative exclusion limit on the WIMP scattering cross-section. Nevertheless, exploiting the spectral information of the observed events will typically yield stronger results than the Feldman-Cousins method; see~\cite{Savage:2015xta} for an example.

The idea of the maximum gap method is to break the signal region into a number of intervals bounded by the observed events. Given an efficiency functions $\phi_k(E)$ for each of these intervals, one can then use Eq.~\ref{eqn:signal} to calculate the expected number of events between any two observed events. The ``maximum gap'' is the interval where this number is largest. By calculating the probability that a gap as large as the observed one could arise from random fluctuations it is then possible to quantify the $p$-value of the assumed model~\cite{Yellin:2002xd}.

\begin{table}[tb]
  \centering
  \begin{tabular}{l l}
    \toprule
    Experiment & Analysis \\
    \midrule
    \multicolumn{2}{l}{\textit{Included in \ddcalc \textsf{1.0.0}}} \\
    XENON100 & 2012 \cite{XENON2013} \\
    SuperCDMS & 2014 \cite{SuperCDMS} \\
    SIMPLE & 2014 \cite{SIMPLE2014} \\
    LUX (run 1) & 2013 \cite{LUX2013}, 2015 \cite{LUX2016} \\
    LUX (run 2) & 2016 \cite{LUXrun2} \\
    PandaX & 2016 \cite{PandaX2016} \\
    PICO-60 & 2016 \cite{PICO60} \\
    PICO-2L & 2016 \cite{PICO2L} \\
    \\
    \multicolumn{2}{l}{\textit{Added in \ddcalc \textsf{1.1.0}}} \\
    Xenon1T & 2017 \cite{Xenon1T2017} \\
    PICO-60 & 2017 \cite{PICO2017} \\
    \bottomrule
  \end{tabular}
  \caption{Experimental analyses included in \ddcalc.}
  \label{tab:ddcalc-analyses}
\end{table}

\begin{table*}[tp]
  \centering
  \begin{tabular}{p{50mm}p{100mm}}
    \toprule
    Mode & Summary
    \\\midrule
    \term{DDCalc_run} & {\justify Calculates the expected number of signal events in the analysis region, the likelihood of the WIMP parameters, and the $p$-value.  The logarithm of the latter two are given.} \\
    \term{DDCalc_run --log-likelihood} & {\justify Calculates the logarithm of the Poisson-based likelihood for the specified WIMP parameters.} \\
    \term{DDCalc_run --log-pvalue} & {\justify Calculates the logarithm of the $p$-value for the specified WIMP parameters. If the analysis includes interval information, this returns the maximum gap $p$ value. If the analysis does not include interval information, the $p$-value calculated from the Poisson likelihood but without background subtraction (i.e.\ setting $b = 0$) to give a conservative upper bound.} \\
    \term{DDCalc_run --spectrum} & {\justify Tabulates the raw recoil spectrum $\frac{dR}{dE}$ for the detector material and given WIMP parameters, by energy.  The tabulation of the energies can be modified using the \term{--E-tabulation} option.} \\
    \term{DDCalc_run --events-by-mass} & {\justify Tabulates the expected signal events for fixed WIMP-nucleon cross-sections, by WIMP mass. The tabulation of masses can be modified using the \term{--m-tabulation} option.} \\
    \term{DDCalc_run --constraints-SI} & {\justify Tabulates the cross-section lower and upper constraints in the spin-independent case, by mass. Constraints are 1D confidence intervals at each mass, determined using a Poisson likelihood with signal plus background and Feldman-Cousins ordering.  The confidence level is given using the \term{--confidence-level} option, with the default set to 0.9 (90\% CL). The ratio of the WIMP--neutron and WIMP-proton couplings is held fixed, with the ratio defined by an angle $\theta$ such that $\tan\theta = G_n/G_p$. The angle can be specified via the \term{--theta-SI} option or, more conveniently, given in units of $\pi$ via \term{--theta-SI-pi}. The default is $G_p=G_n$ ($\theta = \pi/4$). The tabulation of the masses can be modified using the \term{--m-tabulation option}.} \\
    \term{DDCalc_run --constraints-SD} & {\justify As above, but for the spin-dependent case.} \\
    \term{DDCalc_run --limits-SI} & {\justify Tabulates the cross-section upper limits in the spin-independent case, by mass.  Limits are determined using the maximum gap $p$-value method.  Excluded $p$-values are given using the \term{--p} or \term{--lnp} options, with the default being $p=0.10$ (90\% CL exclusion limits).  The ratio of the WIMP--neutron and WIMP--proton couplings is held fixed, with the ratio defined by an angle $\theta$ such that $\tan\theta = G_n/G_p$. The angle can be specified via the \term{--theta-SI} option or, more conveniently, given in units of $\pi$ via \term{--theta-SI-pi}. The default is $G_p=G_n$ ($\theta = \pi/4$).The tabulation of the masses can be modified using the \term{--m-tabulation option}.} \\
       \term{DDCalc_run --limits-SD} & {\justify As above,  but for the spin-dependent case.} \\
    \bottomrule
  \end{tabular}
  \caption{The various run modes of the \ddcalc default executable \protect\term{DDCalc_run}.}
  \label{tab:DDCalcModes}
\end{table*}

\begin{table*}[tp]
  \centering
  \begin{tabular}{p{50mm}p{100mm}}
    \toprule
    General options & \\
    \midrule
    \term{--help} & Displays help information. \\
    \term{--verbosity} & Sets the verbosity level (a higher level gives more output). The data output at different levels is mode-specific. Selection of verbosity level 0 is also available via the flag \term{--quiet}; \term{--verbose} gives verbosity level 2.\\
    \midrule
    Mode-specific options & \\
    \midrule
    \term{--E-tabulation=<min>,<max>[,<N>]} \term{[0.1,1000,-50]} & Specifies the tabulation for the recoil energy $E$.  Energies
         will be tabulated from \term{<min>} to \term{<max>} inclusive, with $N$
         logarithmically spaced intervals between tabulation points.
         A negative \term{<N>} value indicates \term{|<N>|} points per decade. The
         \term{<N>} argument is optional. \\
    \term{--interactive} & Turns on the interactive mode, which will prompt the user for the WIMP parameters. \\
     \term{--m-tabulation=<min>,<max>[,<N>]} \term{[1,1000,-20]} & Specifies the tabulation for the WIMP mass $m$.  Masses will be tabulated from \term{<min>} to \term{<max>} inclusive, with $N$ intervals between tabulation points. The points will be logarithmically spaced unless the fourth comma-separated value is \term{F} or \term{0}, in which case the spacing will be linear. A negative \term{<N>} value indicates \term{|<N>|} points per decade in the logarithmic case. The \term{<N>} and \term{<log>} arguments are optional. \\
     \term{--theta-SD=<val>     [}$\pi/4$\term{]} & Fixes the ratio of the WIMP-nucleon couplings to $\tan\theta=G_p/G_n$ in the spin-dependent case. This option is only used in run modes where the absolute couplings cannot be specified. The similar option \term{--theta-SD-pi} allows the ratio to be specified in units of $\pi$. \\
      \term{--theta-SI=<val>     [}$\pi/4$\term{]} & As above, but for the spin-independent case. \\
      \midrule
      Statistical options & \\
      \midrule
      \term{--confidence-level=<val> [0.9]} & The confidence level to use in determining constraints/limits. \\
      \term{--confidence-level-sigma=<val>} &  The confidence level corresponding to the fraction of the normal distribution within the given number of standard deviations (symmetric).  That is, a value of 3 here would result in a 3$\sigma$ CL constraint/limit. Equivalent to \term{--p-value-sigma}. \\
      \term{--p-value=<val> [0.1]} & The $p$-value to use in determining constraints/limits. The logarithm of the $p$-value can be specified via \term{log-p-value}.\\
      \midrule
      Experiment-specific options & \\
      \midrule
   \term{--XENON100-2012} & Sets the detector and analysis according to the XENON100 2012 result~\cite{XENON2013}.  \\
   \term{--LUX-2013} & Sets the detector and analysis according to the LUX run 1 result from 2013~\cite{LUX2013}. \\
   \term{--LUX-2015} & Sets the detector and analysis according to the reanalysis of the LUX run 1 result from 2015~\cite{LUX2016}. \\
   \term{--LUX-2016} & Sets the detector and analysis according to our implementation of the LUX run 2 result from 2016~\cite{LUXrun2}. \\
   \term{--PandaX-2016} & Sets the detector and analysis according to the PandaX 2016 result~\cite{PandaX2016}. \\
   \term{--SuperCDMS-2014} & Sets the detector and analysis according to the SuperCDMS 2014 low-energy result~\cite{SuperCDMS}. \\
   \term{--SIMPLE-2014} & Sets the detector and analysis according to the SIMPLE 2014 C$_2$ClF$_5$ result~\cite{SIMPLE2014}. \\
   \term{--PICO-2L} & Sets the detector and analysis according to the PICO-2L 2016 result~\cite{PICO2L}. \\
   \term{--PICO-60_F} & Sets the detector and analysis according to the PICO-60 2016 result~\cite{PICO60}, using only the contribution from fluorine. \\
   \term{--PICO-60_I} & Sets the detector and analysis according to the PICO-60 2016 result~\cite{PICO60}, using only the contribution from iodine. \\
   \term{--Xenon1T-2017} (\ddcalc \textsf{1.1.0} only) & Sets the detector and analysis according to the Xenon1T 2017 result~\cite{Xenon1T2017}. \\
   \term{--PICO-60_2017} (\ddcalc \textsf{1.1.0} only) & Sets the detector and analysis according to the PICO-60 2017 result~\cite{PICO60}. \\
   \bottomrule
  \end{tabular}
  \caption{Options of the \ddcalc default executable \protect\term{DDCalc_run}.}
  \label{tab:DDCalcOptions}
\end{table*}

\begin{table*}[tp]
  \centering
  \begin{tabular}{p{60mm}p{90mm}}
    \toprule
    \term{--rho=<val> [0.4]} & The local dark matter density (in units of GeV\,cm$^{-3}$). \\
    \term{--v0=<val> [vrot]} &
         The most probable speed of the Maxwell-Boltzmann
         distribution (in km\,s$^{-1}$).  Related
         to the rms speed by $v_{rms} = \sqrt{3/2}v_0$ and the mean speed
         by $\bar{v} = \sqrt{4/\pi}v_0$.  For the isothermal sphere of the Standard Halo Model, this is equal to the galactic circular velocity (disk rotation speed), so this is set equal to the value supplied with \term{--vrot} (or its default) unless explicitly set here. \\
   \term{--vesc=<val> [550]} & The local Galactic escape speed (in km\,s$^{-1}$).  This is used to
         truncate the Maxwell-Boltzmann distribution as the highest
         speed particles should be depleted due to escape from the
         Galactic potential. \\
   \term{--vlsr=<val>,<val>,<val> [0,vrot,0]} &
         The Local Standard of Rest (in km\,s$^{-1}$), i.e.\ the velocity of the
         Galactic disk relative to the Galactic rest frame in Galactic
         coordinates $UVW$, where $U$ is the anti-radial direction
         (towards the Galactic centre), $V$ is the direction of disk
         rotation, and $W$ is in the direction of the galactic pole.
         Set to (0,\term{--vrot},0) unless specified here. \\
   \term{--vobs=<val> [|vsun|]} &
         The speed of the observer (detector) relative to the galactic
         rest frame (in km\,s$^{-1}$), where the galactic rest frame is the frame
         in which the dark matter exhibits no bulk motion.  This is
         set equal to the magnitude of \term{--vsun}, unless explicitly set
         here. \\
   \term{--vpec=<val>,<val>,<val> [11,12,7]} & The velocity of the Sun
         relative to the Local Standard of Rest in galactic
         coordinates $UVW$. \\
   \term{--vsun=<val>,<val>,<val> [vlsr+vpec]} &
         The motion of the Sun relative the the galactic rest frame (in km\,s$^{-1}$)
         in galactic coordinates $UVW$.  Set equal to \term{--vlsr} + \term{--vpec} unless explicitly set here. \\
   \term{--vrot=<val> [235]} & The local disk rotation speed (in km\,s$^{-1}$). \\
    \bottomrule
 \end{tabular}
  \caption{Dark matter halo distribution options of the \ddcalc default executable \protect\term{DDCalc_run}.}
  \label{tab:DDCalcHaloOptions}
\end{table*}

\begin{table*}[tp]
  \centering
  \begin{tabular}{p{50mm}p{100mm}}
    \toprule
 General detector options & \\
 \midrule
  \term{--background=<val> [0.64]} &
         The average expected number of background events in the
         analysis region.\\

  \term{--events=<N> [1]} &
         The number of observed events in the analysis region. \\
   \term{--exposure=<val> [118}$\times$\term{85.3]} &
         The detector's fiducial exposure (in kg day).  Set equal to
         \term{--mass} $\times$ \term{--time} unless explicitly set here.\\
   \term{--mass=<val> [118]} & The detector's fiducial mass (in kg).\\
   \term{--time=<val> [85.3]} & The detector's exposure time (in days). \\
   \midrule
Isotopic composition options & \\
   \midrule
   \term{--argon} & Sets the isotopic composition to the naturally occurring
         isotopes of argon\\
   \term{--germanium} &  (Ar), germanium (Ge), sodium iodide (NaI),
         silicon (Si), or xenon (Xe). \\
   \term{--sodium-iodide} & \\
   \term{--silicon} & \\
   \term{--xenon} & \\
   \term{--element-Z=<Z1>,<Z2>,<Z3>,...} &  Sets the isotopic composition to the naturally occurring
         isotopes of the com- \\
   \term{--stoichiometry=<N1>,<N2>,<N3>,...} & pound with the given set of elements and
         stoichiometry.  For example, CF$_3$Cl would have \term{Z=6,9,17}
         and a stoichiometry \term{1,3,1}.  If the stoichiometry is not
         given, it is assumed to be 1 for each element.\\
   \term{--isotope-Z=<Z1>,<Z2>,<Z3>,...} &  Provide an explicit list of isotopes to use.  The list of
         atomic numbers (\term{Z}),\\
   \term{--isotope-A=<A1>,<A2>,<A3>,...} &  mass numbers (\term{A}), and mass fractions (\term{f})
         must all be provided and be of\\
   \term{--isotope-f=<f1>,<f2>,<f3>,...} &  the same length or these
         options will be ignored.\\
   \midrule
   Detector efficiency options & \\
   \midrule
    \term{--Emin=<val> [0]} & Applies a minimum energy threshold for rate calculations
         (in keV).  This effectively takes:
             $\phi(E) \rightarrow \Theta(E-E_{min}) \phi(E)$,
         where $\Theta(x)$ is the Heaviside step function.  This allows
         easy removal of low-energy scattering from the signal
         calculations without having to modify efficiency input files. \\
   \term{--file=<file>} &
         A file containing the efficiency $\phi(E)$ tabulated by energy,
         where $\phi(E)$ is defined as the fraction of events at an
         energy $E$ that will be observed in the analysis region after
         factoring in trigger efficiencies, energy resolution, data
         cuts, etc.  The first column is taken to be the energy (in keV).
         The next column should contain only numbers between 0 and 1; this is taken
         to be $\phi(E)$; note that this allows \tpcmc output to be used, as
         columns of \term{<S1>} and \term{<S2>} will be safely ignored.  Blank lines
         and lines beginning with a hash character will be ignored. If the analysis observed any events, the file can optionally
         include additional columns beyond the $\phi(E)$ column
         representing the efficiencies $\phi_k(E)$ for detecting events
         in the sub-intervals for use with the maximum gap method. \\
   \term{--no-intervals} &
         Disables the use of sub-interval calculations, even if
         efficiencies for the sub-intervals are available.  This is
         implied by certain program modes where the sub-intervals
         cannot be used. \\
 \bottomrule
 \end{tabular}
  \caption{Advanced detector options of the \ddcalc default executable \protect\term{DDCalc_run}.}
  \label{tab:DDCalcDetOptions}
\end{table*}

\subsubsection{Experiments}

Event rate calculations in \ddcalc rely on the availability of experimental response functions $\phi(E)$, the predicted number of background events in an analysis $b$, the total number of observed events $N_o$, and the experimental exposure $MT$. Version 1.0.0 of \ddcalc ships with this data already implemented for eight experimental analyses, shown in Table~\ref{tab:ddcalc-analyses} (see also Fig.~\ref{fig:limits_di} for a comparison of our analyses with the published bounds).

The LUX \cite{LUX2013,LUX2016,LUXrun2}, PandaX \cite{PandaX2016}, XENON100 \cite{XENON2013} and SuperCDMS \cite{SuperCDMS} analyses are most useful for constraining SI scattering and SD scattering on neutrons, whereas the PICO-2L \cite{PICO2L}, PICO-60 \cite{PICO60} and SIMPLE \cite{SIMPLE2014} analyses provide good sensitivity to SD scattering on protons. In the following we provide additional details on the implementation of the experimental details.

\textbf{SIMPLE.} We implement the efficiency curve $\phi(E)$ directly as described in Ref.\ \cite{SIMPLE2014}.  We do not include the contribution from carbon, due to its high threshold energy for nuclear recoils. SIMPLE expected 12.7 events and observed 8.

\textbf{PICO-2L.} The efficiency curve $\phi(E)$ for flourine is provided in Ref.~\cite{PICO}. Again, we do not include the contribution from carbon. The background expectation of 1.0 events agrees well with the one event observed.

\textbf{PICO-60.} This experiment has a time-dependent energy threshold, so the efficiency curve $\phi(E)$ is obtained by convolving the exposure as a function of threshold with the (appropriately rescaled) efficiency curve for fixed threshold. The total exposure is reduced by a trial factor of 1.8 as proposed by the collaboration. Because the efficiency curves for fluorine and iodine differ, we implement the two target materials as independent experiments. This is possible only because PICO-60 has observed no signal events and does not perform a background subtraction, in which case the likelihood reduces to $\mathcal{L} = e^{-N_{\mathrm{p}}}$, so that the contributions for the individual target nuclei factorise. Again, we do not include the contribution from carbon.

\textbf{SuperCDMS.} We implement two different efficiency curves including gap information, with nuclear recoil energies converted from phonon energies assuming the Lindhard presecription \cite{Lewin:1995rx}.  The first is based directly on the published efficiencies and event energies of all detectors included in the experimental run \cite{SuperCDMS}.  During this run, the ionisation guard of one detector (T5Z3) was inoperative, allowing additional background events to enter the analysis region and reduce the overall sensitivity of the experiment.  We therefore implement a second efficiency curve and corresponding set of analysis parameters, where the T5Z3 detector is excluded from the analysis.\footnote{The additional efficiency information for T5Z3 was kindly provided by the SuperCDMS Collaboration.}  This alternative parameterisation is the default within \ddcalc for this analysis.

\textbf{PandaX.} We implement the efficiency curve $\phi(E)$ provided by the collaboration \cite{PandaX2016}, with an additional factor of 2 to account for the fact that only events below the mean of the nuclear recoil band are considered. Three events were observed in this search window compared to a background expectation of 4.8.

\textbf{XENON100.} While XENON100 does provide efficiency curves as a function of the nuclear recoil energy \cite{XENON2013}, this information is insufficient to make use of the spectral information, i.e.\ the energy of the observed events. This spectral information is however very helpful, because events at high recoil energies ($E > 30$ keV) as well as events close to the threshold have a higher probability to result from backgrounds than from WIMP scattering. To be able to use this spectral information (for example in the context of the maximum gap method), we have simulated the detector using the Time Projection Chamber Monte Carlo (\tpcmc) code~\cite{Savage:2015tpcmc}, which in turn relies on \nest for modelling the physics of recoiling heavy nuclei~\cite{NEST:url}.  The \tpcmc output is included in the public release of \ddcalc, and provides the full gap information for XENON100.

\textbf{LUX.} The \tpcmc code has also been used to obtain the efficiencies for the first run of LUX \cite{LUX2013} and these are included in \ddcalc. For the reanalysis of the first run~\cite{LUX2016}, which improves the sensitivity to low-mass WIMPs, only the total efficiency curve provided by the collaboration is included (again scaled down by a factor of 2). In its second run, LUX saw a total of six events (compared to a background expectation of 3.5 events). This makes the use of spectral information for both signal and background an important part of the analysis. The maximum gap method is not suitable for this task, as it includes only the spectral information of the signal but treats the background as unknown. To approximately reproduce the LUX analysis, we exclude those parts of the $S1$-$S2$ plane where events are likely to have arisen from the background. Specifically we impose the requirement $3\:\text{phe} \leq S1 \leq 33\:\text{phe}$ as well as an $S2$ signal below the mean of the nuclear recoil band. We calculate the efficiency curve $\phi(E)$ for this search window using similar methods as for the first LUX run. To estimate the expected background, we assume that backgrounds due to leakage from the electron recoil band are equally distributed in $S1$, while all other backgrounds have a shape that resembles neutron recoils from calibration data. Using these assumptions we predict 2.3 background events in the reduced search window, whereas LUX observed one event. While the definition of this reduced search window contains some degree of arbitrariness and the background estimation is rather crude, this approach enables us to reproduce the published LUX bound to good accuracy (see Fig.~\ref{fig:limits_di}). Once the LUX collaboration has released more details on the analysis, the run 2 implementation can be updated accordingly.

In version 1.1.0 of \ddcalc two additional experimental analyses have been implemented. The Xenon1T experiment~\cite{Xenon1T2017} now gives the world-leading limit on SI scattering and SD scattering on neutrons, whereas the 2017 run of PICO-60~\cite{PICO2017} analysis provides the strongest constraints on SD scattering on protons.

\textbf{Xenon1T.} To implement the first results from Xenon1T~\cite{Xenon1T2017}, we consider events with $3\,\mathrm{phe} \geq S1 \geq 70\,\mathrm{phe}$ and an $S2$ signal below the mean of the nuclear recoil band. We calculate the corresponding acceptance function by simulating fluctuations in the $S1$ and $S2$ signal. For this purpose we take the scintillation and ionization yields from~\cite{LUX2016} and determine the detector response from a fit to the nuclear recoil band in~\cite{Xenon1T2017}. No events were observed in the signal region, compared to an expected background of 0.36 events.

\textbf{PICO-60 (2017).} The most recent results from PICO-60 make use of the same target material as previously employed in PICO-2L~\cite{PICO2L}. We therefore use the same acceptance function for scattering on fluorine as described above and again neglect the contribution from carbon. We include an additional factor of 0.851 to account for the selection efficiency for single scatters. No such events have been observed and we do not perform a background subtraction.

We emphasise that the accurate implementation of likelihood functions for WIMP masses below $10\:\text{GeV}$ is very challenging, as the experimental sensitivity results largely from upward fluctuations and is furthermore very sensitive to astrophysical uncertainties. A precision study of the experimental constraints on low-mass WIMPs will likely require the implementation of additional experimental information in order to refine the likelihood functions.

\subsubsection{Command-line usage}

By default, compiling \ddcalc produces static and shared libraries as well as a default executable \term{DDCalc\_run}. This latter can be run via either a command-line interface, or via an interactive mode that is entered automatically if the user supplies no arguments to the executable.

A complete list of arguments for \term{DDCalc\_run} can be obtained with:
\begin{lstterm}
DDCalc_run --help
\end{lstterm}
The command-line signature of the program is:
\begin{lstterm}
DDCalc_run [mode] [options] [WIMP parameters]
\end{lstterm}
The \term{mode} flag switches between a variety of run modes, summarised in Table~\ref{tab:DDCalcModes}. \term{WIMP parameters} is a list of arguments that can take one of four forms (in units of GeV for \term{m}, and pb for the cross-sections):
\begin{lstterm}
m
m sigmaSI
m sigmaSI sigmaSD
m sigmapSI sigmanSI sigmapSD sigmanSD
\end{lstterm}
In the first case, all WIMP-nucleon cross-sections are set to 1 pb. In the second case, only spin-independent couplings are turned on, and \term{sigmaSI} and \term{sigmaSD} are used as common cross-sections for both WIMP-proton and WIMP-neutron interactions (in the SI and SD cases, respectively).  Negative cross-sections can be given in order to indicate that the corresponding coupling should be negative (the actual cross-section will be taken as the magnitude of the value supplied).

Various general options for \ddcalc are summarised in Table~\ref{tab:DDCalcOptions}. In Table~\ref{tab:DDCalcHaloOptions}, we list the options that can be used to change the dark matter halo distribution.

The \ddcalc package includes a number of advanced detector options for defining
the precise isotopic composition of a detector, and the efficiency functions.
These are listed in Table \ref{tab:DDCalcDetOptions}, but should not be necessary for the general user, as the existing analysis flags set the defaults correctly for the experimental analyses that are included in the release.

\subsubsection{Library interface (API)}
\label{ddcalc_api}

For reference, here we include a summary of the main \ddcalc functions. These can be accessed from a \Fortran calling program with
\begin{lstfortran}
USE DDCalc
\end{lstfortran}
and from a \plainC/\Cpp program with
\begin{lstcpp}
#include "DDCalc.hpp"
\end{lstcpp}
Usage of these functions by \GB is documented in the following subsection.

\paragraph{Derived types}
\ddcalc defines three types of object. These are the bedrock of the code; almost every calculation must be provided with an instance of each of these to do its job.
\begin{description}
\item\fortran{WIMPStruct} A WIMP model, containing values of the WIMP mass and couplings.
\item\fortran{HaloStruct} A halo model, containing values of the local DM density and the parameters of the truncated Maxwell-Boltzmann velocity distribution.
\item\fortran{DetectorStruct} A detector/analysis object, containing efficiencies, the background model, energy threshold, exposure and observed events.
\end{description}

\paragraph{WIMP model}
\begin{description}
\medskip\item{\footnotesize\textbf{Initialisation}}\\
  \fortran{TYPE(WIMPStruct) FUNCTION DDCalc_InitWIMP()}\\
  Creates a new \fortran{WIMPStruct} and initialises it with a mass of 100\,GeV and couplings of 1\,pb.
\medskip\item{\footnotesize\textbf{Parameter setting}}\\
  \fortran{SUBROUTINE DDCalc_SetWIMP_mfa(WIMP,m,fp,fn,ap,an)}\\
  \fortran{SUBROUTINE DDCalc_SetWIMP_mG(WIMP,m,GpSI,GnSI,GpSD,}\\
  \fortran{\ GnSD)}\\
  \fortran{SUBROUTINE DDCalc_SetWIMP_msigma(WIMP,m,sigmapSI,}\\
  \fortran{\ sigmanSI,sigmapSD,sigmanSD)}\\
  Sets the internal parameters of the \fortran{WIMP} object.  Here \fortran{m} is $m_\chi$ in GeV, \fortran{fp} and \fortran{fn} are $\fpSI$ and $\fnSI$ in GeV$^{-2}$, \fortran{ap} and \fortran{an} are (dimensionless) $\apSD$and $\anSD$, the \fortran{G} parameters are $\GpSI$, $\GnSI$, $\GpSD$ and $\GnSD$, and the \fortran{sigma} parameters are $\sigmapSI$, $\sigmanSI$, $\sigmapSD$ and $\sigmanSD$ in pb. Negative cross-sections indicate that the corresponding coupling should be negative.  In all cases, `p' refers to proton and `n' to neutron.
\medskip\item{\footnotesize\textbf{Parameter retrieval}}\\
  \fortran{SUBROUTINE DDCalc_GetWIMP_mfa(WIMP,m,fp,fn,ap,an)}\\
  \fortran{SUBROUTINE DDCalc_GetWIMP_mG(WIMP,m,GpSI,GnSI,GpSD,}\\
  \fortran{\ GnSD)}\\
  \fortran{SUBROUTINE DDCalc_GetWIMP_msigma(WIMP,m,sigmapSI,}
  \fortran{\ sigmanSI,sigmapSD,sigmanSD)}\\
  As per \fortran{SetWIMP}, but retrieves the WIMP parameters from \fortran{WIMP}. The only difference here is that returned cross-sections are always positive, regardless of the sign of the corresponding coupling.
\medskip\item{\footnotesize\textbf{Advanced setters and getters}}\\
  See \fortran{DDCalc_SetWIMP()} and \fortran{DDCalc_GetWIMP()} in \term{DDWIMP.f90}.
\end{description}

\paragraph{Halo model}
\begin{description}
\medskip\item{\footnotesize\textbf{Initialisation}}\\
  \fortran{TYPE(HaloStruct) FUNCTION DDCalc_InitHalo()}\\
  Creates a new \fortran{HaloStruct} and initialises it to the Standard Halo Model ($\rho_\chi = 0.4$\,GeV\,cm$^{-3}$, $\vrot = 235$\,km\,s$^{-1}$, $\vmp = 235$\,km\,s$^{-1}$, $\vesc=550$\,km\,s$^{-1}$).
\medskip\item{\footnotesize\textbf{Parameter setting}}\\
  \fortran{SUBROUTINE DDCalc_SetSHM(Halo,rho,vrot,v0,vesc)}\\
  Sets the internal parameters of the \fortran{Halo} object.  Here \fortran{rho} is $\rho_\chi$ in GeV\,cm$^{-3}$, \fortran{vrot} is $\vrot$ in km\,s$^{-1}$, \fortran{v0} is $\vmp$ in km\,s$^{-1}$, and \fortran{vesc} is $\vesc$ in km\,s$^{-1}$.
\medskip\item{\footnotesize\textbf{Advanced setters and getters}}\\
  See \fortran{DDCalc_SetHalo()} and \fortran{DDCalc_GetHalo()} in \term{DDHalo.f90}.
\end{description}

\paragraph{Experiments and analysis}
\begin{description}
\medskip\item{\footnotesize\textbf{Initialisation}}\\
  \fortran{TYPE(DetectorStruct) FUNCTION}\\
  \phantom{grrrrrrrrrrrrrrrrrrrrr}\fortran{DDCalc_InitDetector(intervals)}\\
  \phantom{grrrrrrrrrrrrrrrrrrrrr}\metavar{analysis\_name}\fortran{_Init(intervals)}\\
  The first of these functions initialises an object carrying the information about the default
  experimental analysis; in \ddcalc \textsf{1.0.0} this is the LUX 2013 analysis \cite{LUX2013}.  Here \fortran{intervals} is a flag
  indicating whether calculations should be performed for intervals between observed events or not.  This is
  only necessary for maximum gap calculations and can be set to \fortran{FALSE}
  for likelihood analyses. Non-default analyses can be obtained with the specific functions listed second,
  where \metavar{analysis\_name} is one of the analyses given in Table\ \ref{tab:ddcalc-analyses}, e.g.\ \fortran{SuperCDMS_2014}. See \term{DDExperiments.f90} and \term{analyses/}\metavar{analysis\_name}\term{.f90} for more details.
  Note that these specific analysis constructors are only available directly by declaring
  \begin{lstfortran}
  USE DDExperiments
  \end{lstfortran}
  or
  \begin{lstcpp}
  include "DDExperiments.hpp"
  \end{lstcpp}
  in the calling program, in addition to the regular \fortran{USE DDCalc}/\cpp{include "DDCalc.hpp"}.  Both versions of these functions return a \fortran{DetectorStruct} containing the
  analysis details.
\medskip\item{\footnotesize\textbf{Set threshold for nuclear recoils}}\\
  \fortran{SUBROUTINE DDCalc_SetDetectorEmin(Detector,Emin)}\\
  Manually sets the minimum nuclear recoil energy discernible by a given \fortran{Detector} to \fortran{Emin}, in keV. The default value is 0, meaning that the detector response is determined by the pre-calculated efficiency curves, which account for detector and analysis thresholds. Setting \fortran{Emin} to a non-zero value allows to obtain more conservative exclusion limits or to study the dependence of the experimental results on the assumed low-energy cut-off.
\medskip\item{\footnotesize\textbf{Do rate and likelihood calculations}}\\
  \fortran{SUBROUTINE DDCalc_CalcRates(Detector,WIMP,Halo)}
  Perform the rate calculations used for likelihood and confidence
  intervals, using the analysis \fortran{Detector} on the specified \fortran{WIMP} and \fortran{Halo} models.  The results are saved internally in the \fortran{Detector} analysis
  object, and can be accessed using the following routines.
\medskip\item{\footnotesize\textbf{Retrieve results of calculations}}\\
  \vspace{-3mm}
  \begin{enumerate}
  \item Number of observed events in the analysis:\\
    \fortran{INTEGER FUNCTION DDCalc_Events(Detector)}
  \item Expected number of background events:\\
    \fortran{REAL*8 FUNCTION DDCalc_Background(Detector)}
  \item Expected number of signal events:\\
    \fortran{REAL*8 FUNCTION DDCalc_Signal(Detector)}\\
    Alternatively, for the the separate spin-independent and spin-dependent contributions:
    \fortran{REAL*8 FUNCTION DDCalc_SignalSI(Detector)}\\
    \fortran{REAL*8 FUNCTION DDCalc_SignalSD(Detector)}
  \item Log-likelihood:\\
    \fortran{REAL*8 FUNCTION DDCalc_LogLikelihood(Detector)}\\
    Uses the Poisson distribution Eq.\ \ref{eqn:Poisson}.
  \item Logarithm of the p-value:\\
    \fortran{REAL*8 FUNCTION DDCalc_LogPValue(Detector)}\\
    Uses the maximum gap method if \fortran{Detector} was initialised with
    \fortran{intervals = TRUE} and the analysis contains the necessary
    interval information to allow such a method; otherwise uses a Poisson
    distribution in the number of observed events, assuming zero background
    contribution (Eq.\ \ref{eqn:Poisson} with $b=0$).
  \item Factor by which the WIMP cross-sections must be multiplied to achieve a given p-value:\\
    \fortran{REAL*8 FUNCTION DDCalc_ScaleToPValue(lnp)}\\
    Calculates the factor $x$ by which the cross-sections must be scaled
    ($\sigma \rightarrow x\sigma$) to achieve the desired $p$-value, given via \fortran{lnp} as $\ln(p)$.
    See \fortran{DDCalc_LogPValue} above for a description of the statistics.
  \end{enumerate}
\end{description}

\paragraph{\plainC/\textsf{C{\smaller ++}} interface}
For ease of use in linking to these routines from \plainC/\Cpp code, a second
(wrapper) version of each of the interface routines described above
is defined within a \plainC namespace \cpp{DDCalc}.  These use \plainC-compatible types only,
and access interfaces to the main \ddcalc library through explicitly-specified symbol names, to get around
name-mangling inconsistencies between different compilers when dealing with \Fortran modules.  These
functions work just like the ones above, but neither accept nor
return \fortran{WIMPStruct}, \fortran{HaloStruct} nor \fortran{DetectorStruct} objects directly.  Instead,
they return and accept integers corresponding to entries in an internal
array of \Fortran objects held in trust for the \plainC/\Cpp calling program
by \ddcalc. The routines
\begin{lstcpp}
void DDCalc::FreeWIMPs();
void DDCalc::FreeHalos();
void DDCalc::FreeDetectors();
void DDCalc::FreeAll();
\end{lstcpp}
can be used to delete the objects held internally by \ddcalc.

%%%%%%%%%%%%%%%%%%%%%%%%%%%%%%%%%%%%%%%%%%%%%%%%%%%%%%%
\subsection{Direct detection implementation in \DB}

\subsubsection{WIMP-nucleon couplings}

As shown in Table\ \ref{tab:darkbitcentral}, \GB contains three functions capable of calculating effective WIMP-nucleon couplings: \cpp{DD_couplings_DarkSUSY}, \cpp{DD_couplings_MicrOmegas}, and \cpp{DD_couplings_SingletDM}. Each has capability
\yaml{DD_couplings} and returns a \cpp{DM_nucleon_couplings} object, which
contains the parameters $\GpSI$, $\GnSI$, $\GpSD$ and $\GnSD$. \cpp{DD_couplings_DarkSUSY} calculates these couplings for a generic MSSM model using \ds, while \cpp{DD_couplings_SingletDM} calculates them for the scalar singlet model internally. \cpp{DD_couplings_MicrOmegas} calculates the couplings using \micromegas for \textit{either} model -- the appropriate version of the \micromegas backend is chosen by \GB depending on which model is being scanned.

For the MSSM \cpp{DD_couplings_DarkSUSY} and \cpp{DD_couplings_MicrOmegas} compute the 
couplings by passing MSSM \cpp{Spectrum} information and nuclear parameters (described in 
more detail in the next section) to the external codes. By default, \ds does not take into 
account loop corrections to the DM-nucleon scattering process, aside of course from the 
one-loop coupling of the Higgs to gluons via the triangle diagram involving heavy quarks. 
On the other hand, certain classes of one-loop corrections are taken into account by 
default when using \micromegas, such as SUSY QCD corrections to the Higgs-exchange 
diagrams or box diagrams involving external gluons (for details we refer the reader 
to~\cite{Belanger:2008sj}). Similar corrections have been implemented in \ds 
(see~\cite{Drees:1993bu} for details), and in \GB \textsf{1.1.0} we enable these by 
default. If the user instead wishes to use the tree level \ds cross sections, they can do 
so by setting the \YAML option \yaml{loop} (default: \yaml{true}) for the function 
\cpp{DD_couplings_DarkSUSY} to \yaml{false}. Also added to this function is the option 
\yaml{pole} (default: \yaml{false}), which, when set to \yaml{false}, causes \ds to 
approximate the squark propagators in the calculation of the SI and SD couplings as $1/
m_{\tilde q}^2$ to avoid poles (this only applies when \yaml{loop} is \yaml{false}, 
otherwise this option is ignored). A final change in \GB \textsf{1.1.0} is addition of the 
\yaml{box} option (default: \yaml{true}) to \cpp{DD_couplings_MicrOmegas}, which 
determines whether the box diagrams for DM-gluon scattering are calculated for MSSM-like 
models.

In the case of the scalar singlet model, \cpp{DD_couplings_SingletDM} calculates the effective Higgs-nucleon coupling $f_N$ internally, using the nuclear matrix elements relevant for SI scattering, and combines it with the scalar mass and Higgs-portal coupling as described in Ref.\ \cite{Cline13b} to determine the effective SI couplings (there is no SD scattering in this model). \cpp{DD_couplings_MicrOmegas} uses \micromegas with our \textsf{CalcHEP} implementation of the scalar singlet model for a similar calculation.

\subsubsection{Nuclear uncertainties}
\label{matrix_elements}

When the interaction between DM and quarks can be described by a scalar operator, the
spin independent effective couplings $\GpSI$ and $\GnSI$ depend heavily on both the
sea and valence quark content of the proton and neutron respectively. These are
parameterised by the 6 nuclear matrix elements
\begin{equation}
m_N f^{(N)}_{T_q} \equiv \langle N | m_q \bar{q} q | N \rangle \, ,
\end{equation}
where $N \in \{p,n\}$ and $q \in \{u, d, s\}$. The equivalent quantities for the heavy quarks $Q \in \{c, b, t\}$ are related to these parameters via the formula \cite{Ellis:2008hf, Cline13b}
\begin{equation}
f^{(N)}_{TQ} = \frac{2}{27} \left(1 - \sum_{q = u,d,s} f^{(N)}_{T_q} \right) \, .
\end {equation}
The 6 lighter quark matrix elements are part of the \textbf{nuclear\_params\_fnq} model in \GB. They can be calculated from other quantities more closely related to lattice and experimental results. These are often the light and strange quark contents of the nucleon, defined as
\begin{align}
\sigma_l &\equiv m_l \langle N | \bar{u}u + \bar{d}d | N \rangle \\
\sigma_s &\equiv m_s \langle N | \bar{s}s | N \rangle \, ,
\end{align}
where $m_l \equiv (1/2) (m_u + m_d)$. In the
\textbf{nuclear\_params\_sigmas\_sigmal} model, these 2 parameters replace the
6 $f^{(N)}_{T_q}$ values, with the conversion between the two parameter sets
described in Ref.\ \cite{Cline13b}.
An equivalent parameterisation replaces $\sigma_s$
with the parameter $\sigma_0$:
\begin{equation}
\sigma_0 \equiv m_l \langle N | \bar{u}u + \bar{d}d - 2 \bar{s}s | N \rangle \label{eq:sigma_0} \, .
\end{equation}
By comparing the forms of the above equations, we see that $\sigma_0$ and $\sigma_l$ are related by the formula
\begin {equation}
\sigma_0 = \sigma_l - \sigma_s \left( \frac{2 m_l}{m_s} \right) \, .
\end {equation}
The \GB model \textbf{nuclear\_params\_sigma0\_sigmal} contains $\sigma_0$ and $\sigma_l$.

For the spin-dependent effective couplings $\GpSD$ and $\GnSD$, the relevant nuclear parameters are $\Delta^{(N)}_q$, which describe the spin content of the nucleons, where again $N \in \{p,n\}$ and $q \in \{u, d, s\}$. Here there are only three parameters since the values for the proton and neutron are directly related: $\Delta^{(n)}_u = \Delta^{(p)}_d$, $\Delta^{(n)}_d = \Delta^{(p)}_u$, and $\Delta^{(n)}_s = \Delta^{(p)}_s$. All of the \textbf{nuclear\_params} models contain the three $\Delta^{(p)}_q$.

All of the nuclear parameters are set in the appropriate backend when calculating the WIMP-nucleon couplings using the functions \cpp{DD_couplings_DarkSUSY} and \cpp{DD_couplings_MicrOmegas}. \cpp{DD_couplings_SingletDM}, which does not make use of an external backend, also uses the nuclear parameters in its calculation of the couplings.

A combined likelihood for $\sigma_s$ and $\sigma_l$ is implemented in \darkbit as the capability \yaml{lnL_SI_nuclear_parameters}. This capability is provided by the function \yaml{lnL_sigmas_sigmal} (Table \ref{tab:ddcap}), in which the two likelihoods take the form of simple Gaussian distributions (Eq.~\ref{eq:Gaussian}), with the default values $\sigma_s = 43 \pm 8 \textrm{ MeV}$, based on a global fit of lattice calculations \cite{Lin:2011ab}, and $\sigma_l = 58 \pm 9 \textrm{ MeV}$. This last value is based on the range of recent determinations of $\sigma_l$ in the literature from analyses of pion-nucleon scattering data \cite{Pavan:2001wz,Alarcon:2011zs,RuizdeElvira:2017stg}. $\sigma_l$ has also been extracted from lattice QCD results \cite{Alvarez-Ruso:2013fza,Bali:2016lvx,Abdel-Rehim:2016won,Shanahan:2016pla}, and the more recent analyses based on this approach point to a lower preferred value of $\sigma_l \sim 40$ MeV.

For the $\Delta^{(N)}_q$, we provide likelihoods for the following 2 combinations of parameters:
\begin{eqnarray}
a_3 &=& \Delta^{(p)}_u - \Delta^{(p)}_d \\
a_8 &=& \Delta^{(p)}_u + \Delta^{(p)}_d - 2 \Delta^{(p)}_s
\end{eqnarray}
and $\Delta^{(p)}_s$ itself.
A combined likelihood for all of these parameters is given by the capability \yaml{lnL_SD_nuclear_parameters}, which can currently only be fulfilled by the function \yaml{lnL_deltaq} (Table \ref{tab:ddcap}). The likelihoods again take the form of Gaussian distributions, with $a_3 = 1.2723 \pm 0.0023$, determined via analysis of measurements of neutron $\beta$ decays \cite{PDB}, and $a_8 = 0.585 \pm 0.023$ based on hyperon $\beta$ decay results \cite{Goto:1999by}. $\Delta^{(p)}_s = -0.09 \pm 0.03$, based on a a measurement of the spin-dependent structure function of the deuteron from the COMPASS fixed target experiment \cite{Alexakhin:2006oza}.

For both the SI and SD nuclear parameters, the central values and widths of the Gaussians can be adjusted with the \YAML options \metavar{param}\yaml{_obs} and \metavar{param}\yaml{_obserr} respectively, where \metavar{param} refers to the parameter name (see Table \ref{tab:ddcap} for the list of options).

\subsubsection{Event rates and likelihoods}

\GB \textsf{1.0.0} includes rate and likelihood functions for the eight direct detection experiments supported by \ddcalc \textsf{1.0.0}: LUX (run 1) \cite{LUX2013,LUX2016}, LUX (run 2) \cite{LUXrun2}, PandaX \cite{PandaX2016}, XENON100 \cite{XENON2013}, SuperCDMS \cite{SuperCDMS}, PICO-2L \cite{PICO2L}, PICO-60 \cite{PICO60} and SIMPLE \cite{SIMPLE2014}. As discussed above, the two dominant target elements of PICO-60 (fluorine and iodine) are implemented as independent experiments, but should always be used together. \GB \textsf{1.1.0} includes in addition rate and likelihood functions for the two new experiments included in \ddcalc \textsf{1.1.0}: Xenon1T~\cite{Xenon1T2017} and the 2017 analysis of PICO-60~\cite{PICO2017}.

At the beginning of a scan, \GB creates \ddcalc~\fortran{WIMP} and \fortran{Halo} objects, as well as a separate \fortran{Detector} object for each supported experimental analysis (see Sec.\ \ref{ddcalc_api} for explanation of these classes).  For each point in a scan, it updates the \fortran{WIMP} object with newly-calculated nuclear couplings from \darkbit, and the \fortran{Halo} object with any new halo parameters.  \darkbit provides a series of `getter' routines for different quantities that the user may be interested in, for each supported analysis: the observed and predicted number of events, the predicted number of background and signal events (broken down into SI and SD components if desired), and the final Poissonian log-likelihood for the given model and experiment.  These functions are detailed in Table\ \ref{tab:ddcap}.  Each of them depends on a single calculation routine, which passes the \fortran{WIMP}, \fortran{Halo} and relevant \fortran{Detector} object (or rather, their internal indices) to \ddcalc's own master \fortran{DDCalc_CalcRates} function (cf.\ Sec.\ \ref{ddcalc_api}).  That function then computes the rates and likelihoods for the given analysis and model point, and provides them to the getter functions.

%%%%%%%%%%%%%%%%%%%%%%%%%%%%%%%%%%%%%%%%%%%%%%%%%%%%%%%
%%%%%%%%%%%%%%%%%%%%%%%%%%%%%%%%%%%%%%%%%%%%%%%%%%%%%%%
\section{Indirect detection}
\label{sec:id}

%%%%%%%%%%%%%%%%%%%%%%%%%%%%%%%%%%%%%%%%%%%%%%%%%%%%%%%
\subsection{Background}
\label{phys_id}

Besides collider and direct searches, the third traditional way of looking for DM is to test the
particle hypothesis {\it in situ}, by
identifying the (Standard Model) products that result from DM annihilation or decay
at places with large DM densities. {\it Locally}, the injection rate of a Standard Model
particle type $f$, per volume and energy, is given by

\begin{equation}
\frac{d\mathcal{Q}(E_f,\mathbf{x})}{dE_f}= \frac1{N_\chi}
\frac{\rho_\chi^2(\mathbf{x})}{m_\chi^2} \sum_i \left\langle \sigma_i
v \frac{dN_i}{dE_f}\right\rangle\,.
\label{eqn:annihilationSource}
\end{equation}
Here, $\sigma_i$ is the annihilation cross section into final state $i$,
$v$ the relative
velocity of the annihilating DM particle pairs, $\langle...\rangle$ denotes an ensemble average over the DM
velocities, and $dN_i/dE_f$ is the (differential) number of particles $f$ that result per
annihilation into final state $i$. The dark matter mass density is given by $\rho_\chi$  and its
particle mass by $m_\chi$. $N_\chi$ depends on the nature of the DM particle, e.g.~$N_\chi=2$ for Majorana fermions, and $N_\chi=4$ for Dirac fermions.
For {\it decaying} DM, one simply would have to replace
$\langle \sigma_i v\rangle \rho_\chi^2/N_f\to m_\chi \Gamma_i \rho_\chi$ in the above
expression, where $\Gamma_i$ is the partial decay width.

The yields (i.e.~the number and energy distribution of final state
particles) are typically not significantly affected by the ensemble average, allowing
to write $\langle \sigma_i v$ $ {dN_i}/{dE_f}\rangle=\langle \sigma_i v \rangle \, {dN_i}/{dE_f}$
(and correspondingly for the decaying case),
and therefore to tabulate $\left.{dN_i}/{dE_f}\right|_{v=0}$ for a pre-defined set of centre-of-mass
energies for e.g.\ annihilation into quarks (given by the DM mass for highly non-relativistic DM).  Interpolating between these
tables rather than running event generators such as {\sf Pythia} \cite{Sjostrand:2006za} for
every model point constitutes a significant gain in performance.

The DM density enters quadratically into Eq.~\ref{eqn:annihilationSource}.
This implies that substructures in
the DM distribution (usually in form of subhalos within larger halos) in
general enhance the observable annihilation flux significantly with respect to what one
would expect in absence of substructures.  In practice, one hence often
replaces the DM density squared as follows
\begin{equation}
  \rho_\chi^2(\mathbf{x}) = [1+B_\chi(\mathbf{x})]
  \rho_{\chi, \rm nosub}^2(\mathbf{x})\;,
  \label{eqn:boost}
\end{equation}
where $\rho_{\chi, \rm nosub}^2(\mathbf{x})$ is the DM distribution smoothed
over substructures (describing the general distribution of DM in the main
halo), whereas $B_\chi(\mathbf{x})$ is the boost factor that parameterises the
enhancement due to substructure.

Once produced, those particles $f$ then propagate through a often significant
part of the Galaxy, before they reach the observer.  The details of this
process depend strongly
on the type of messenger, as well as on the type of the source. {\it Gamma rays} (Section
\ref{phys_ga}) play a pronounced role in this context as they propagate completely
unperturbed through the Galaxy, for energies below a few TeV, thus offering distinct
spatial and spectral features to look for \cite{Bringmann:2012ez}. While much harder to detect,
this property is
shared by {\it neutrinos} (Section \ref{phys_nu}); they are furthermore unique in that
they can easily leave even very dense environments and hence are the only probes of
expected high DM concentrations in celestial bodies like the Sun and the
Earth \cite{Bergstrom:1998xh}.

{\it Charged cosmic ray particles}, on the
other hand, are deflected by randomly distributed inhomogeneities in the Galactic
magnetic field such that (almost) all directional information is lost. In particular, antiprotons
provide a powerful tool to constrain DM annihilating or decaying to hadronic channels
\cite{Cirelli:2013hv,Bringmann:2014lpa,Cirelli:2014lwa}, while cosmic-ray positron data
strongly constrain leptonic channels
\cite{Bergstrom:2013jra}. While charged cosmic rays are not
included in this first release of \DB, the implementation of both propagation and  relevant
experimental likelihoods for these channels is high on the priority
list for planned extensions of the code (see Section \ref{sec:out}).

\subsubsection{Gamma rays}
\label{phys_ga}

Gamma-ray spectra from dark matter annihilation or decay can be broadly
separated in two components (see Ref.~\cite{BringmannWeniger} for a review).
(1) \textit{Continuous}
gamma-ray spectra are generated in annihilations into quarks, massive gauge
bosons and $\tau$-leptons.  The gamma-ray photons come here mostly from
the decay of neutral pions ($\pi^0\to\gamma\gamma$), which are produced
in fragmentation and hadronisation processes.
Most of the gamma-ray energy is deposited into photons with energies about an
order of magnitude below the dark matter mass.  (2) \textit{Prompt} photons are
directly produced in the hard part of the annihilation process and lead to much
sharper features in the gamma-ray spectrum, typically with energies close to
the DM mass.  The most extreme example is the annihilation into
photon pairs ($\chi\chi\to\gamma\gamma$), which gives rise to mono-energetic
gamma rays~\cite{Bergstrom:1997fj}.
Virtual internal bremsstrahlung \cite{Bringmann:2007nk} or
box-like features from cascade-decays~\cite{Ibarra:2012dw} can also play a
significant role.  Such sharp spectral features are usually much simpler to
discern from astrophysical backgrounds, and hence play a central role in
indirect dark matter searches.

Various target objects are interesting for dark matter searches with gamma
rays.  The predicted annihilation flux is largest from the centre of the Milky
Way.  However, the diffuse gamma-ray emission caused by cosmic rays along the
line-of-sight towards the Galactic bulge makes detecting a signal from the
Galactic centre subject to large systematic uncertainties.  Simpler and
basically background-free targets are dwarf spheroidal galaxies, which are dark
matter-dominated satellite galaxies of the Milky Way.  We will discuss the
results from various targets and instruments that we use in \DB in
Sec.~\ref{sec:gamLikeTargets}.

In all cases, the morphology and intensity of the gamma-ray annihilation signal
depends on the spatial distribution and clumping of dark matter in the target
object, according to Eq.~\ref{eqn:boost}.  For the various likelihoods that we
discuss below, in most cases the user can choose to either employ the spatial distribution of dark matter
in the Milky Way as defined in the halo model used in the corresponding scan (see Sec.~\ref{sec:halo}), or to
make the same assumptions on the target halos as in the corresponding
publications from which the likelihoods were extracted. In both cases, we neglect
the substructure boost, which can be of $\mathcal{O}(1)$ for reasonable
assumptions~\cite{Sanchez-Conde:2013yxa}.   It is however straightforward to
change the halo properties, if so desired.

\subsubsection{Neutrinos}
\label{phys_nu}

Like other indirect probes, searches for high-energy astrophysical neutrinos can be used to constrain the DM annihilation cross-section, by looking in directions with high DM densities such as the Galactic centre and dwarf galaxies.  Unlike other indrect searches however, neutrinos can also probe nuclear scattering cross-sections.  This is because a cross-section with nuclei implies that DM can scatter on nuclear matter in stars and other compact objects, losing sufficient energy to become gravitationally bound to the object's potential well \cite{Steigman78,Gaisser86,Gould87b,DanningerPDU}. Being on a bound orbit, the DM then returns to scatter repeatedly, eventually settling to an equilibrated, concentrated distribution within the stellar body. If it is of a variety that is able to annihilate, DM therefore does so in the stellar core, producing high-energy annihilation products.  Many of these products are stopped very quickly by interactions with nuclei, forming unstable intermediaries such as $B$ mesons, which go on to decay to other SM particles including high-energy neutrinos \cite{Blennow08}.

Neutrinos produced this way, and any produced directly in the annihilation, are able to escape the stellar body because they interact so weakly with nuclei.  They can then propagate to Earth, and be detected by neutrino telescopes.  The most promising target for this type of search by far is the Sun, owing to its proximity and deep potential well, allowing it to accumulate large amounts of DM.\footnote{We point out that such a population of DM in the Sun and other stars can have a raft of other observable consequences beyond high-energy neutrinos, which can be highly relevant for some models \cite{Scott09,Frandsen:2010yj,Taoso10,Iocco12,Vincent14,Vincent15,Vincent16}; these are slated for inclusion in future releases of \darkbit.}

At present, neutrino telescope limits on the total DM annihilation cross-section are weak \cite{ICGC, AntaresGC}, and cannot compete with gamma rays.  In \darkbit we therefore currently focus exclusively on solar searches for high-energy neutrinos, and their implications for annihilating DM models with significant nuclear scattering cross-sections.  Here `high energy' means GeV-scale neutrino energies; MeV-scale neutrinos are more difficult to distinguish from regular solar (fusion) neutrinos, and neutrinos with energies in the TeV range and above suffer from significant nuclear opacities in the Sun, making their escape difficult and substantially lowering their fluxes at Earth.

For scattering cross-sections that do not depend on the relative velocity or momentum transfer in the DM-nucleus system, the capture rate in the Sun is given by \cite{Gould87b}
\begin{equation}
\label{capture_rate}
C(t) = 4\pi\int^{R_\odot}_0 r^2\int^\infty_0 \frac{f_\odot(u)}{u}w\Omega_v^-(w)\,\mathrm{d}u\,\mathrm{d}r,
\end{equation}
where $r$ is the height from the solar centre, $u$ is the DM velocity before being influenced by the Sun's gravitational potential, and $f_\odot(u)$ is the distribution of $u$ in the Sun's rest frame.  The quantity $\Omega_v^-(w)$ is the scattering rate of DM particles from velocity $w$ to less than $v$, where $v$ is the local escape velocity at height $r$ in the Sun, and $w = w(u,r,t)\equiv\sqrt{u^2+v(r,t)^2}$ is the velocity an infalling DM particle obtains by the time it collides with a nucleus at height $r$.  $\Omega_v^-(w)$ thus describes the local capture rate at height $r$ in the Sun, from the part of the velocity distribution corresponding to incoming velocity $u$.  The total population $N_\chi$ of DM in the Sun can then be determined at any point in its lifetime by solving the differential equation
\begin{equation}
\frac{d N_\chi(t)}{dt} = C(t) - A(t) - E(t),
\label{population}
\end{equation}
where $A$ and $E$ are the annihilation and evaporation rates, respectively.  Except where DM is lighter than a few GeV, $E$ is generally negligible \cite{Gould87a, Busoni13, BusoniVincent}.  Assuming $E=0$, and that $C$ and $A$ are constant, the solution to Eq.\ \ref{population} approaches a steady state on characteristic timescale $t_\chi$, the equilibration time between capture and annihilation.  When this steady state is reached, the rate-limiting step in the whole process is capture rather than annihilation.  In this regime, the annihilation rate is identical to the capture rate, and the annihilation cross-section has no further bearing upon the number of neutrinos coming from the Sun.  Many previous analyses have assumed that capture and annihilation are in equilibrium in the Sun, so that the annihilation rate can be obtained directly from the capture rate; in \darkbit we instead solve Eq.\ \ref{population} and determine $N_\chi$ explicitly for each each model.

Knowing the DM population (and therefore annihilation rate) by solving Eq.\ \ref{population}, the annihilation branching fractions can then be used to determine the spectra of high-energy particles injected into the Sun, on a model-by-model basis.  The stopping of these annihilation products, ensuing particle production and decay, and subsequent propagation and oscillation of neutrinos through the Sun, space and Earth, have been studied by extensive Monte Carlo simulations \cite{Blennow08,Cirelli_nu}.  The resulting neutrino yield tables at Earth are included in \ds \cite{darksusy}, \pppc \cite{pppc} and \micromegas \cite{micromegas_nu}.  In \darkbit, the canonical method to obtain these fluxes is to compute the capture and annihilation rates using \ds, and to then obtain neutrino yields at Earth from the \textsf{WimpSim} \cite{wimpsimweb} tables contained therein (although getting the same from \pppc or \micromegas would also be straightforward).

Although SI scattering of DM by e.g.\ oxygen, helium and iron can dominate the capture rate for some models, the differing strength of direct limits on SI and SD scattering cross-sections, and the fact that hydrogen possesses nuclear spin, mean that typically, solar neutrinos are most interesting for SD scattering.

Neutrino telescopes are presently responsible for the strongest limits on the SD scattering cross-section with protons, with IceCube providing the tightest limits above masses of $\sim$100\,GeV \cite{IC79_SUSY,IC86}, Super Kamiokande (Super-K) dominating at lower masses \cite{SuperK15}, and ANTARES and Baksan providing weaker constraints.  We implement the IceCube search likelihood on a model-by-model basis, using the \nulike package \cite{IC22methods, IC79_SUSY} to compute the likelihood function for each model, given its predicted neutrino spectrum at Earth.  \textsf{Nulike} computes a fully unbinned likelihood based on the event-level energy and angular information contained in the three independent event selections of the 79-string IceCube dataset \cite{IC79}.  We do not implement a Super-K likelihood for now, as a) unlike IceCube, Super-K have not publicly released their event data, b) IceCube does have some sensitivity at low mass, and c) Super-K data only become more constraining for relatively light DM particles (at least in the context of SUSY and scalar singlet models, given other constraints).

%%%%%%%%%%%%%%%%%%%%%%%%%%%%%%%%%%%%%%%%%%%%%%%%%%%%%%%
\subsection{\gamlike}
\label{sec:gamlike}

\subsubsection{Overview}

\begin{table*}[t]
  \centering
  \label{tab:label}
  \begin{tabular}{lllll}
    \toprule
    Instrument & Target(s) & Notes & Energy range [GeV] & Reference \\
    \midrule
    \textit{Fermi}-LAT & 15 dSphs pass7 & composite likelihood; $J$-factor profiling & 0.5 GeV -- 500 GeV & \cite{Ackermann:2013yva} \\
    \textit{Fermi}-LAT & 15 dSphs pass8 & composite likelihood; $J$-factor profiling & 0.5 GeV -- 500 GeV & \cite{LATdwarfP8} \\
    H.E.S.S. & Galactic Halo & single $E$-bin; $J$-factor fixed; NFW & 265 GeV -- 30 TeV & \cite{Abramowski:2011hc} \\
    H.E.S.S. & Galactic Halo & single $E$-bin; $J$-factor from halo model & 265 GeV -- 30 TeV & \cite{Abramowski:2011hc} \\
    H.E.S.S. & Galactic Halo & multiple $E$-bins; $J$-factor fixed; NFW & 230 GeV -- 30 TeV &\cite{Abramowski:2011hc} \\
    H.E.S.S. & Galactic Halo & multiple $E$-bins; $J$-factor from halo model & 230 GeV -- 30 TeV &\cite{Abramowski:2011hc} \\
    \textit{Fermi}-LAT & GCE & $J$-factor fixed; contr.~NFW & 0.3 GeV -- 500 GeV & \cite{Calore:2014xka} \\
    \textit{Fermi}-LAT & GCE & $J$-factor marginalised & 0.3 GeV -- 500 GeV & \cite{Calore:2014xka, Caron:2015wda} \\
    \textit{Fermi}-LAT & GCE & $J$-factor marginalised + 10\% HEP syst.& 0.3 GeV -- 500 GeV & \cite{Calore:2014xka, Caron:2015wda} \\
    \textit{Fermi}-LAT & GCE & $J$-factor from halo model & 0.3 GeV -- 500 GeV & \cite{Calore:2014xka} \\
    \midrule
    CTA & Galactic Halo & Morphological analsiys; Einasto & 25 GeV -- 10 TeV & \cite{Silverwood:2014yza} \\
    %CTA & Galactic Halo & Morphological analsiys; $J$-factor from halo model & 25 GeV -- 10 TeV & \cite{Silverwood:2014yza} \\
    \bottomrule
  \end{tabular}
  \caption{Likelihoods and $J$-factor treatment currently implemented in
    \gamlike, for dwarf spheroidals, the Galactic halo, the \textit{Fermi} Galactic
    centre GeV excess (GCE). $J$-factors are either calculated for the halo model employed in the scan (``$J$-factor from halo model'') or
    derived based on the dark matter profiles in the respective references
    (usually regular NFW profiles, and a contracted NFW in the case of the GCE).
    The CTA likelihood is a future projection.}
  \label{tab:gamlike}
\end{table*}

Constraints on dark matter annihilation come from gamma-ray observations of
various targets using various instruments.  The experimental collaborations usually present their
results as constraints on particular annihilation
channels, using particular dark matter profiles.  This makes the limits not only
often difficult to compare, but also makes it hard to directly use the
experimental results in scans over dark matter models with complex final
states.  In order to simplify and unify the adoption of gamma-ray indirect
detection results in global scans and beyond, we present \gamlike.\footnote{We note that a tool with a similar purpose, \textsf{LikeDM} \cite{LikeDM}, which deals with both gamma rays and charged cosmic rays, has recently been released.}

The \gamlike code is released under the terms of the MIT license\footnote{\href{http://opensource.org/licenses/MIT}{http://opensource.org/licenses/MIT}} and maintained as a standalone backend
by the GAMBIT Dark Matter Workgroup. It can be obtained from
\url{http://gamlike.hepforge.org}.

The present version of \gamlike ships with the likelihood functions discussed
in Sec.~\ref{phys_ga}, which are also listed in Table \ref{tab:gamlike}.
It is written in \Cpp and can be linked either as shared library (this is how it
is used in \GB), or just as a static library.  All experimental likelihoods are
called in the same way: with a function that takes as its argument a tabulated
version of the quantity
\begin{equation}
  \frac{d\Phi}{dE} = \frac{\sigma v}{8\pi\, m_\chi^2} \frac{dN_\gamma}{dE} \;,
  \label{eqn:gamLikePhiDiff}
\end{equation}
which is the differential version of Eq.~\ref{eqn:gamLikePhi}.  The
integration over energy bins happens within \gamlike according to the energy
bins used in the various analyses.  Eq.~\ref{eqn:gamLikePhiDiff} holds for
self-conjugate dark matter particles, but can be easily adapted to
e.g.~Dirac fermion dark matter by using the appropriate prefactors as discussed in
Sec.~\ref{phys_id}.

Various options for the so-called $J$-factors, which describe the effective DM
content of the targets, are included in \gamlike as well.  These make it possible
to a marginalise or profile over $J$-factor uncertainties.  The
implementation of the combined dwarf limits from Ref.~\cite{LATdwarfP8}, for
example, performs a profiling over the $J$-factors of all 15 adopted
dwarfs separately for determining the combined likelihood value.  The various
implemented treatments are listed in Table \ref{tab:gamlike}.

\subsubsection{\gamLike targets}
\label{sec:gamLikeTargets}

\paragraph{Dwarf spheroidal galaxies with \textit{Fermi}-LAT}

Some of the most stringent and robust limits on the dark matter annihilation
cross-section come from the non-observation of gamma-ray emission from dwarf
spheroidal Galaxies (dSphs).  The most recent and stringent constraints on
gamma-ray emission from dSphs from six years of data of the \textit{Fermi} Large Area
Telescope (LAT) were derived in Ref.~\cite{LATdwarfP8}, based on the new
\texttt{Pass 8} event-level analysis.  They performed a search for gamma-ray
emission from the sources and presented upper limits that in general disfavour
s-wave dark matter annihilation into hadronic final states at the thermal rate for dark matter masses
below around 100 GeV.

The results from Ref.~\cite{LATdwarfP8} are available as tabulated binned
Poisson likelihoods.\footnote{\url{https://www-glast.stanford.edu/pub\_data/1048/}}
The composite likelihood from the dSph analysis is given by
\begin{equation}
  \ln\mathcal{L}_\text{exp} = \sum_{k=1}^{N_\text{dSph}}\sum_{i=1}^{N_\text{ebin}}
  \ln\mathcal{L}_{ki}(\Phi_i \cdot J_k)\;.
  \label{eqn:FermiTabLike}
\end{equation}
Here $N_\text{dSph}$ is the number of dwarf spheroidal galaxies included in the analysis, and $N_\text{ebin}$ is the number of energy bins to be considered.  The partial likelihoods
$\mathcal{L}_{ki}$ depend on the predicted signal flux $\Phi_i\cdot J_k$.  This
is a product of the particle physics factor
\begin{equation}
  \Phi_i = \frac{(\sigma v)_0}{8\pi m_\chi^2}\int_{\Delta E_i} dE \,
  \frac{dN_\gamma}{dE} \;,
  \label{eqn:gamLikePhi}
\end{equation}
which depends only on the gamma-ray yield per
annihilation $dN_\gamma/dE$ and the zero-velocity limit of the annihilation
cross-section ($(\sigma v)_0\equiv \sigma v|_{v\to0}$),
and the astrophysics factor
\begin{equation}
  J_k = \int_{\Delta \Omega_k} d\Omega\int_\text{l.o.s.} ds \, \rho_\chi^2\;.
  \label{eqn:J}
\end{equation}
Here, $\Delta E_i$ and $\Delta \Omega_k$ denote the energy bin $i$ and solid
angle $k$ over which the signal is integrated, $m_\chi$ is the mass of dark
matter particles and $\rho_\chi$ is the dark matter mass density in the target
object.

Our knowledge of the distribution of dark matter in dSphs relies on Jeans
analyses of the kinematics of member stars.  Following Refs.~\cite{LATdwarfP8, Chiappo:2016xfs},
to a good approximation the corresponding uncertainties for the $J$-factors can be
modelled by a log-normal distribution,
\begin{equation}
  \ln\mathcal{L}_J = \sum_{k=1}^{N_\text{dSph}} \ln\mathcal{N}(\log_{10} J_k |
    \log_{10} \hat J_k ,
  \sigma_k)\;,
\end{equation}
with parameters taken from Ref.~\cite{LATdwarfP8}, and $\mathcal{N}(x|\mu,
\sigma)$ being a normal distribution with mean $\mu$ and standard deviation
$\sigma$.

The halo and gamma-ray likelihoods can be combined by profiling over the
nuisance parameters $J_k$.  The corresponding profile likelihood is given by
\begin{equation}
  \ln \mathcal{L}_\text{dwarfs}^\text{prof.}(\Phi_i) = \max_{J_1\dots
  J_k}\left(\ln\mathcal{L}_\text{exp} + \ln\mathcal{L}_J\right)\;.
\end{equation}
An alternative is to marginalise over the nuisance parameters, which yields the
marginalised likelihood function
\begin{equation}
  \mathcal{L}_\text{dwarfs}^\text{marg.}(\Phi_i) = \int dJ_1\dots dJ_k\, \mathcal{L}_\text{exp} \cdot \mathcal{L}_J\;.
\end{equation}

The main results from Ref.~\cite{LATdwarfP8} are derived using the composite
profile likelihood for 15 dwarfs, and this is what we implemented in \gamLike.
Furthermore, for comparison, we also implemented the older results from
Ref.~\cite{Ackermann:2013yva}, which are based on four years of pass 7 data.
In both cases, the energy range spans from 500 MeV to 500 GeV.  The implemented
likelihoods are listed in Table \ref{tab:gamlike}.

Note that the effects of energy dispersion are neglected when evaluating the
constraints.  Given that current constraints from a dedicated line search are
much more constraining than the line limits that one can derive from dSph
observations, this limitation is of little practical consequence.  Implementing
\textit{Fermi}-LAT line search results is high on the priority list for future versions
of $\gamLike$ (see Sec.~\ref{sec:out}).

\paragraph{The `\textit{Fermi} Galactic centre excess'}

Gamma-ray observations of the Galactic centre with the \textit{Fermi}-LAT identified an
extended excess emission at GeV energies, which can be interpreted as a dark
matter annihilation signal (see e.g.~\cite{Hooper:2010mq,
Macias:2013vya, Abazajian:2014fta, Daylan:2014rsa, Zhou:2014lva,
Calore:2014xka, TheFermi-LAT:2015kwa}).  Although the case for millisecond
pulsars as explanation for the excess emission was strengthened in recent
analyses~\cite{Bartels:2015aea, Lee:2015fea}, it remains interesting to
consider the consequences of various DM explanations.

Ref.~\cite{Calore:2014xka} presented a spectral characterisation of the GeV
excess that included estimates of systematic uncertainties, which were derived
from residuals along the Galactic disk.  These uncertainties are correlated
over the different energy bins.  The corresponding likelihood function was
approximated to be Gaussian, and has the form
\begin{equation}
  \ln\mathcal{L}_\text{GC} = -\frac12\sum_{ij}
  (J_\text{GC}\Phi_i-f_i)\Sigma_{ij}^{-1}(J_\text{GC}\Phi_j-f_j)\;,
\end{equation}
where $f_i$ denotes the measured flux in energy bin $i$, $\Sigma_{ij}$ is the
covariance matrix, and $J_\text{GC}$ denotes the $J$-factor of the Galactic
centre Region-Of-Interest (ROI).  The considered energy range spans from 300
MeV to 500 GeV, and the ROI covers Galactic latitudes $20^\circ>|b|>2^\circ$
and longitudes $|\ell|<20^\circ$.

For the $J$-factors, the default behaviour is to employ the value calculated from the
halo model used in the corresponding scan (see Sec.~\ref{sec:halo}).\footnote{Note that here only the overall flux within the region $|\ell|<20^\circ$ and $2^\circ<|b|<20^\circ$ is rescaled.  Variations in the signal morphology are not taken into account in the current treatment.} Alternatively, one can
choose a fixed $J$-factor, $J=2.07\times10^{23}\,\rm GeV^2cm^{-5}$, corresponding to a contracted NFW profile with inner
slope $\gamma=1.2$ (see Ref.~\cite{Calore:2014xka} for details), or derive a marginalised likelihood function assuming a log-normal distribution of
$J$-factors with mean $1.96\times10^{23}\,\rm GeV^2cm^{-5}$ and standard
deviation of $0.37$ (as done in
Ref.~\cite{Caron:2015wda}, motivated by Ref.~\cite{Calore:2014nla}).

Finally, we also included a likelihood function that (on top of
astrophysical uncertainties) includes an estimated $10\%$ uncorrelated
systematic accounting for possible uncertainties in the modelling of the DM
signal spectrum. (This scenario was considered in the work of
Ref.~\cite{Caron:2015wda}, and we include it here for completeness.)  All the
available likelihoods are listed in Table \ref{tab:gamlike}.

\paragraph{Galactic centre observations with H.E.S.S.}
For dark matter masses above several hundred GeV, the current best limits on
dark matter annihilation come from observations of the Galactic centre with the
Air Cherenkov Telescope H.E.S.S.~\cite{Abramowski:2011hc}, based on 112 hours
of data.  The limits are derived from a comparison of the measured gamma-ray
fluxes in a search region within $1^\circ$ of the Galactic centre, and a
background region just outside the inner $1^\circ$.  Limits are derived from the
non-observation of an excess of gamma-ray emission in the signal region over the
flux measured in the background region.

We model the corresponding likelihood function as a Gaussian,
\begin{equation}
\begin{split}
  \ln\mathcal{L}_\text{HESS} &= \\
  -\frac12 & \sum_{i=1}^{N_\text{ebins}}
  \frac{(\Phi_i(J_\text{sig}-J_\text{bg}R) - f_i^\text{sig} - f_i^\text{bg}R)^2}{\Delta f_i^2} \, ,
\end{split}
\end{equation}
where $J_\text{sig(bg)}$ denotes the $J$-factors in the signal (background)
regions, $f_i^\text{sig(bg)}$ the corresponding fluxes in energy bin $i$, and
$R\equiv\Omega_\text{sig}/\Omega_\text{bg}$ is a geometrical rescaling factor
that depends on the angular size $\Omega_\text{sig(bg)}$ of the signal
(background) region.  For the dark matter profile, we assume by default the halo model employed in
the corresponding scan. Alternatively, one can use the NFW profile used in
Ref.~\cite{Abramowski:2011hc} (local dark matter density of
0.39\,GeV\,cm$^{-3}$ at 8.5\,kpc distance from the Sun).

In Ref.~\cite{Abramowski:2011hc} limits are derived from a combination of all
energy bins into one single wide energy bin from 265 GeV to 30 TeV.  Using this
same energy bin, we can reproduce the results from that work.  However,
Ref.~\cite{Abramowski:2011hc} also provides enough information for a spectral
analysis with 35 energy bins in the range 230 GeV to 30 TeV, which we
implemented as well (see Table \ref{tab:gamlike}).  It provides more accurate
results in cases where the DM signal has a pronounced spectral structure (like
the annihilation spectrum of $\tau$-leptons).  However, because the effects of
energy dispersion are not included in the present version of the likelihood,
results obtained with this likelihood should be interpreted with care.
Spectral features like gamma-ray lines should be constrained by results
from a dedicated line search.

\paragraph{Projected Galactic centre searches with CTA}
As discussed in Ref.~\cite{Silverwood:2014yza}, the sensitivity of the future
Cherenkov Telescope Array (CTA) to diffuse emission from DM annihilation will
be not limited by statistics, but mainly by systematic uncertainties in the
differential detector acceptance within the Field-of-View (FoV), and by the
modelling of the Galactic diffuse emission.  Ref.~\cite{Silverwood:2014yza}
addressed these issues and proposed a combined morphological analysis of the
fluxes measured in different segments of the FoV.  To this end, it is
(optimistically) assumed that the Galactic diffuse emission can be modelled well
up to an overall unconstrained normalisation.  The results of that work can
be represented as tabulated likelihood functions~\cite{Silverwood:2013tables}.

The corresponding likelihood function takes essentially the form of
Eq.~\ref{eqn:FermiTabLike} above, and covers energies between 25\,GeV and 10\,TeV.
The halo model is fixed to the Einasto profile with local density $\rho_\chi =
0.4\,\rm GeV\, cm^{-3}$, used in the original analysis (since this analysis was
taking into account morphological information, it is for the projected CTA
limits currently not possible to adopt the general halo model that is used in
the scan).  Further details about the adopted detector response,
Galactic diffuse emission model and DM profile can be found
in~\cite{Silverwood:2014yza}.

\subsubsection{Library Interface (API)}

The \gamlike library interface is the same for all implemented experiments.
There are four relevant functions that can currently be called:
\begin{itemize}
  \item \cpp{void set_data_path(const std::string & path)} \\
    Sets the path to the data files
  \item \cpp{void init(experiment_tag)} \\
    Initialise the likelihood for \cpp{experiment_tag}
  \item \cpp{double lnL(experiment_tag,}\\\cpp{\ const std::vector<double>& E,}\\\cpp{\ const std::vector<double>& dPhidE)} \\
    Retrieve log-likelihood $\ln \mathcal{L}$ for \cpp{experiment_tag}, given the tabulated $d\Phi/dE$ as in Eq.~\ref{eqn:gamLikePhiDiff}
  \item \cpp{void set_halo_profile(int mode,}\\\cpp{\ const std::vector<double>& r,}\\\cpp{\ const std::vector<double>& rho, double dist)} \\
    Initialise the Galactic halo model
\end{itemize}

The \cpp{experiment_tag} is a \Cpp\ \cpp{enum} that corresponds to one of the likelhoods listed in
Table \ref{tab:gamlike}.  The relevant functions and the enum are declared in
\term{gamLike.hpp}, and a \plainC-compliant API for e.g.\ linkage with \Fortran
code is available in \term{gamLike_C.hpp}.

The range over which $d\Phi/dE$ is tabulated should cover the energy range of
the activated experiments as shown in Table \ref{tab:gamlike}.  Outside of the
tabulated range it is assumed to be zero.  The integration over energy bins is
a simple trapezoidal integration based on the tabulation grid.  This has the
advantage that spectral features can be arbitrarily well resolved,
provided the user chooses a fine enough grid in the critical energy range (but we
note again that energy dispersion effects are not currently included in
\gamlike).

%%%%%%%%%%%%%%%%%%%%%%%%%%%%%%%%%%%%%%%%%%%%%%%%%%%%%%%
\subsection{Implementation of indirect detection in \DB}

\subsubsection{The Process Catalogue}
\label{process_catalog}
% (note non-US spelling -- catalog is fine for class names etc but not in literate text)

One of the central structures in \DB
is the `Process Catalogue'.  This is an object of \Cpp type \cpp{DarkBit::TH_ProcessCatalog}.  Functions able to generate the Process Catalog (Table~\ref{tab:darkbitcentral}) have \cross{capability} \cpp{TH_ProcessCatalog}.  The Process Catalogue carries all relevant information about
the decay and annihilation of particles.  This information is mainly used to
calculate dark matter annihilation rates, gamma-ray and neutrino yields for
indirect searches.  It can also be used for relic density calculations, although
in this case coannihilation processes are currently not supported.  The
information from the Process Catalogue is also used in the cascade annihilation
Monte Carlo, which we discuss in Sec.~\ref{sec:Cascades}.

The Process Catalogue is a simple \Cpp \space \cpp{struct}.  The internal structure
is most easily summarised using the following nested list.\\

\noindent
\fbox{\parbox{0.97\columnwidth}{
    \textbf{Process Catalogue:}
\begin{itemize}
  \item Processes
    \begin{itemize}
      \item Initial state: $\chi\chi$
        \begin{itemize}
          \item Channels
            \begin{itemize}
              \item $\bar bb$: \cpp{genRate}
              \item $\mu^+\mu^-\gamma$: \cpp{genRate}
              \item \dots
            \end{itemize}
          \item \cpp{genRateMisc}
        \end{itemize}
      \item Initial state: $h_2^0$
        \begin{itemize}
          \item \dots
        \end{itemize}
      \item \dots
    \end{itemize}
  \item Particle properties
    \begin{itemize}
      \item $\chi$: mass, spin
      \item $h_2^0$: mass, spin
      \item \dots
    \end{itemize}
\end{itemize}
}}\\

\noindent For a detailed example of how to set up the catalogue and access data within, we refer the reader to the source code of the \term{DarkBit\_standalone\_WIMP} example program described in Sec.~\ref{sec:simpleWIMP}.

The Process Catalogue carries a list of annihilation and decay processes.  Each
process has one (decays) or two (annihilations) initial states, and a list
of decay/annihilation channels.  Each channel consists of a list of two or
three final state particles (more than three final state particles are
currently not supported), as well as the specific rate for that channel given
by \yaml{genRate}, which has type \cpp{daFunk::Funk} (see Sec.~\ref{sec:dafunk}
for details).  It provides information about the decay width or the
annihilation cross section of the described process.

\begin{table}
  \begin{tabular}{lrrr}
    \toprule
    \cpp{genRate} & Units & Parameters & Process \\\midrule
    $\Gamma$ & $\GeV^{-1}$ & -- & 2-body decay \\
    $\frac{d^2\Gamma}{dEdE_1}$ & $\GeV^{-3}$ & \cpp{"E","E1"} & 3-body decay \\
    $(\sigma v)$ & $\rm cm^3\,s^{-1}$ & \cpp{"v"} & 2-body ann. \\
    $\frac{d(\sigma v)}{dEdE_1}$ & $\rm cm^{3}\,s^{-1}\, GeV^{-2}$ & \cpp{"E","E1","v"} & 3-body ann. \\
    \bottomrule
  \end{tabular}
  \caption{Overview of the various definitions of the generalised rate
  \cpp{genRate}, for different possible processes in the Process Catalog.
  Note that \cpp{genRate} is a \cpp{daFunk::Funk} object with the indicated
  parameters.  See main text for details.}
  \label{tab:genRate}
\end{table}

In the case of two-body decay, \yaml{genRate} is simply a constant that equals
the decay width $\Gamma$ in GeV (cf.~Table \ref{tab:genRate}).  In the
case of two-body annihilation, it is the annihilation
cross-section $(\sigma v)$, given as a function of the relative velocity
\cpp{"v"}.  For three-body decays, \yaml{genRate} refers
to the differential decay width, which is a function of the two kinematic
variables \cpp{"E"} and \cpp{"E1"}, corresponding to the energy of the first and second final state particles, respectively.  In the case of
three-body annihilation final states, \yaml{genRate} refers to the differential
annihilation rate, and has an additional dependence on the relative velocity
\cpp{"v"} (as in the two-body case).

Lastly, each process also has a \yaml{genRateMisc}, which describes additional
invisible contributions to the decay width or annihilation cross-section that
are not associated with a specific channel, but can affect relic density
calculations.  The different roles of \yaml{genRate} are summarised in
Table \ref{tab:genRate}.  Note that \yaml{genRateMisc} enters the
calculation of $W_\text{eff}$ from the Process Catalogue (if this how the user has chosen to obtain $W_\text{eff}$), but does not directly affect annihilation yields.

Besides the list of processes, the catalogue also comes with a list of particle
properties relevant for its processes.  This section of the catalogue maps particle IDs onto particle
masses and spin.  Masses are required for calculating decay or annihilation kinematics.  The
remaining information is currently unused, but has obvious potential future applications.

We stress that channels involving three-body final states are conceptually
different from those with two-body final states, because they cannot be implemented
independently from the two-body states, unless the contribution from any of the associated
two-body processes is absent or strongly suppressed. (An example of the latter situation is virtual internal bremsstrahlung
from neutralinos annihilating to light fermions \cite{Bringmann:2007nk}.)
In general, three-body final states  provide a \textit{correction} to the tree-level result,
and \yaml{genRate} hence returns the \textit{difference} between the full NLO
spectrum and the spectrum at tree level.  The output can therefore be positive or negative.
This implies that setting up many-body final states in the Process Catalogue requires detailed knowledge
of how the tree-level annihilation or decay spectrum is defined.\footnote{%
   Final state radiation of photons and gluons, for example, is often argued to contribute to the
   `model-independent part' of the three-body spectrum, and to therefore typically be
   included already in tabulated yields from two-body final states obtained by Monte Carlo simulation. Sometimes even electroweak final state radiation
   is added following a similar philosophy \cite{pppc}. In general, however, all these contributions are highly
   model-dependent and can differ substantially from the `model-independent' results
   \cite{Bringmann:2015cpa,Bringmann:2013oja,Bringmann:2017sko}.  For $\bar qqg$ final states, for
   example, the change in the normalisation of $\bar qq$ final states due to loop corrections
   at the same order in $\alpha_s$ must also be included for consistency \cite{Bringmann:2015cpa}.
   For three-body final states involving Higgs or  electroweak gauge bosons, it is
   challenging to even define the contribution from a single annihilation or decay channel in
   a consistent way \cite{Bringmann:2013oja,Bringmann:2017sko}. Although this can of course
   always be done \textit{formally} for the sake of fitting into the structure of the Process Catalogue, no particular physical
   significance should be associated with any individual channel in that case.  Rather, only the \textit{sum}
   over \textit{all} three-body channels provides a meaningful correction (to the \textit{total}
   tree-level yield resulting from the sum over all assocaited two-body channels).
}

While the structure of the Process Catalogue in principle allows one to take into
account all possible radiative corrections (with the above caveats in mind),
currently only three-body final states involving hard photons are included explicitly in
the Process Catalogue.  Contributions from the decay and/or fragmentation of (on-shell) final state
particles can be obtained either from tabulated yield tables (Sec.\ \ref{code_ga}) or via \darkbit's own Fast Cascade Monte Carlo (FCMC; Sec.\ \ref{sec:Cascades}).

\subsubsection{Gamma rays}
\label{code_ga}

As discussed above, the calculation of annihilation or decay spectra often
involves tabulated results from event generators like \pythia \cite{Sjostrand:2014zea}.  In order to allow
maximal flexibility with the adopted yield tables, \DB provides access to
tabulated yields using a general structure called \cpp{SimYieldTable}.
Currently, this structure makes it possible to import spectra from the \ds and \micromegas
backends\footnote{If the \YAML option \yaml{allow_yield_extrapolation} is set to \yaml{true}, the spectra from these backends are extrapolated to dark matter masses that have not been covered by the corresponding \pythia runs. By default, extrapolation is not performed, in which case the DM mass is limited to values below 5\,TeV both for \ds and \micromegas, while there is a lower limit of 10\,GeV for \ds and 2\,GeV for \micromegas. Setting the flag to \yaml{true} allows for DM masses of up to 1\,PeV and down to the mass of the final state particle both for \ds and \micromegas.}, but can easily be extended to other sources.  The information carried by
a \cpp{SimYieldTable} is summarised in the following.\\

\noindent
\fbox{\parbox{0.97\columnwidth}{
    \textbf{SimYieldTable:}
\begin{itemize}
  \item Channel list
    \begin{itemize}
      \item $\bar bb$: \cpp{dNdE}, \cpp{Ecm_min}, \cpp{Ecm_max}
      \item $\mu^+\mu^-$: \cpp{dNdE}, \cpp{Ecm_min}, \cpp{Ecm_max}
      \item \dots
    \end{itemize}
\end{itemize}
}}\\

Each one- or two-particle spectrum is defined by the ID(s) of the initial
particle(s), the ID of the tabulated final-state particle (currently only
gamma-rays are implemented), and the
centre-of-mass energy range for which the tabulated spectrum is available.  The
spectrum itself is conveniently wrapped into a \cpp{daFunk::Funk} object (like
\cpp{genRate} above, see Appendix \ref{sec:dafunk} for more details).  We
emphasise that this
object does not, in most cases, directly carry the tables from \ds or
\micromegas, but is merely a convenient and flexible wrapper that directly
calls the corresponding backend functions.

In extreme cases, the tabulated yields implemented in different dark matter
codes can differ substantially -- mostly due to being based on different
(versions of) event generators, but also due to different methods (or the
absence of a method) for including contributions from higher-order processes.  With the above
structure, \DB offers a flexible and convenient way to select the desired
yields and switch between them for detailed comparisons.

The gamma-ray yield is calculated by module functions with the capability
\cpp{GA_AnnYield},  based on information from both the \cpp{ProcessCatalog} and the
\cpp{SimYieldTable}.  These are outlined in Table \ref{tab:darkbitgammacap}. Note that the resulting spectra are in general only
partially based on the \cpp{SimYieldTable}, and can also make use of analytic
expressions for e.g.~three body final states, or include results from
the FCMC (Sec.\ \ref{sec:Cascades}). The result of
\cpp{GA_AnnYield} is a \cpp{daFunk::Funk} object.  It refers to the
physical expression $m_\chi^{-2}\cdot \sigma v\cdot dN/dE$,
which is equivalent to Eq.~\ref{eqn:gamLikePhiDiff} up to a factor of $8\pi$ .  It is a
function of the photon energy \cpp{"E"} and of the relative velocity \cpp{"v"} (currently, only the $v=0$ case is used for
actual likelihood evalulations; adding velocity-dependent effects is planned
for the near future).  Note that the function object is in general a
composite object that wraps various \ds and \micromegas functions
providing e.g.~tabulated yields or differential cross-sections for
three-body final states.  Calling this generalised function at different
values of $E$ or $v$ calls all of these backend functions behind the scenes, sums up
and rescales their results, and performs phase space integrations if necessary.
Note that the calculation of gamma-ray spectra often involves internal
integrations (over 3-body phase space, or for boosting spectra into the lab
frame).  In cases where the integrations fail, a warning is issued, and the
integration returns zero (see appendix \ref{sec:dafunk}).  This implies that
derived upper limits are in all cases conservative.

If results from the FCMC are required, the Monte Carlo simulation is
automatically run before the \cpp{GA_AnnYield} module function (see
Sec.~\ref{sec:Cascades}).  It is the job of module functions with capability
\cpp{GA_missingFinalStates} to determine, by comparing Process Catalogue
entries and the \cpp{SimYieldTable}, for which final states the cascade
annihilation Monte Carlo is necessary.

\begin{figure}[t]
  \centering
  \includegraphics[width=0.95\linewidth]{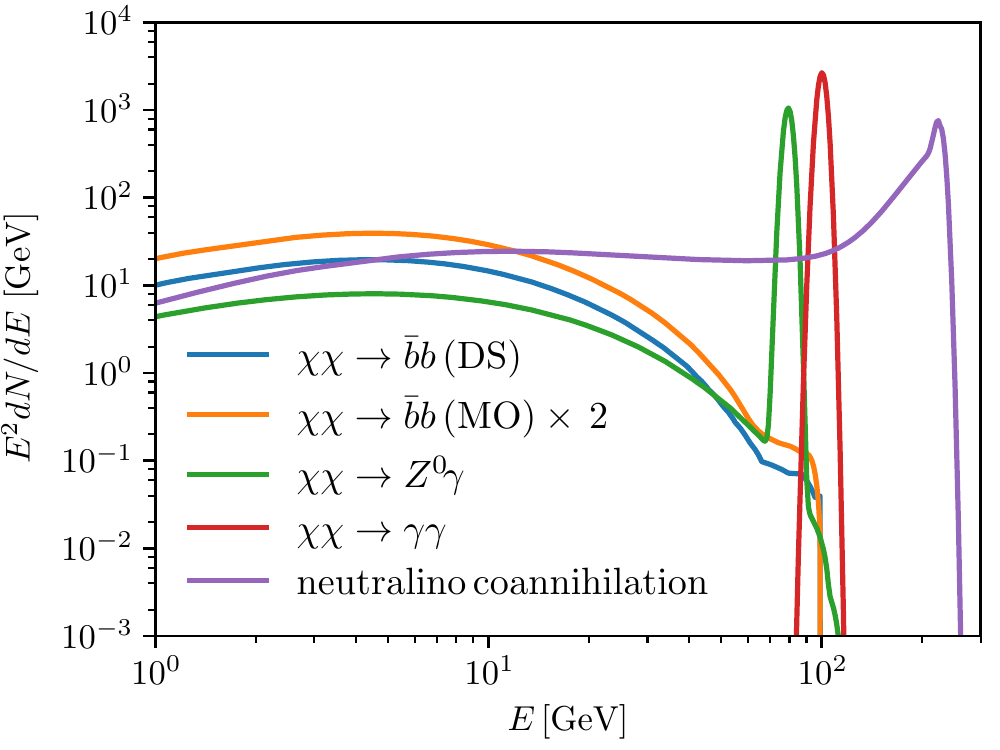}
  \caption{Example spectra generated with \DB, using tabulated 2-body
  final states from \ds and \micromegas, line-like spectra
  assuming an energy resolution of $3\%$, and a benchmark case from MSSM
  neutralino annihilation in the stau-coannihilation region.  The DM mass is
  $m_\chi = 226\rm\, GeV$ (neutralino) or $m_\chi = 100\rm\, GeV$ (all other cases).}
  \label{fig:tab_spectra}
\end{figure}

In Fig.~\ref{fig:tab_spectra}, we show a number of annihilation spectra
generated with \cpp{GA_AnnYield}, comparing the yields obtained from
various backends, including line features.  The processes are indicated in the
legend of the figure.  One of the spectra shown in Fig.~\ref{fig:tab_spectra}
corresponds to annihilation into $Z^0\gamma$ final states.  The actual spectrum
is calculated as combination of a monochromatic $\gamma$ and the gamma-ray yield from decay of a single $Z^0$.
The latter is approximated by the tabulated spectrum from threshold decay into
$Z^0Z^0$ final states, with the photon yield divided by two.
%We use tabulated spectra from on-shell
%$Z^0$ decays, in order to avoid unphysical super-threshold photons that would
%occur when constructing single $Z^0$ spectra from tabulated $Z^0Z^0$ yields.

Gamma-ray likelihood functions in \DB make use of the
backend \gamLike (see Sec.~\ref{sec:gamlike} and Table \ref{tab:gamlike}).  As detailed in Table \ref{tab:darkbitIDlike}, the resulting
likelihoods are wrapped in various \DB module
functions, with one capability for each experiment-target pair.  The different
options concerning $J$-factors or the version of a measurement can be selected with
the run-time option \yaml{version} in the \YAML file.  The list of available
capabilities and modes is:\begin{itemize}
\item\cpp{lnL_FermiLATdwarfs} (\yaml{version = "pass7"}, \yaml{"pass8"}),
\item\cpp{lnL_FermiGC} (\yaml{version = "fixedJ"}, \yaml{"margJ"}, \\\yaml{"margJ_HEP"}, \yaml{"externalJ"}),
\item\cpp{lnL_HESSGC} (\yaml{version = "integral_fixedJ"},\\ \yaml{"spectral_fixedJ"}, \yaml{"integral_externalJ"},\\ \yaml{"spectral_externalJ"}),
\item\cpp{lnL_CTAGC} ().
\end{itemize}

\subsubsection{Neutrinos}
\label{code_nu}

The different neutrino indirect detection capabilities of \darkbit are summarised in Table \ref{tab:darkbitnucap}.  The capabilities that describe the relevant WIMP properties are listed in Table~\ref{tab:darkbitdirect}.

The neutrino routines in \darkbit use the DM mass and nuclear scattering cross-sections to first calculate the DM capture rate in the Sun (capability \cpp{capture_rate_Sun}).  The canonical way to do this is to call the corresponding function from \ds, which (at least in v\textsf{5.1}) assumes the AGSS09ph solar density profile \cite{AGSS, Serenelli09}, and does not distinguish between scattering on protons and neutrons.  \darkbit uses the SI and SD cross-sections on protons for this purpose.

The capture rate is then used together with the late-time annihilation cross-section to solve Eq.\ \ref{population} for the equilibration time $t_\chi$ and annihilation rate (capabilities \cpp{equilibration_time_Sun} and \cpp{annihilation_rate_Sun}).  Together with the contents of the \cpp{ProcessCatalog}, the annihilation rate is used to prime the calculation of the neutrino spectrum at Earth (essentially by setting appropriate common blocks in \ds with annihilation and Higgs decay information), producing a pointer to a function in \ds that can return the neutrino yield at the IceCube detector (capability \cpp{nuyield_ptr}).\footnote{The neutrino yield can also be calculated using routines in \micromegas, but these functions are currently not backended in \GB. We plan to add the ability to use them as an alternative to the \ds calculation in a future version of \DB.}

This pointer is passed to \nulike \cite{IC79_SUSY}, which uses it to convolve the predicted differential neutrino flux with the various IceCube detector response functions, and evaluate the overall neutrino telescope likelihood for the model.  \darkbit provides individual likelihoods from each of the three independent event selections included in the original 79-string IceCube analysis \cite{IC79}: winter high-energy (WH), winter low-energy (WL) and summer low-energy (SL).  The combined likelihood from all three of these searches is provided under the capability \cpp{IC79_loglike}; the individual likelihoods correspond to \cpp{IC79WH_loglike}, \cpp{IC79WL_loglike} and \cpp{IC79SL_loglike}.  The earlier 22-string likelihood \cite{IceCube09,IC22methods} is also available as \cpp{IC22_loglike}, and combined with all 79-string likelihoods as simply \cpp{IceCube_likelihood}.  More information about these capabilities is available in Table \ref{tab:darkbitnulikecap}.

When combining different search regions like this, we work with an effective log-likelihood defined as the difference between the actual log-likelihood and the log-likelihood of the background-only model.\footnote{This is the same strategy as employed by \colliderbit in combining different LHC searches; more information on the procedure can be found in that paper \cite{ColliderBit}.  This case is somewhat simpler than the \colliderbit one, as we do not allow different signal region combinations for different model parameter values, given that the IceCube event selections are statistically independent by construction.  Also unlike the LHC searches implemented in \colliderbit, here the different signal regions do not come with different numbers of counting bins --- each has exactly one such `bin', the Poisson counting term at the front of its likelihood function --- but different event selections \textit{do} bring different numbers of event terms into the unbinned part of the likelihood function \cite{IC79_SUSY}.}  We also impose a hard upper limit of zero to this effective likelihood, to prevent overfitting of spurious features at low energies, where the IceCube limits degrade steeply and the instrumental response functions (especially in energy) are least well understood.  This has the impact of preventing exclusion of the background hypothesis, in much the same way as the $CL_s$ method \cite{CLs1,CLs2}.

\begin{figure}[t]
  \centering
  \includegraphics[width=\linewidth]{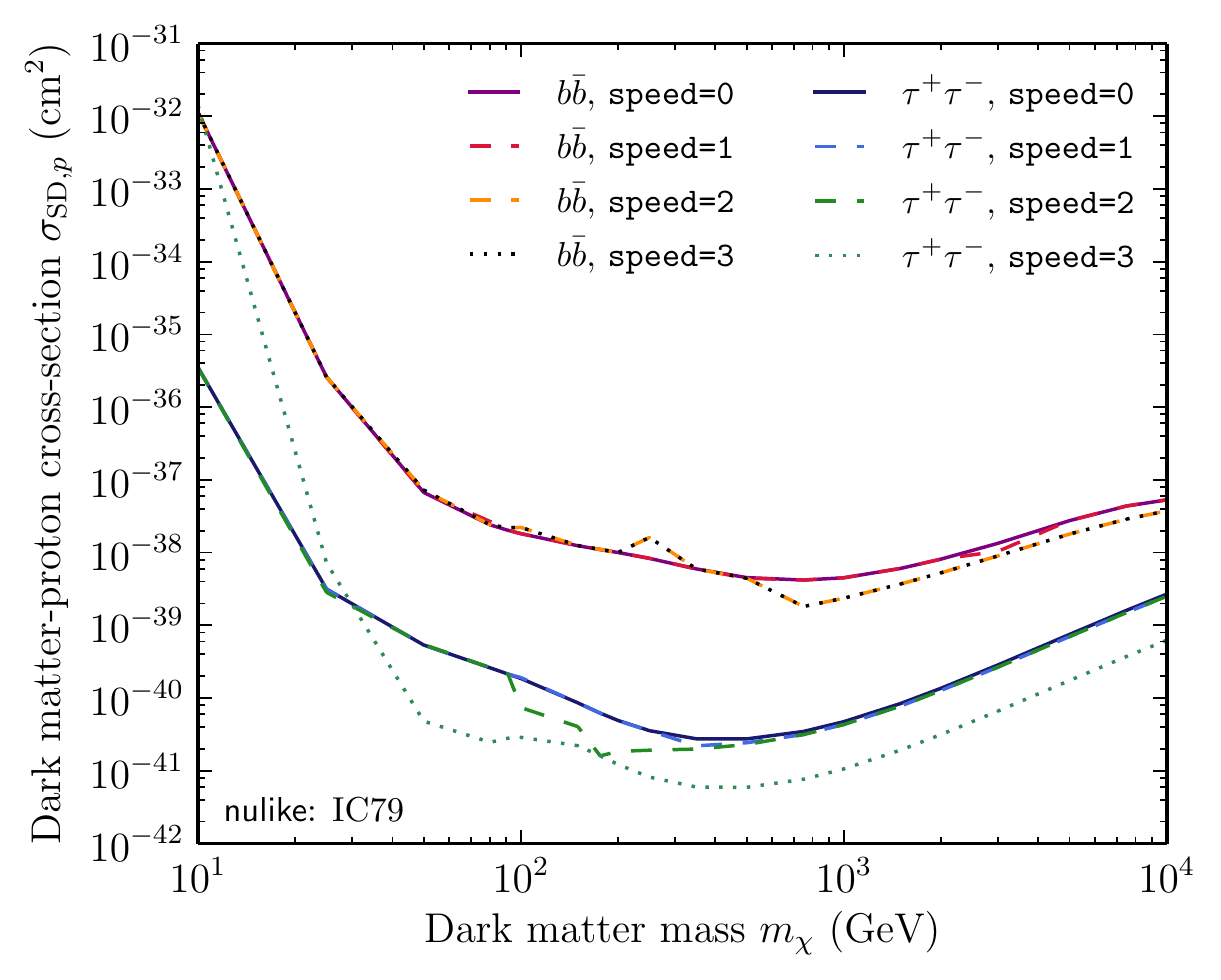}
  \caption{Comparison of different limits obtained from 79-string IceCube data using \nulike, with different \protect\fortran{speed} settings.  The default is \protect\fortran{3}, but the \protect\fortran{speed} setting is configurable from the master \YAML file via the module function option \protect\yaml{nulike_speed}.}
  \label{fig:nulike_speed}
\end{figure}

The \nulike\ \fortran{speed} parameter can be chosen from the master initialisation file, by setting the module function option \yaml{nulike_speed}.  The default is \yamlvalue{3}.  For production scans, we run \nulike with this default, as this allows \omp-enabled neutrino likelihood evaluations for models with appreciable signal fractions to be achieved in walltimes of order one second.  Parameter combinations resulting in negligible signal fractions run much faster, so the mean runtime is well below a second.  This does come with an accuracy cost; Fig.\ \ref{fig:nulike_speed} compares the accuracy of some example limits obtained with the different speed settings.

Following the original \nulike paper \cite{IC79_SUSY}, we assume a flat theoretical error on the predicted neutrino yield of 5\% for DM masses below 100\,GeV, rising logarithmically to 50\% at masses of 10\,TeV, and onward at even higher masses.  This form of the error term is designed to account for neglected higher-order contributions and round off errors, present at all masses, and the increasing error introduced by \ds's interpolation in its \textsf{WimpSim} tables with increasing mass above $\sim$100\,GeV.

As side products of the various likelihood calculations, \darkbit also provides module function access to various related \nulike outputs for the different analyses (Table \ref{tab:darkbitnulikecap}).  For \metavar{X} $\in\{$\cpp{IC22}$,$ \cpp{IC79WH}$,$ \cpp{IC79WL}$,$ \cpp{IC79SL}$\}$, these are:
\begin{description}
\item[\CPPidentifierstyle\metavar{X}\_signal] The predicted number of signal events (from DM annihilation in the Sun).
\item[\CPPidentifierstyle\metavar{X}\_bg] The predicted number of background events.
\item[\CPPidentifierstyle\metavar{X}\_nobs] The total number of observed events.
\item[\CPPidentifierstyle\metavar{X}\_bgloglike] The likelihood for the background-only model.
\item[\CPPidentifierstyle\metavar{X}\_pvalue] A simple $p$-value for the model, based only on the number of events observed in the individual analysis, i.e.\ discarding event-level information.
\end{description}
These are all extracted from \cpp{nudata} objects, which are returned by functions with capabilities \metavar{X}\cpp{_data}.

\subsubsection{Fast Cascade Monte Carlo (FCMC)}
\label{sec:Cascades}

In \DB, the calculation of gamma-ray yields from cascade decays is implemented
as a Monte Carlo, as the kinematics of long decay chains are too
complicated to handle analytically. Codes like \ds, for example, take simplified
angular averages over decay phase spaces.

The cascade decay code has two main parts: a decay chain Monte Carlo and an
accompanying analysis framework.  The Monte Carlo code generates random decay
chains based on relative branching fractions for individual decays.  Currently,
all particles in the decay chain are assumed to be on-shell, and only two-body
decays are allowed in each step of the chain.  Furthermore, spin correlations are neglected.
The analysis framework
interpolates the resulting histograms and creates wrapper functions for these
interpolated spectra, which can be used in e.g.~\cpp{GA_AnnYield} in order
to derive the total annihilation yield.
The generation of decay chains, and the following analysis steps are fully
\omp-enabled.  Each available CPU core independently generates and analyses
events.

As mentioned in Sec.\ \ref{code_ga}, it is the responsibility of each module function with
capability \cpp{GA_missingFinalStates} to deterime, by comparing Process
Catalogue entries with the \cpp{SimYieldTable}, for which initial state particles
gamma-ray yields are required.
The FCMC then generates decay chains for each of the identified initial states by Monte
Carlo.

The decay chains in \DB are implemented as a doubly-linked tree:
each particle in the decay chain is represented by an instance of a class named
\cpp{ChainParticle}, which contains pointers to its parent particle, and any
child particles (decay products).  A decay chain is generated by first creating
an initial state \cpp{ChainParticle}, which is initialised using a decay table
containing relevant masses and decays for all particles that can occur in the
cascade.  The \cpp{ChainParticle} class features a member function named \cpp{
generateDecayChainMC}, which recursively generates a decay chain.  The function
uses the FCMC-internal decay table (see below) to select a decay from the list of possible
processes, using probabilities given by the relative branching fractions for
the allowed decays.\footnote{Relative, as the branching ratios of the two-body
decays may not always sum up to one.}

The recursive decay continues until particles that are stable or have
pre-computed decay spectra are reached, or until one of the pre-defined cutoff conditions
is reached.  These cutoff conditions are the maximum number of allowed decay
steps, as specified by the \YAML option \yaml{cMC_maxChainLength}, and a cut on the lab frame\footnote{`Lab frame' means the rest frame of the initial state in the cascade.} energy of
the decaying particle, specified by the \YAML option \yaml{cMC_Emin}.  The cutoff is triggered by whichever of these two conditions is
reached first.

Once a full decay chain has been generated, the final state particles of the
chain are collected and analysed, and final states of interest are
histogrammed.  Tabulated final state spectra
are boosted to the lab frame. To this end, we sample photon
energies from the tabulated spectra, boost these to the lab frame, and add the
corresponding box spectra to the result histogram.

In the remaining part of this section, we provide information about the
implementation in \DB.  The relevant functions and capabilities are summarised in Tables \ref{tab:darkbitcascadecap} and \ref{tab:darkbitcascadeloopcap}.
The module function \cpp{cascadeMC_FinalStates} provides
a list of string identifiers (\cpp{"gamma"}, \cpp{"e+"}, \cpp{"pbar"}, etc)
that indicate which final states need to be calculated.  This list can be set
in the master \YAML file using the option \yaml{cMC_finalStates}.  The default
(and currently only supported) option
is a list with the single entry \mbox{\cpp{"gamma"}}.  The function
\cpp{cascadeMC_DecayTable} generates an FCMC-internal list of all relevant decays
(based on the content of the Process Catalogue).  Next, the loop
manager \cpp{cascadeMC_LoopManagement} runs a sequence of module functions with
capabilities \yaml{cascadeMC_InitialState} $\to$ \yaml{cascadeMC_ChainEvent}
$\to$ \yaml{cascadeMC_Histograms} $\to$ \yaml{cascadeMC_EventCount} that takes
care of generating the Monte Carlo samples and histograms.
Finally, \cpp{cascadeMC_gammaSpectra} uses the histogram results to generate interpolating \cpp{daFunk::Funk}
objects, which are used in \yaml{GA_AnnYield}.

Various options are available to tune the FCMC.  We list them here, with default values in square brackets.
\begin{description}
  \item\yaml{cascadeMC_LoopManagement}:\begin{description}
    \item\cpp{int} \yaml{cMC_maxEvents[20000]}: sets the maximum number of MC runs per point.
  \end{description}
  \item\cpp{cascadeMC_GenerateChain}:\begin{description}
    \item\cpp{int} \yaml{cMC_maxChainLength[-1]}: the maximum number of decay steps to consider.  A value of \yaml{-1} indicates that no maximum should be applied.
    \item\cpp{double} \yaml{cMC_Emin[0.0]}: the minimum lab-frame energy of particles to be tracked, in GeV.
  \end{description}
  \item \cpp{cascadeMC_Histograms}:\begin{description}
    \item Histogramming options:\begin{description}
      \item\cpp{int} \yaml{cMC_NhistBins[140]}: number of logarithmic bins of the generated histogram.
      \item\cpp{double} \yaml{cMC_binLow[0.001]}: the lowest energy in GeV of the generated histogram.
      \item\cpp{double} \yaml{cMC_binHigh[10000]}: the highest energy in GeV of the generated histogram.
      \item\cpp{int} \yaml{cMC_numSpecSamples[25]}: the number of samples to draw from tabulated spectra.
    \end{description}
    \item Convergence criteria:\begin{description}
      \item\cpp{double} \yaml{cMC_gammaRelError[0.20]}: the maximum allowed relative error in the bin with the highest expected signal-to-background ratio.
      \item\cpp{double} \yaml{cMC_gammaBGPower[-2.5]}: power-law slope to assume for astrophysical background when calculating the position of the bin with the highest expected signal-to-background ratio.
      \item\cpp{int} \yaml{cMC_endCheckFrequency[25]}: the number of events to wait between successive checks of the convergence criteria.
    \end{description}
  \end{description}
\end{description}

Note that if \yaml{cMC_maxEvents} is exceeded in \cpp{cascadeMC_} \cpp{LoopManagement} before convergence is reached in \cpp{cascadeMC_Histograms}, the Monte Carlo will terminate before any convergence criteria are met.

\begin{figure}[t]
  \centering
  \includegraphics[width=0.95\linewidth]{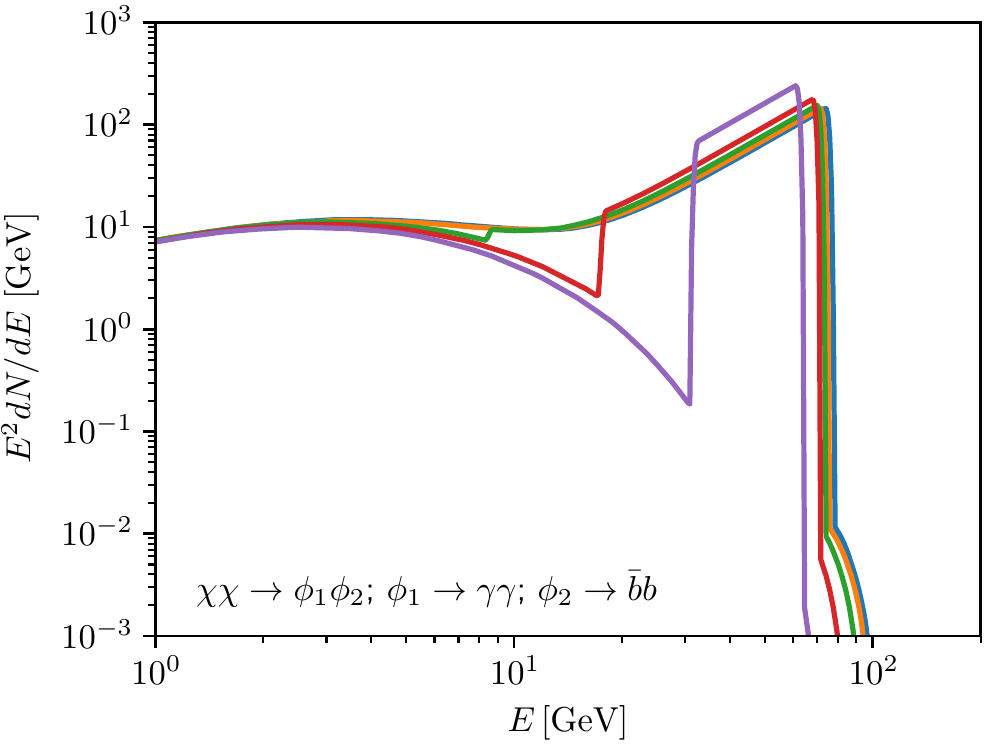}
  \caption{Gamma-ray spectra from DM cascade annihilation, generated with the \DB FCMC. DM particles, $\chi$, annihilate into bosons $\phi_1$ and $\phi_2$, which decay, respectively, in photon pairs and bottom quark pairs.  For all lines, we assume $m_\chi = 100\,\GeV$ and $m_{\phi_2} = 100\,\GeV$.  The mass of $\phi_1$ varies between $m_{\phi_1} = 10, 30, 50, 70, 90\,\GeV$ (with the lowest mass corresponding to the magenta line with the pronounced box-like feature on the right).}
  \label{fig:MC_spectra}
\end{figure}

We show example spectra generated with the FCMC in Fig.~\ref{fig:MC_spectra}.  These
which were set up using the \DB WIMP standalone discussed in Sec.\ \ref{sec:simpleWIMP}. Specifically,
we set up a pair of DM particles annihilating to two scalars,
$\chi\chi\to\phi_1\phi_2$, where $\phi_1$ decays to a pair of photons and $\phi_2$
to $\bar b b$. In the rest frame of $\phi_1$, the resulting photons are monochromatic;
in the galactic rest-frame of the annihilating DM particles, this leads to a flat ``box" feature
with the following spectrum (see  e.g.\ \cite{Ibarra:2012dw}):
\begin{equation}
\label{eq:box}
 \frac{dN_{\phi\to\gamma\gamma}}{dE_\gamma}=\frac{2}{\Delta E}\theta(E_\gamma-E_\mathrm{min})
 \theta(E_\mathrm{max}-E_\gamma)\,.
\end{equation}
Here, $\theta$ is a step function, $\Delta E \equiv E_\mathrm{max}-E_\mathrm{min}$ and
\begin{equation}
E_\mathrm{max,min}=(E_{\phi_1}/2)\left(1\pm\sqrt{1-m^2_{\phi_1}/E_{\phi_1}^2} \right),
\end{equation}
where
\begin{equation}
E_{\phi_1}=m_\chi\left[1+(m_{\phi_1}^2-m_{\phi_2}^2)/(4m_\chi^2)  \right]
\end{equation}
is the energy of $\phi_1$.  These boxes are clearly seen in Fig.~\ref{fig:MC_spectra}, with endpoints and normalisation
of the numerical results agreeing nicely with the above analytical expression.

The decay $\phi_2\to\bar bb$ produces a continuum spectrum of photons from the
tabulated yields of $\bar bb$ final states. Compared to the direct annihilation of DM to
 $\bar bb$, as in $\chi\chi\to\bar bb$, the form of the resulting photon spectrum is roughly
 retained but the peak
 (in $E_\gamma^2 dN/dE_\gamma$) is shifted down by a factor of about
 2 in energy \cite{Elor:2015bho} -- essentially because each of the quarks now has on average
 only half the kinetic energy at its disposal. This part of the spectrum is clearly visible in
Fig.~\ref{fig:MC_spectra} as the ``background'' of the box feature discussed above.
At high energies, this part of the FCMC-produced spectrum is also seen to be affected
by the mass of $\phi_1$ (for constant $m_{\phi_2}=m_\chi=100$\,GeV). The reason is that
the largest photon energy kinematically available from $\phi_2\to\bar bb$ is given by
$E_\mathrm{max}$ as provided in the expression after Eq.~\ref{eq:box}, with
${\phi_1}$ and ${\phi_2}$ interchanged.
Again, this is in agreement with the location of the cutoff visible in the figure.

We finally note that the cascade code in the current form does not handle
off-shell decay, and neglects the finite widths of particles in the kinematics.

%%%%%%%%%%%%%%%%%%%%%%%%%%%%%%%%%%%%%%%%%%%%%%%%%%%%%%%
%%%%%%%%%%%%%%%%%%%%%%%%%%%%%%%%%%%%%%%%%%%%%%%%%%%%%%%
\section{Examples}
\label{examples}

In this Section we present a few selected examples that illustrate the scope
and potential applications of \DB.  At the same time, these examples serve as
validation tests of the code.

\begin{table}[t!]
  \centering
  \begin{tabular}{l c c}
  \toprule
  \textbf{Model} & \textbf{Parameter} & \textbf{Value} \\
  \midrule
  \multirow{2}{*}{Singlet DM} & $\lambda_{hS}$                & 0.03 \\
                              & $m_S$ [GeV]                   & 90 \\
  \midrule
  \multirow{5}{*}{CMSSM}      & $M_0$ [GeV]                   & 3075 \\
                              & $M_{1/2}$ [GeV]               & 465 \\
                              & $\tan \beta$                  & 51         \\
                              & $A_0$ [GeV]                   & 1725       \\
                              & sign($\mu$)                   & $+$        \\
  \midrule
  \multirow{8}{*}{MSSM 7}     & $M_2$ [GeV]                   & 690 \\
                              & $m_{H_d}^2$ ${\rm [GeV^2]}$ & $9.86 \times 10^7$ \\
                              & $m_{H_u}^2$ ${\rm [GeV^2]}$ & $1.4 \times 10^4$  \\
                              & $\tan \beta$                & 23   \\
                              & $m_f^2$ ${\rm [GeV^2]}$     & $3.8 \times 10^6$  \\
                              & $A_d$ [GeV]                 & 1000  \\
                              & $A_u$ [GeV]                 & 2680  \\
                              & sign($\mu$)                 & $+$    \\
  \bottomrule
  \end{tabular}
  \caption{\label{tab:central} Central points of spokes plotted in Fig.~\ref{fig:spokes}. All of the parameters of the MSSM 7 are defined at an energy scale of 1 TeV.}
\end{table}

\begin{table}[t!]
  \begin{tabular}{l c c}
  \toprule
  \textbf{Model} & \textbf{Parameter} & \textbf{Range} \\
  \midrule
  \multirow{2}{*}{Singlet DM} & $\lambda_{hS}$                & [0.01, 0.05] \\
                              & $m_S$ [GeV]                   & [70, 110] \\
  \midrule
  \multirow{2}{*}{CMSSM}      & $M_0$ [GeV]                   & [2840, 3310] \\
                              & $M_{1/2}$ [GeV]               & [340, 590] \\
  \midrule
  \multirow{2}{*}{MSSM 7}     & $M_2$ [GeV]                   & [450, 850] \\
                              & $m_{H_d}^2$ ${\rm [GeV^2]}$ & [$9.2 \times 10^7$, $1.03 \times 10^8$] \\

  \bottomrule
  \end{tabular}
\caption{\label{tab:params} Ranges that parameters are varied over in Fig.~\ref{fig:spokes}. All of the parameters of the MSSM 7 are defined at an energy scale of 1 TeV.}
\end{table}

\begin{figure*}[t]
\centering
\includegraphics[width=0.49\linewidth]{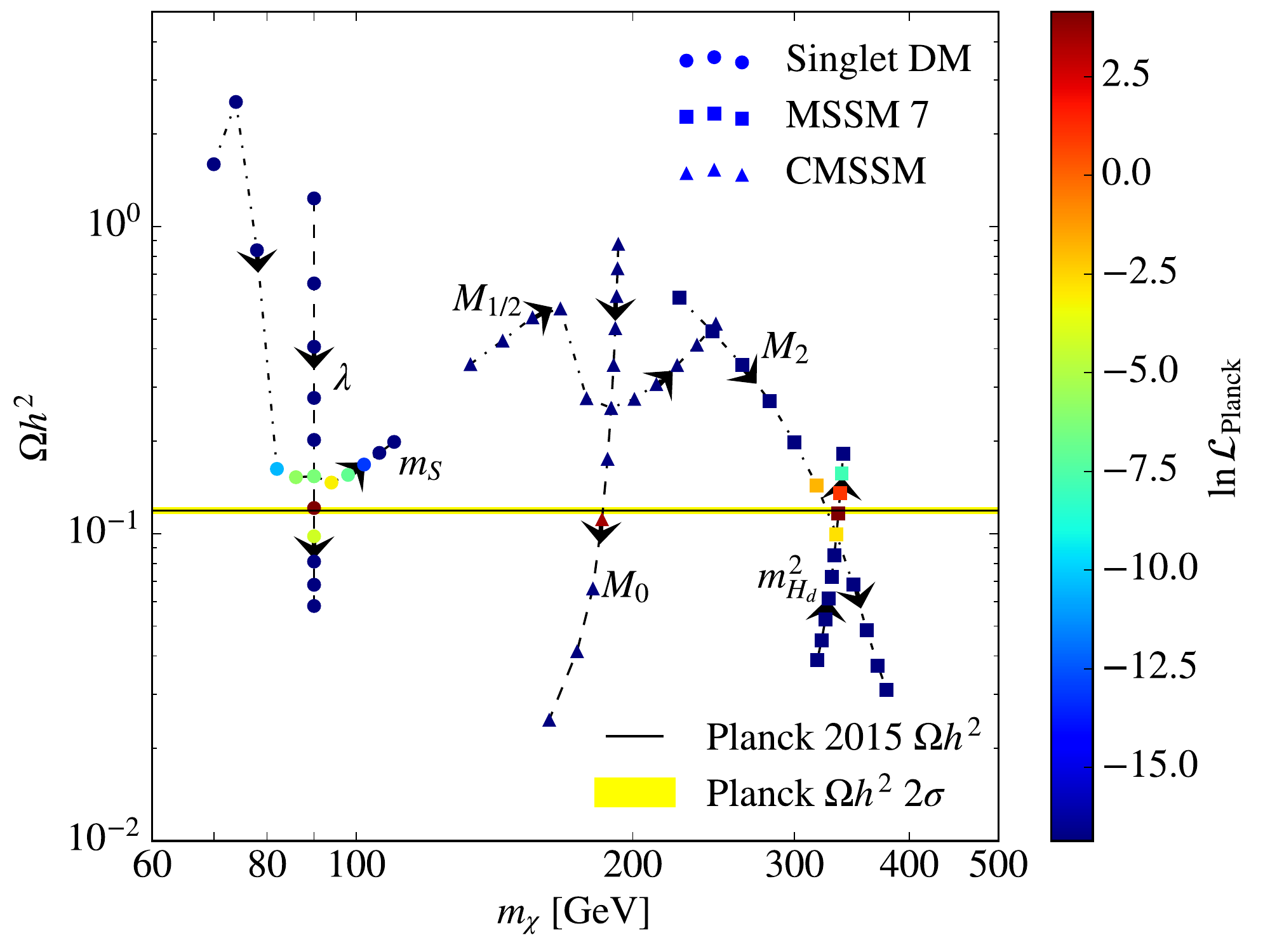}
\includegraphics[width=0.49\linewidth]{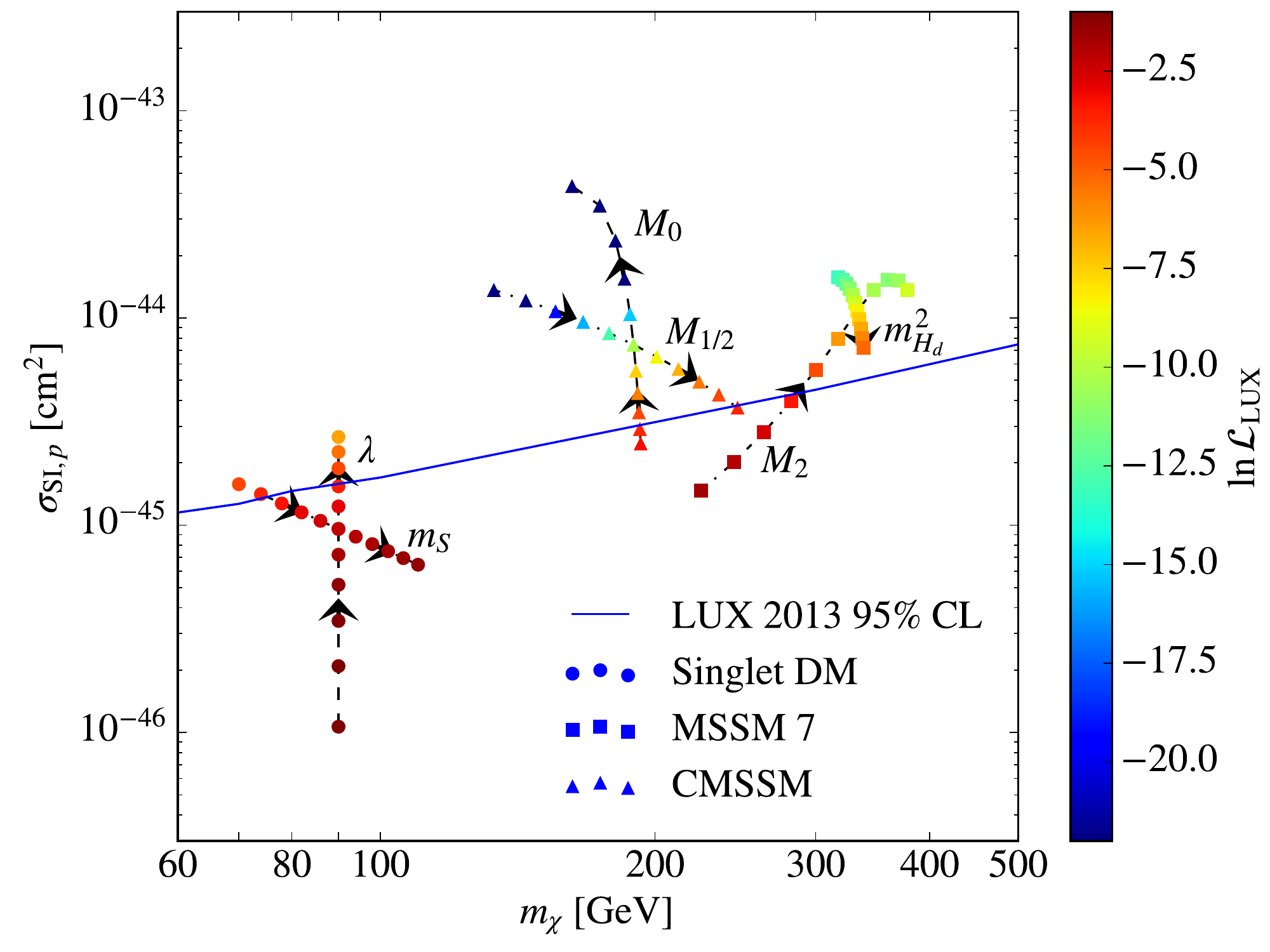}
\includegraphics[width=0.49\linewidth]{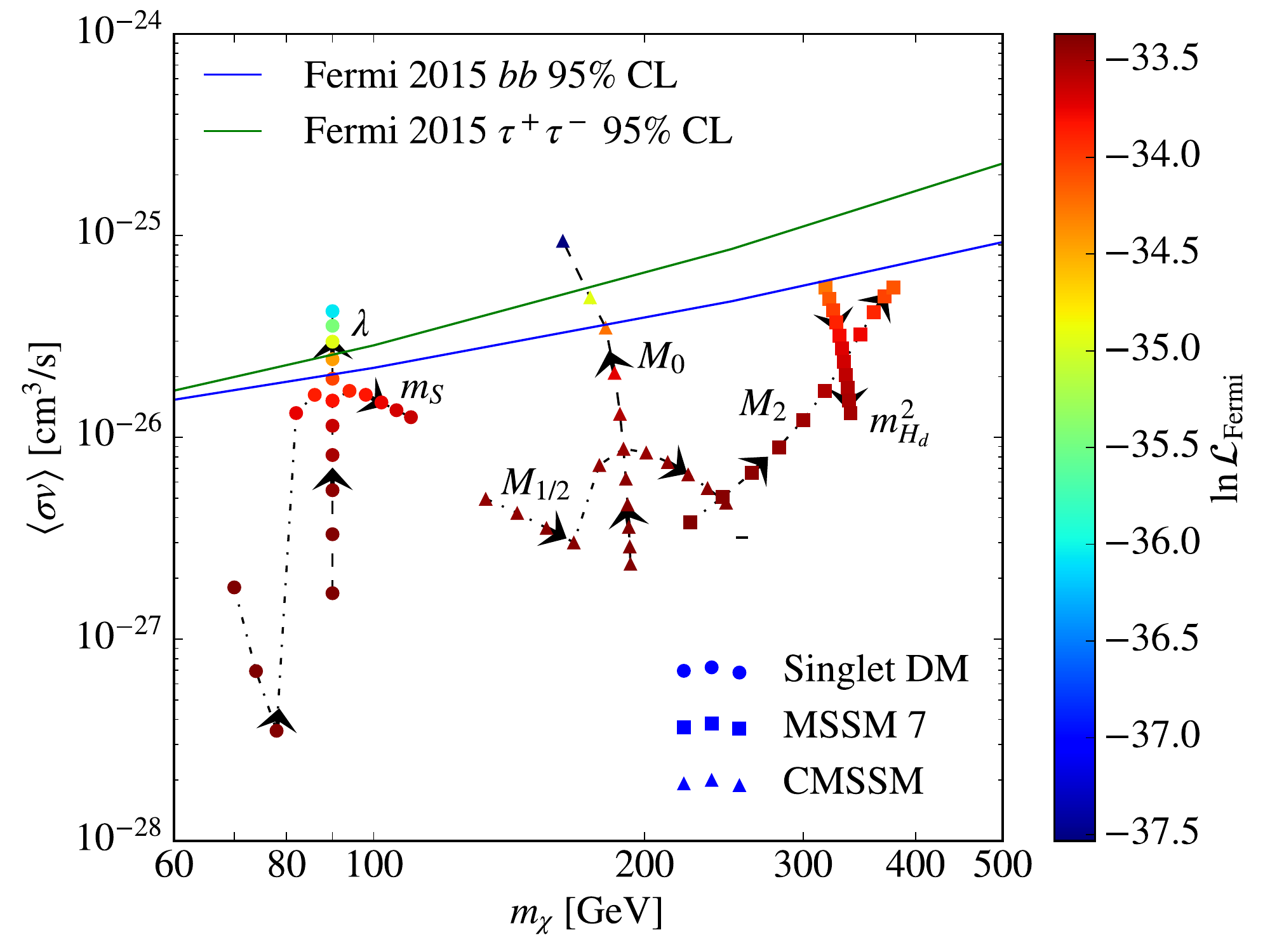}
\includegraphics[width=0.49\linewidth]{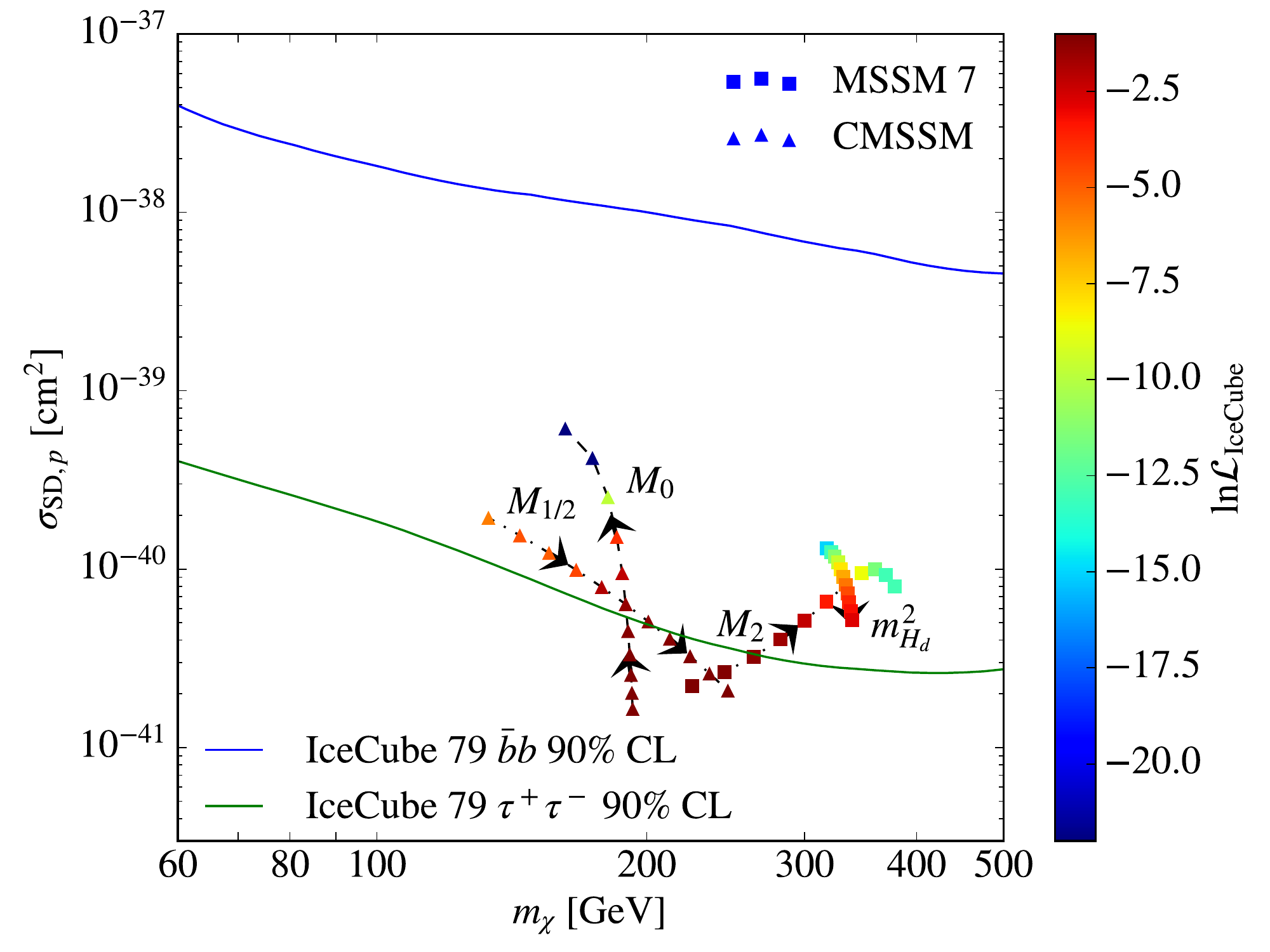}
\caption{\label{fig:spokes} Tracks in observable space corresponding to spokes in model parameter space.  The arrows point in the direction of increasing parameter values.  Clockwise from the top-left, panels show the DM relic density, spin-independent and spin-dependent DM-proton scattering cross sections, and the velocity-averaged DM annihilation cross section.  Points are colour-coded with a Gaussian likelihood based on the \textit{Planck} 2015 analysis, the LUX 2013 likelihood from \ddcalc, the IceCube 79-string likelihood from \nulike, or the \textit{Fermi} dwarf spherodial likelihood from \gamlike respectively. The darkest blue points correspond to likelihoods below the smallest value shown in the colour bars. For comparison, we also plot the limit from the corresponding analyses done by the experimental collaborations. For the relic density, we show the \textit{Planck} best-fit value for $\Omega h^2$ and its 2$\sigma$ (experimental) uncertainty.}
\end{figure*}

%%%%%%%%%%%%%%%%%%%%%%%%%%%%%%%%%%%%%%%%%%%%%%%%%%%%%%%
\subsection{CMSSM, MSSM and Singlet DM}

To demonstrate the ability of \DB to calculate observables and likelihoods, we undertake a number of simple grid scans using the \textsf{grid} scanner \cite{ScannerBit}.  For these demonstrations, we consider the parameter spaces of the CMSSM \cite{CMSSM}, MSSM7 \cite{MSSM} and scalar singlet DM \cite{SSDM} models.  For each model, we choose 2 parameters that are particularly relevant for dark matter phenomenology. The parameters and their ranges are shown in Table \ref{tab:params}. We vary only one parameter at a time, whilst keeping all others fixed at the values shown in Table \ref{tab:central}, leading to a scan over two ``spokes" in parameter space for each model (Fig.\ \ref{fig:spokes}).

For each point in the scan, we calculate spin-independent and spin-dependent DM-proton scattering cross sections, velocity-averaged DM annihilation cross sections at late times, and the DM relic density. For all of these observables, we also calculate a corresponding experimental likelihood using an appropriate backend code included with \DB: the LUX 2013 likelihood from \ddcalc (Sec.~\ref{sec:ddcalc}), the IceCube 79-string likelihood for WIMP annihilation in the Sun from \nulike (Sec.~\ref{code_nu}), and the the stacked dwarf spherodial likelihood based on six years of \textit{Fermi} data from \gamlike (Sec.~\ref{sec:gamLikeTargets}). For the relic density, we calculate the simple Gaussian likelihood based on the best fit value from the \textit{Planck} analysis \cite{Planck15cosmo} described in Sec.~\ref{code_rd}. The results of these scans are shown in Fig.~\ref{fig:spokes}, where the colour-coding of the points represents the likelihood value. For comparison, we plot the limits from corresponding analyses from the LUX \cite{LUX2013}, IceCube \cite{IC79_SUSY}, and \textit{Fermi} \cite{LATdwarfP8} collaborations. For the relic density, we plot the \textit{Planck} best fit value and its $2 \sigma$ uncertainty. In all cases, the likelihoods calculated by \DB and its associated backends agree with the results from the experimental collaborations.

\begin{figure}[t]
  \centering
  \includegraphics[width=0.95\linewidth]{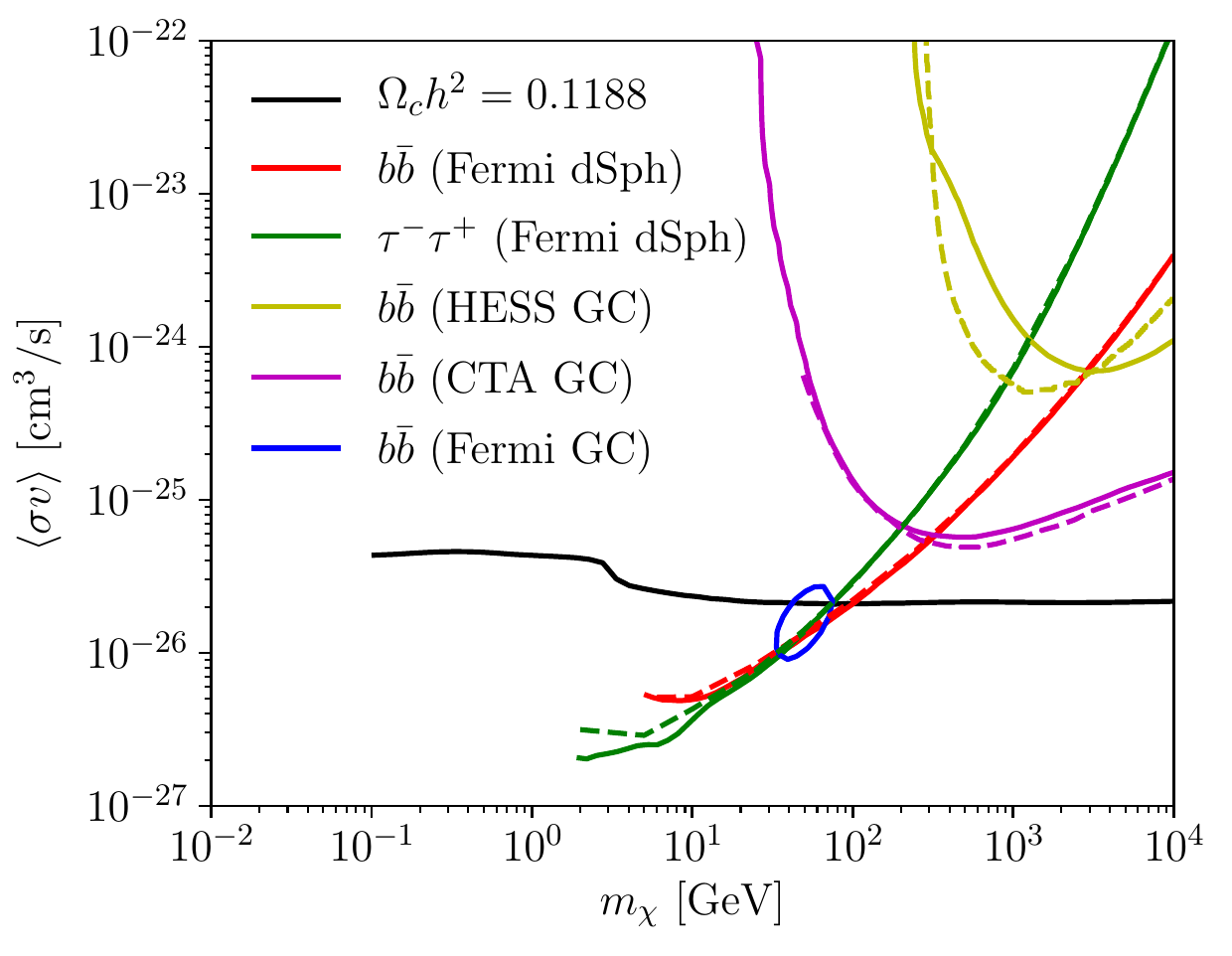}
  \caption{Simple WIMP results obtained with \DB. The green and red solid lines represent $95\%$ CL upper
    limits on $b \bar b$ and $\tau^+ \tau^-$ final states from \textit{Fermi} \texttt{pass 8} observations of dwarf
    spheroidal galaxies as calculated by \DB, while the dashed lines represent the
    corresponding limits reported by the \textit{Fermi} collaboration \cite{LATdwarfP8}.  The magneta and yellow solid lines show the HESS and projected CTA GC limits, and the dashed lines the published results from Refs.~\cite{Abramowski:2011hc, Silverwood:2014yza} (differences are due to differences in the adopted photon yields).  The blue solid line shows the $99.7\%$ CL contour for our \textit{Fermi} GC likelihood.  Finally, the solid black
    line represents the values of $\sigma v$ that reproduce (for $s$-wave annihilation) a DM
    density of $\Omega_{c} h^2 = 0.1188$.
  }
  \label{fig:limits_in}
\end{figure}

\begin{figure}[t]
  \centering
  \includegraphics[width=0.95\linewidth]{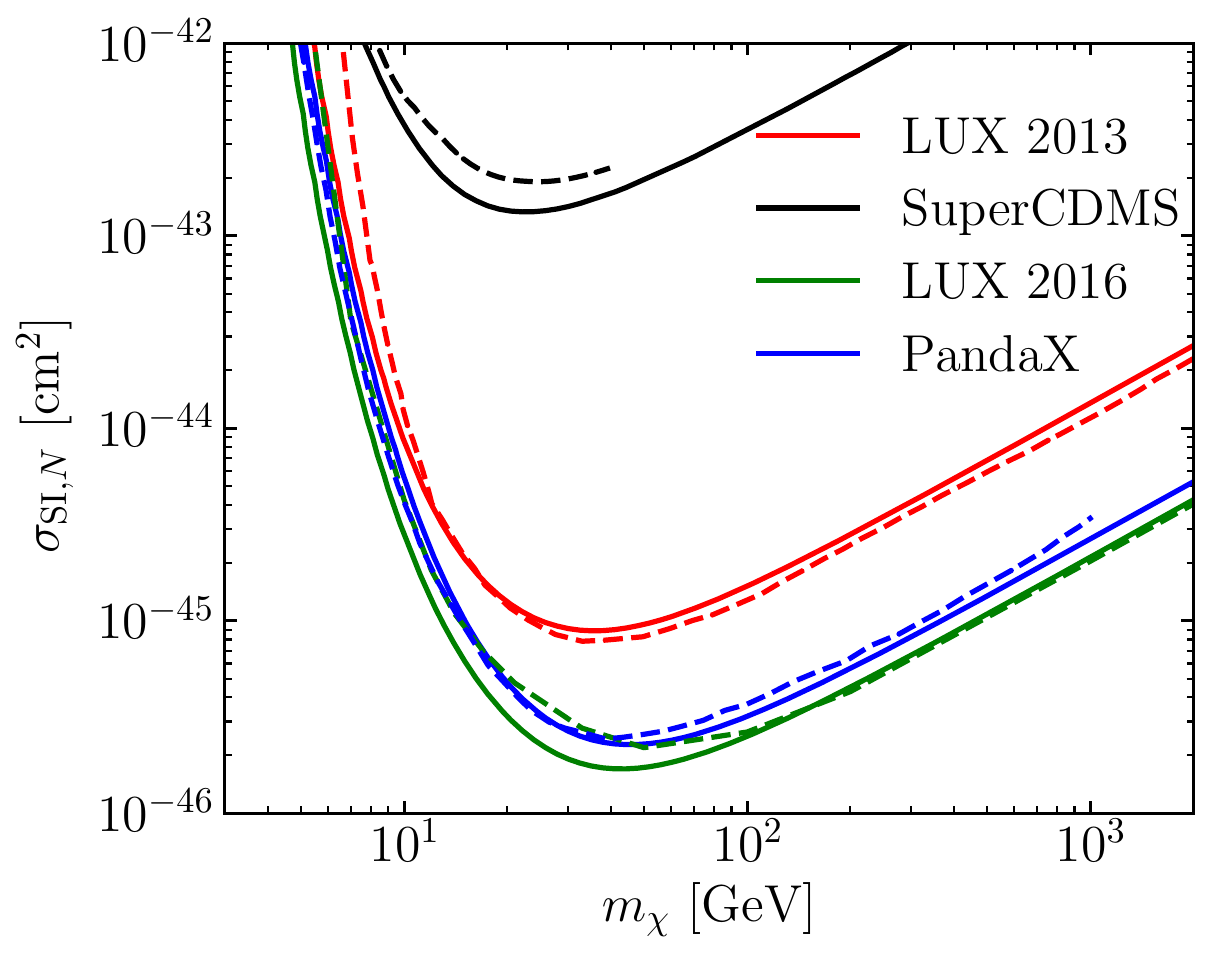}
  \caption{Limits on the spin independent DM-nucleon scattering cross section from SuperCDMS, LUX, and PandaX at 90\% CL. The solid curves are the limit determined using \DB and the dashed curves are the official limits from the collaborations \cite{SuperCDMS, LUX2013, LUX2016, PandaX2016}.}
  \label{fig:limits_di}
\end{figure}

\begin{figure}[t]
  \centering
  \includegraphics[width=0.95\linewidth]{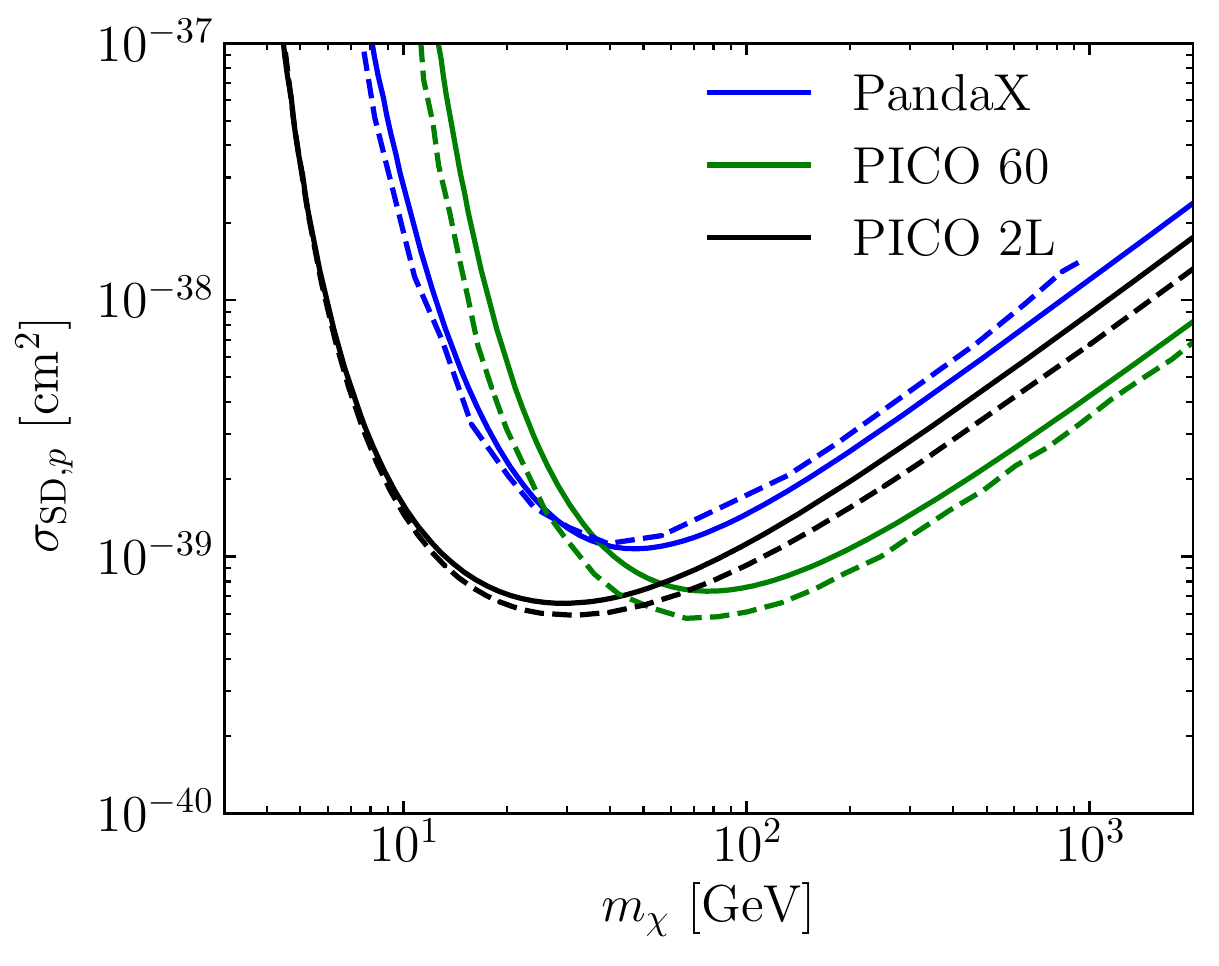}
  \caption{Limits on the spin dependent DM-proton scattering cross section from PICO-2L, PICO-60L, and PandaX at 90\% CL. The solid curves are the limits determined using \DB and the dashed curve are the official limits from the collaborations \cite{PandaXSD, PICO2L, PICO60}.}
  \label{fig:limits_sd}
\end{figure}

%%%%%%%%%%%%%%%%%%%%%%%%%%%%%%%%%%%%%%%%%%%%%%%%%%%%%%%
\subsection{Effective WIMPs}
\label{sec:simpleWIMP}

In order to further illustrate some of the functionality of \DB, and to show how \DB
can be used without a full scan in \GB, \DB ships with three example
standalone programs.  One of them, \term{DarkBit\_standalone\_WIMP}, shows
how to set up and calculate various observables for a simple WIMP model, in which the three parameters are the WIMP mass and the cross sections for WIMP self-annihilation and WIMP-nucleon scattering.  We
will discuss this example here in some detail.  Further examples specific to
singlet dark matter and the MSSM can be found as well; the MSSM example will be
discussed in the next subsection.

All of the model specifics for the standalone example are specified in only
three module functions.  These are defined as \cpp{QUICK_FUNCTIONS} at the
beginning of the source file of the example.  One function, \cpp{DarkMatter_ID_WIMP}, simply returns
the string identifier for the WIMP particle.  Another function,
\cpp{DD_couplings_WIMP}, sets up the direct detection couplings.  In the
present example, these are entirely determined by module function options.  The
most complex function is the function that sets up the Process Catalog for the
given example, \cpp{TH_ProcessCatalog_WIMP}.  Besides the relevant processes,
the masses and spins of the participating particles also have to be defined.
Furthermore, we define a few functions to dump gamma-ray annihilation spectra
into ASCII tables.

In the main part of the code, different options are available that illustrate how to
calculate annihilation yields for various final states and DM masses. Furthermore, the
standalone example has the ability to calculate tables of likelihoods for \textit{Fermi}-LAT/HESS/CTA indirect
detection and direct detection experiments as functions of the dark matter mass and the annihilation or scattering cross
section; we show the corresponding upper limits in Figs.\ \ref{fig:limits_in}--
\ref{fig:limits_sd}. These are $95\%$\,CL upper limits (obtained at $\Delta 2
\ln\mathcal{L} =2.71$) in Fig.~\ref{fig:limits_in}, as is customary for indirect
searches, and at $90\%$\,CL in Figs.\ \ref{fig:limits_di} and \ref{fig:limits_sd}, as is
typical in direct detection. For the direct detection limits, we determine the $90\%$\,CL
value for $\Delta 2 \ln\mathcal{L}$ from the Poisson upper limit on the expected number
of events. For low statistics this value can be substantially larger than $\Delta 2
\ln\mathcal{L} = 1.64$, the value obtained in the asymptotic limit. The agreement between
the official limits and those calculated with the standalone show that known
results can be easily reproduced. The standalone example can also be used to calculate
similar tables for the relic density. With this output, in Fig.\
\ref{fig:limits_in} we indicate the cross-section for which the relic density
reaches $\Omega_\chi h^2 = 0.1188$, the preferred value from \textit{Planck}~
\cite{Planck15cosmo}.

%%%%%%%%%%%%%%%%%%%%%%%%%%%%%%%%%%%%%%%%%%%%%%%%%%%%%%%
\subsection{Comparing \ds and \micromegas}

\DB offers the unique possibility to easily compare different numerical codes
for the computation of DM properties in a well-defined and consistent way.
For illustration, here we focus on \ds
\cite{darksusy} and \micromegas \cite{micromegas}. We stress, however, that it
is straightforward for users to perform similar comparisons for essentially any
other numerical code, simply by adding it as a backend to \GB.

The ability of \DB to facilitate these comparisons for the MSSM is demonstrated in the example program
\term{DarkBit\_standalone\_MSSM}. This program takes an SLHA file (including a \term{DECAY} block, if present) as input and
calculates the relic density and DM-nucleon
scattering cross sections using both \ds and \micromegas. Analogously to \term{DarkBit\_standalone\_WIMP}, it also calculates likelihoods for
the relic density, direct detection experiments, and indirect searches in neutrinos and
gamma rays. The standalone shows how it is possible to change the source of the theoretical inputs for these likelihood calculations (such as the DM-nucleon coupling in the case of direct detection) by just changing a single line of code.

As a demonstration of the sorts of comparisons possible, we have chosen some benchmark MSSM\footnote{Here we use the MSSM7; see Ref.\ \cite{gambit} for the definition of this model, and Ref.\ \cite{MSSM} for a thorough treatment of its phenomenology and present status.} model points that
can cause difficulties in the calculation of the relic density due
to coannihilations or resonances.  Details of the points are given in
Table \ref{tab:benchmarks}.  We generated SLHA files for each of these model points (including \term{DECAY}
blocks) using \specbit and the standalone example \threebit \cite{SDPBit}, which we then fed into \term{DarkBit\_standalone\_MSSM}. These SLHA files can be found in the \term{DarkBit/data/benchmarks/} directory of the \GB distribution. In \GB \textsf{1.1.0}, which was used to find the results presented here, \term{DarkBit\_standalone\_MSSM} by default calculates the relic density with the \YAML option \yaml{fast} set to 0 (corresponding to a more accurate calculation) in both \cpp{RD_oh2_MicrOmegas} and \cpp{RD_oh2_DarkSUSY}. The options \yaml{loop} and \yaml{box} are set to true in \cpp{DD_couplings_DarkSUSY} and \cpp{DD_couplings_MicrOmegas} respectively, so that all available loop corrections are used in each backend.

Results of the calculations can be seen in Table \ref{tab:MO_DS}.  While the values of $\Omega h^2$ from the two backends agree well for some of the benchmarks, in others there are significant differences. In particular for benchmarks 1-4, where there is resonant annihilation via the $A^0$ or $h$, the relic density is substantially higher when the calculation is done using \ds. This can be traced to the fact that the \micromegas $\langle \sigma v_{\rm eff} \rangle$ at temperatures around freeze-out for these points is consistently larger than the \ds result. The fractional differences are largest on the resonance; adjusting $m_{A_0}$ or $m_h$ away from $2m_{\chi_0}$ leads to much better agreement between the two codes. The ultimate source of this discrepancy should be tracked down, and to this end we are planning a follow-up study using the \GB framework in which we investigate the reasons behind the differences and look into the effects of adding loop corrections currently not included in both backends.

For some of the benchmarks, there are also significant differences between the nuclear scattering cross-sections computed with \darksusy and \micromegas, specifically in cases where the cross-sections are small.  The discrepancies between the two codes are almost certainly linked to the fact that at tree level, small nuclear scattering cross-sections in the MSSM arise due to cancellations between different diagrams.  The cancellations can easily be spoilt by small changes in the input parameters, or equivalently, different choices of spectrum generator and/or treatments of running parameters in the calculation of the matrix elements, as well as different treatments of loop corrections in the calculations of scattering amplitudes. As in the case of the relic density, we are planning a future study to more precisely understand the source of these differences.

\begin{table*}
  \centering
  \begin{tabular}{c l r r r r r r r}
  \toprule
  \textbf{Model} & \textbf{Description} & \parbox{1cm}{\centering $M_2$ [GeV]} & \parbox{1.5cm}{\centering $m^2_{H_d}$ \\ {[$10^6$ GeV$^2$]}} & \parbox{1.5cm}{\centering $m^2_{H_u}$\\ {[$10^6$ GeV$^2$]}} & $\tan \beta$ & \parbox{1.5cm}{\centering $m^2_f$\\ {[$10^6$ GeV$^2$]}} & \parbox{1cm}{\centering $A_d$ [GeV]} & \parbox{1cm}{\centering $A_u$ [GeV]} \\
  \midrule
  1 & \parbox{4cm}{Resonant annihilation via $A^0$, gaugino-like neutralino}
     & 3442.  & $-$10.86 & 10.07  & 17.25 & 71.45  & 9588   & $-$5886  \\
  \midrule
  2 & \parbox{4cm}{Resonant annihilation via $A^0$, mixed neutralino}
     & 2224   & $-$0.007416                  & $-$9.361 & 42.63 & 87.23  & 3019   & $-$3716  \\
  \midrule
  3 & \parbox{4cm}{Resonant annihilation via $A^0$, Higgsino-like neutralino}
     & 3283   & 6.904     & $-$8.602  & 39.22 & 73.11  & $-$5453  & $-$2963  \\
  \midrule
  4 & \parbox{4cm}{Resonant annihilation via $h$}
     & $-$659.0 & 27.52      & $-$0.4085  & 21.68 & 4.309  & 9870  & 4565  \\
  \midrule
  5 & $\tilde \tau$ coannihilations
     & $-$681.8 & 94.43     & $-$1.667  & 9.798 & 0.4103  & 69.60  & $-$1471  \\
  \midrule
  6 & \parbox{4cm}{$\tilde t$ coannihilations,\\gaugino-like neutralino}
     & 2631  & 4.369      & $-$4.448  & 7.760 & 11.52  & 9993   & $-$5103  \\
  \midrule
  7 & \parbox{4cm}{$\tilde t$ coannihilations,\\mixed neutralino}
     & 2323  & 4.169      & $-$2.222  & 9.283 & 8.899  & 9617   & $-$4472  \\
  \midrule
  8 & \parbox{4cm}{$\tilde t$ coannihilations,\\Higgsino-like neutralino}
     & 2316  & 4.164      & $-$2.072  & 11.44 & 8.560  & 229.2  & $-$3976  \\
  \midrule
  9 & \parbox{4cm}{Chargino coannihilations}
     & 1582  & 8.029     & $-$2.938  & 45.01 & 42.07  & $-$125.5 & $-$768.3 \\
  \bottomrule
  \end{tabular}
  \caption{MSSM7 points used as benchmarks for comparisons between \ds and \micromegas (see Ref.\ \cite{gambit} for the definition of the model). The sign of $\mu$ is positive for all points and the parameters are defined at an energy scale of 1 TeV. As shown in the description column, the points were chosen to have different types of processes contribute to the relic density calculation.}
  \label{tab:benchmarks}
\end {table*}

\begin{table*}
  \centering
  \begin{tabular}{c rr rr rr}
  \toprule
  \multirow{2}{*}{\textbf{Model}} & \multicolumn{2}{c}{$\Omega h^2$} &
    \multicolumn{2}{c}{$\sigma_{{\rm SI}, p}$ [$10^{-46}{ \rm cm^2}$]} &
    \multicolumn{2}{c}{$\sigma_{{\rm SD}, p}$ [$10^{-43}{ \rm cm^2}$]} \\
     & \ds       & \micromegas & \ds     & \micromegas & \ds                    & \micromegas \\
  \midrule
     1 & 0.1545 & 0.09471 & 9.817 & 10.88 & 1.332 & 1.263 \\ %HA funnel gaugino
  \midrule
     2 & 0.01617 & 0.008546 & 166.5 & 186.6 & 61.14 & 58.01 \\ %HA funnel mixed
  \midrule
     3 & 0.05888 & 0.03355 & 189.5 & 211.1 & 32.88 & 31.21 \\ %HA funnel higgsino
  \midrule
     4 & 0.002318 & 0.001480 & 25.50 & 26.28 & 5272. & 5002. \\ %hZ funnel
  \midrule
     5 & 0.1110 & 0.1094 & 10.45 & 7.011 & 0.06781 & 0.06170 \\ %stau coannihilation
  \midrule
     6 & 0.02290 & 0.02410 & 3.367 & 3.745 & 0.7975 & 0.8305 \\ %stop coannihilation gaugino
  \midrule
     7 & 0.004982 & 0.003623 & 198.0 & 218.7 & 41.34 & 39.00 \\ %stop coannihilation mixed
  \midrule
     8 & 0.006317 & 0.004722 & 250.0 & 273.9 & 55.69 & 52.61 \\ %stop coannihilation higgsino
  \midrule
     9 & 0.003008 & 0.003131 & 8.749 & 8.869 & 192.4 & 182.6 \\ %chargino coannihilation
  \bottomrule
  \end{tabular}
  \caption{The dark matter relic density and proton-scattering cross-sections, both spin-independent and spin-dependent, for a range of MSSM model points. The model points are defined in Table \ref{tab:benchmarks}. All quantities were calculated with \protect\term{DarkBit_standalone_MSSM} using both the \ds and \micromegas backends.}

  \label{tab:MO_DS}
\end {table*}

%%%%%%%%%%%%%%%%%%%%%%%%%%%%%%%%%%%%%%%%%%%%%%%%%%%%%%%
%%%%%%%%%%%%%%%%%%%%%%%%%%%%%%%%%%%%%%%%%%%%%%%%%%%%%%%
\section{Outlook}
\label{sec:out}
As detailed in the preceding sections, \DB is equipped with sophisticated tools
for calculating observables and likelihoods for the DM relic density, direct detection
experiments and indirect searches with neutrinos and gamma rays. Each of these
cases demonstrates the modularity of the code, and the ease
with which external codes can interfaced with \DB. This modularity also implies that extensions of  \DB
in all possible directions are straight-forward to implement, and do not in general require the
expertise of GAMBIT Collaboration members or highly experienced external users of the code.
The focus of future developments will thus be steered largely by the needs (and indeed, contributions) of the community.
Nevertheless, here we list a few obvious code extensions that we expect to include in future releases (aside from
obvious additions of new experimental likelihoods to existing components like \gamlike, \ddcalc and \nulike).

The combination of the process catalogue and the \ds Boltzmann solver currently allows us to calculate the relic density for simple arbitrary DM models. We intend to expand this framework so that coannihilations can be included in the process catalogue and the relic density can be calculated in the case of semi-annihilating \cite{D'Eramo:2010ep} and asymmetric DM \cite{Petraki:2013wwa}. This would complement the existing capabilities of \micromegas to handle these scenarios. We also plan to backend \textsf{MadDM} \cite{Backovic:2013dpa, Backovic:2015cra} in a future version of \darkbit, which with its interface to \textsf{MadGraph} \cite{Alwall:2011uj} will be a useful alternative to \micromegas for quickly implementing new DM models. To enhance the accuracy of our relic density calculations, we also intend to add the effects of Sommerfeld enhancement on the relevant (co)annihilation cross sections \cite{Hryczuk:2010zi, Feng:2010zp}.
Moving beyond the standard relic density calculation, we have plans to include the ability to deal with non-standard cosmological expansion histories. This capability is available in \textsf{SuperIso Relic} \cite{Arbey:2009gu}, to which we will provide a frontend.
A final natural
extension in this area is to calculate kinetic freeze-out of DM from the thermal bath rather than only
chemical freeze-out as is done now.  This would lead to an additional observable: a cutoff in
the power spectrum of matter density perturbations \cite{Bringmann:2009vf, Cornell:2013rza}.

For direct detection experiments, the implementation of velocity- and momentum-dependent
cross sections in \ddcalc will be a high priority extension, allowing one to systematically
study the full set of available non-relativistic operators \cite{Fitzpatrick:2012ix}, for example. Helio- and astroseismological probes
of DM-nucleon couplings (see e.g.~\cite{Vincent16}) are another expected extension.

The most important extension relevant for indirect DM searches, given the high precision
expected from the AMS-02 experiment on board the international
space station, is charged cosmic rays. Indeed, positron data already put one of
the most stringent limits on leptophilic DM models \cite{Bergstrom:2013jra}, and constraints from
antiprotons can likely be improved considerably \cite{Cirelli:2013hv,Bringmann:2014lpa}. Another extension that we
aim for in the near future is to fully allow for velocity-dependent annihilation cross sections, such as in the case of resonances or the
Sommerfeld effect \cite{Hisano:2004ds,Iengo:2009ni}.  These can be relevant for e.g.\ DM
annihilation in subhalos \cite{Bovy:2009zs}, or close to the black hole at the Galactic centre \cite{Arina:2014fna, Choquette:2016xsw}.

%%%%%%%%%%%%%%%%%%%%%%%%%%%%%%%%%%%%%%%%%%%%%%%%%%%%%%%
%%%%%%%%%%%%%%%%%%%%%%%%%%%%%%%%%%%%%%%%%%%%%%%%%%%%%%%
\section{Conclusions}
\label{conc}

The particle nature of DM is one of the most perplexing puzzles in present-day
particle physics and cosmology.  Despite decades of research, only
non-gravitational signals of DM have been identified so far.  However,
rapid experimental developments in recent years have provided a wealth of new
data that can be exploited in the search for DM signals, and used to constrain DM
models.  In order to facilitate a systematic study of a large number of DM
scenarios, in this paper we presented \DB, a new numerical tool for DM calculations.

\DB is designed to allow DM observables to be included in global scanning tools like
\GB.  It can also be used as standalone module. The first release
of \DB ships with a large number of likelihood functions for various
experiments.  These are implemented more accurately than what is usually done in
the literature.  The overarching design goals are reusability,
self-consistency and modularity, which are achieved in a number of ways.
First, by allowing seamless integration of existing numerical tools like \ds
and \micromegas, using the \GB method of abstracting backend function-handling
for cross-language coding environments.  Second, by providing internal structures for particle and
astrophysical DM properties that are consistently used in all calculations, and
passed to external codes if necessary.  Third, by splitting up
calculations into their most elemental building blocks wherever possible.

The modular implementation of the DM relic density calculations in \DB allows it to solve the
Boltzmann equation independently of the actual particle model chosen for DM (Fig.~\ref{fig:RD_flow}).
Alternatively, the user can directly call relic density routines provided by backend codes for specific models,
allowing, for example, a systematic comparison between the results of \darksusy and \micromegas.

The new backend code \ddcalc provides a general solution to the problem of testing
DM models against direct detection data, including detailed likelihoods for
many of the most important experiments.  This allows both spin-dependent and spin-independent signals
of the same model to be calculated and combined self-consistently across the full
range of relevant experiments.  For liquid noble gas detectors, the sensitivities in \ddcalc are based on the output of
\tpcmc \cite{Savage:2015tpcmc}, a dedicated detector Monte Carlo simulation.

We have implemented likelihood functions for gamma-ray indirect DM searches in the new backend \gamlike.  We included \textit{Fermi}-LAT and HESS
observations of dwarf spheroidal galaxies and the Galactic centre, as well as
projections for the future with CTA.  The likelihood functions in \gamlike are
pre-tabulated for fast evaluation, but based on event-level (mock) data where
possible.  \DB also includes a new Monte Carlo code for the fast simulation of cascade annihilation spectra, and an interface to the event-level neutrino telescope
likelihood tool \nulike for calculating neutrino indirect detection likelihoods.

The first release of \DB ships with the essentials of DM indirect searches.
Extensions planned for the near future include charged cosmic rays, accurate
treatment of velocity-dependent annihilation cross-sections and Sommerfeld enhancement,
and inclusion of new experimental analyses in \gamlike.  In direct detection, we plan
to implement velocity- and momentum-dependent cross-sections in \ddcalc, as well
as new experimental results as they become available.  Furthermore, new classes
of likelihoods, like helio/astroseismological probes for DM and limits from
radio and CMB observations, will be included in future releases.

\begin{acknowledgements}
We wish to thank Lauren Hsu for contributing to the \ddcalc treatment of SuperCDMS, and Ankit Beniwal and Andre Scaffidi for helpful conversations regarding \ddcalc. \gambitacknosplus
\end{acknowledgements}

%%%%%%%%%%%%%%%%%%%%%%%%%%%%%%%%%%%%%%%%%%%%%%%%%%%%%%%
%%%%%%%%%%%%%%%%%%%%%%%%%%%%%%%%%%%%%%%%%%%%%%%%%%%%%%%
\appendix

\setcounter{table}{0}
\renewcommand\thetable{A\arabic{table}}

\label{quickstart}

%%%%%%%%%%%%%%%%%%%%%%%%%%%%%%%%%%%%%%%%%%%%%%%%%%%%%%%
\section{Getting started}
\label{code_init}

As described in Sec.~\ref{sec:overview}, \DB is a standalone and complimentary module of the \GB software, which can be downloaded from the official \GB website\footnote{\href{http://gambit.hepforge.org}{gambit.hepforge.org}}. In the following, we describe the content of the \DB standalone download, the installation of the \DB software as standalone or \GB module, and the running of the example program.

%%%%%%%%%%%%%%%%%%%%%%%%%%%%%%%%%%%%%%%%%%%%%%%%%%%%%%%
\subsection{Content of \DB download \& installation}

Each \GB module contains the
\begin{itemize}
  \item \textsf{Backends} (utility functions used for backend interfaces)
  \item \textsf{Models} (predefined BSM models and utility functions used for model definitions)
  \item \textsf{Logs} (general \GB logging system);
  \item \textsf{Utils} (\GB utility functions);
  \item \textsf{Elements} (general \GB macro and type definitions)
\end{itemize}
folders in addition to the specific module folder (here the \DB folder). A
detailed description of the \GB functionalities can be found in the \GB main
paper~\cite{gambit}. Each standalone module requires all of these folders to work.
If the \DB module is used in conjunction with other \GB modules, only the \DB folder
is needed and should be placed into the main
folder containing the other \GB modules.

\GB uses the open-source cross-platform build system
\textsf{CMake}\footnote{\url{www.cmake.org}}. \textsf{CMake} supports in-source and out-of-source
builds, but we recommend the latter to keep the source directory unchanged and
enable multiple builds. To do such a build, run the following commands in the directory that contains the \GB module folders:
\begin{lstterm}
mkdir build
cd build
cmake ..
make
\end{lstterm}
For further details we refer to the \GB main paper~\cite{gambit}.

The \DB standalones can be found in \term{Darkbit/} \term{examples}, and can be built
with
\begin{lstterm}
make DarkBit_standalone_@\metavar{X}@
\end{lstterm}
where \metavar{X} can be \term{MSSM}, \term{WIMP} or \term{SingletDM}.

%%%%%%%%%%%%%%%%%%%%%%%%%%%%%%%%%%%%%%%%%%%%%%%%%%%%%%%
\subsection{Running the example program}

\label{sec:example}

To demonstrate how \DB can be used in a
fully-fledged scan, we provide 2 annotated examples of of a \darkbit~\term{yaml}
file: \term{yaml_files/DarkBit_SingletDM.yaml} and \term{yaml_files/DarkBit_MSSM7.yaml}. These
examples show how to specify a model, prior ranges over which to sample its parameters, a scanner and an output device (`printer'), then
run the relic density, gamma-ray, neutrino and direct search likelihoods. The
user can also select the parameters of the halo model and the nuclear
parameters relevant for direct detection. The examples require the
\micromegas, \gamlike, \ddcalc, \ds, \nulike, \feynhiggs and \SUSYHIT
backends to be present, which can be accomplished by running the following
commands in the \gambit build directory
\begin{lstterm}
make micromegas_MSSM
make micromegas_SingletDM
make gamlike
make ddcalc
make darksusy
make nulike
make feynhiggs
make susyhit
\end{lstterm}

The \term{yaml} file is complete, \textit{i.e.}~all options of all module
functions available in \DB are mentioned and documented there.

%%%%%%%%%%%%%%%%%%%%%%%%%%%%%%%%%%%%%%%%%%%%%%%%%%%%%%%
\section{Handling Fortran/C/C++ functions with \daFunk}
\label{sec:dafunk}

\subsection{Design goals and philosophy}

One of the major technical challenges when combining a large number of
different codes but trying to maintain maximum portability is wrapping and
manipulating \Fortran, plain \plainC and \Cpp object member functions in a systematic and
coherent way.  In \DB, quite commonly functions of the type
$\mathbb{R}^n\to\mathbb{R}$ (which describe e.g.~annihilation yields,
dark matter profiles, velocity distributions, differential cross sections, or
effective annihilation rates) need to be wrapped in a generic structure so they can be shared amongst different backends.

Typically, the results of these functions need to be manipulated before they
can be used, in order to comply with conventions and requirements in
the subsequent codes.  Sometimes this means using basic arithmetic operations,
sometimes passing them through complex trigonometric functions or performing
variable substitution.  Further common operations are partial integrations,
sometimes with non-constant boundaries and singularity handling, or checks
of parameter domains.  Often, these manipulated functions need to be wrapped
back into plain \plainC functions in order to be able to pass them back into the
backend codes (like e.g.~the \ds Boltzmann solver).

In order to facilitate all these operations, we present the new lightweight
\Cpp header-only library \daFunk (\textit{dynamisch allokierbare Funktionen}).
Despite the complex function handling that it allows, the
\daFunk API is relatively simple.  This is achieved with recursive variadic templates, polymorphism and shared pointers.  The
most relevant features are:
\begin{itemize}
  \item Interpolation in linear- and log-space
  \item Multidimensional integration with complex boundaries
  \item Handling of (1-dimensional) singularities
  \item Parameter substitution and chaining of functions
  \item Wrapping of \Fortran functions, plain \plainC/\Cpp functions and \Cpp object
    member functions
  \item Reverse wrapping of \cpp{daFunk::Funk} objects into \Fortran and plain
    \plainC/\Cpp functions
  \item Overloading of all basic arithmetic and trigonometric functions for
    easy manipulation and combination of \cpp{daFunk::Funk} objects
  \item Flexible function handling based on shared pointers
  \item Checks of parameter domains
  \item Basic \cpp{if...else} constructions
  \item \omp-enabled calculations
\end{itemize}

\subsection{Selected examples}

The atomic object in \daFunk is \cpp{daFunk::Funk},  which is a shared pointer
to an instance of the \cpp{daFunk::FunkBase} class.  Importantly, the \cpp{daFunk::FunkBase}
class is an \textit{abstract} base class:  it leaves the virtual member function
responsible for all actual calculations, \cpp{virtual} \cpp{double value(...)},
undefined.  The main purpose of \cpp{daFunk::Funk} is to provide a unified
interface and a flexible \Cpp type, independent of whatever calculation it
actually performs.  The actual computations are implemented in classes
derived from \cpp{daFunk::FunkBase}.

Each \cpp{daFunk::Funk} object provides a list of names of variables on which
it depends.  This list is simply a list of \cpp{std::string} tags.  The specific
content of this list depends on the implementation of \cpp{value}.
The most basic implementations are variables and constants, shown in the
following simple example:
\begin{lstcpp}
daFunk::Funk x = daFunk::var("x");  // variable @\cpppragma{$x$}@
daFunk::Funk y = daFunk::var("y");  // variable @\cpppragma{$y$}@
daFunk::Funk c = daFunk::cnst(2.);  // constant @\cpppragma{$2$}@
daFunk::Funk f = c*x+3*cos(y); //@\cpppragma{$f(x,y) \equiv 2x + 3\cos(y)$}@
\end{lstcpp}
The name of a variable is specified as a \cpp{std::string}, and \cpp{daFunk::Funk}
objects can be combined into new \cpp{daFunk::Funk} objects using normal arithmetic or
common (appropriately overloaded) \cpp{std::math} operations.  In the above
example, \cpp{x} is a function of variable list \cpp{["x"]}, \cpp{y} is a function of
the variable list \cpp{["y"]}, \cpp{f} is a function of the variable list
\cpp{["x", "y"]}, and \cpp{c} is a function of the variable list \cpp{[]}.

For performance reasons, the evaluation of \cpp{daFunk::Funk} objects is split
into two steps.  First, the positions of the variables are `bound', using the
\cpp{bind} member function.  This generates an object of the type
\cpp{daFunk::FunkBound}, which can then be evaluated using the
\cpp{eval} member function.  This is shown in the following example:
\begin{lstcpp}
  // bind variable positions
  daFunk::FunkBound fb = f->bind("y", "x");
  // return @\cpppragma{$f(4, 3) = 2 \times 4 + 3\cos(3)$}@
  fb->eval(3., 4.);
\end{lstcpp}
This two-step procedure separates the overhead related to the dynamical
construction of nested functions with tagged variables (which in general
includes string matching, various consistency checks and needs to be done
only once) from the possibly large number of actual function evaluations.

A more complex situation that involves (1) the wrapping of plain \plainC functions into
\daFunk objects, (2) the wrapping of \daFunk objects in plain \plainC functions, and
(3) variable substitution, is shown in the following example:
\begin{lstcpp}
// Declaration of a plain @\cpppragma{\plainC}@ function
double dNdE(double E, double m, double v);

// Define traits class for daFunk function pointer
DEF_FUNKTRAIT(T)

// daFunk variables
int main()
{
  daFunk::Funk v = daFunk::var("v");
  daFunk::Funk E = daFunk::var("E");
  daFunk::Funk m = daFunk::var("m");
  daFunk::Funk Ecm = daFunk::var("Ecm");

  // Wrap plain C function
  daFunk::Funk f = daFunk::func(dNdE, E, m, v);

  // Variable substitution
  daFunk::Funk g =
   f->set("v", 0.0001)->set("m", Ecm/2);

  // Wrap daFunk in plain function
  double (*h)(double&, double&) =
   g->plain<T>(g, "Ecm", "E");

  // Returns ann. yield for @\cpppragma{$E_\text{cm} = 100$ at $E = 10$}@
  double Ecm_d = 100;
  double E_d = 10;
  std::cout << (*h)(Ecm_d, E_d) << std::endl;
}
\end{lstcpp}

Here, \cpp{dNdE} is a plain \plainC or \Fortran function (e.g.\ the
annihilation yield as function of final state energy $E$, initial state
relative velocity $v$, and for dark matter mass $m$), which is wrapped into a
\daFunk object \cpp{f}.  The function \cpp{g} is derived from \cpp{f} by fixing
$v=10^{-4}$, and substituting the dark matter mass by the centre-of-mass frame
energy $m=E_\text{cm}/2$.  The function \cpp{g} is then wrapped back into a plain
\plainC function \cpp{h} that can be evaluated as usual, or passed back into some of
the backend codes (note that the \cpp{bind} step happens here behind the scenes
when calling \cpp{plain<T>}).  Note that the pointer to the \cpp{daFunk::Funk}
object that is wrapped in $h$ is stored in the traits class \cpp{T}.
Furthermore, the generated \plainC function takes arguments by reference, which is an
implicit convention in \Fortran.  This makes it possible to pass $h$ directly
back to a \Fortran backend.

Lastly, we give an example for 1-dimensional integration with non-trivial boundaries.
\begin{lstcpp}
daFunk::Funk x = var("x");  // variable @\cpppragma{$x$}@
daFunk::Funk a = var("a");  // variable @\cpppragma{$a$}@
daFunk::Funk f = x*x;       // @\cpppragma{$f(x) = x^2$}@

// @\cpppragma{$g(a) \equiv \int_0^a f(x)\,dx = \int_0^a x^2\,dx$}@
daFunk::Funk g = f->gsl_integration("x", 0., "a");
daFunk::FunkBound gb = g->bind("a");
// print @\cpppragma{$g(\frac52)$}@ to stdout
std::cout << gb->eval(2.5) << std::endl;
\end{lstcpp}
Note that in cases where the integration fails, a warning message is printed
to \cpp{stderr}, and zero is returned.  For more examples we refer the reader the \DB
code.

\startglossary

\gitem{backend}\input{"glossary/backend.glossentry"}
\gitem{backend function}\input{"glossary/backend_function.glossentry"}
\gitem{backend requirement}\input{"glossary/backend_requirement.glossentry"}
\gitem{backend variable}\input{"glossary/backend_variable.glossentry"}
\newcommand{\seecompdatabase}{see Sec.\ 10.7 of Ref.\ \cite{gambit}}
\gitem{capability}\input{"glossary/capability.glossentry"}
\gitem{dependency}\input{"glossary/dependency.glossentry"}
\gitem{dependency resolution}\input{"glossary/dependency_resolution.glossentry"}
\newcommand{\deptreefig}{Fig.\ 5 of Ref.\ \cite{gambit}}
\gitem{dependency tree}\input{"glossary/dependency_tree.glossentry"}
\gitem{frontend}\input{"glossary/frontend.glossentry"}
\gitem{frontend header}\input{"glossary/frontend_header.glossentry"}
\gitem{module}\input{"glossary/module.glossentry"}
\gitem{module function}\input{"glossary/module_function.glossentry"}
\gitem{physics module}\input{"glossary/physics_module.glossentry"}
\gitem{rollcall header}\input{"glossary/rollcall_header.glossentry"}
\gitem{type}\input{"glossary/type.glossentry"}

\finishglossary

\section{Capability overview}

For reference, we provide a complete list of the \darkbit capabilities,
dependencies and options.
%in Tables~\ref{tab:darkbitcentral}--\ref{tab:darkbitmisc}
These include the complete
process and coupling capabilities (Table~\ref{tab:darkbitcentral}), some
simple informative capabilities (Table~\ref{tab:darkbitdirect}),
capabilities related to DM halo properties (Table~\ref{tab:halocap}),
relic density capabilities
(Tables~\ref{tab:darkbitrelcap} and \ref{tab:darkbitmainrelcap}),
direct detection capabilities
(Table~\ref{tab:ddcap}),
gamma-ray yield capabilities (Table~\ref{tab:darkbitgammacap}),
gamma-ray likelihoods (Table~\ref{tab:darkbitIDlike}),
neutrino capabilities
(Tables~\ref{tab:darkbitnucap} and~\ref{tab:darkbitnulikecap}),
cascade decay capabilities
(Tables~\ref{tab:darkbitcascadecap} and~\ref{tab:darkbitcascadeloopcap}),
and various
miscellaneous capabilities
(Table~\ref{tab:darkbitmisc}).
%\balance

\renewcommand\metavar\metavars
\newcommand\descwidth{4.2cm}
\newcommand\bewidth{2cm}
\newcommand\optwidth{2cm}

\begin{table*}[h!]
  \scriptsize
  \centering
 \begin{tabular}{l|p{\descwidth}|l|l|l}
   \toprule
  \textbf{Capability}
      & \multirow{2}{*}{\parbox{\descwidth}{\textbf{Function} (\textbf{Return Type}):
             \\  \textbf{Brief Description}}}
          & \textbf{Dependencies}
          & \multirow{2}{*}{\parbox{\bewidth}{\textbf{Backend}
             \\  \textbf{requirements}}}
          & \multirow{2}{*}{\parbox{\optwidth}{\textbf{Options}
             \\  (\textbf{Type})}}
          \\ & & &
  \\ \hline
 \cpp{TH_ProcessCatalog}
      & \multirow{2}{*}{\parbox{\descwidth}{\cpp{TH_ProcessCatalog_MSSM}
              (\cpp{DarkBit::TH_ProcessCatalog}):
              \\  Generates process catalogue for the MSSM, based
          on the \darksusy backend.}}
 & \cpp{DarkSUSY_PointInit} &  \cpp{setMassesForIB} & \cpp{ignore_three_body} (\cpp{bool})
  \\ &  & \cpp{MSSM_spectrum}  &  \cpp{dssigmav} & \cpp{ProcessCatalog_}
  \\ & & \cpp{decay_rates} & \cpp{dsIBffdxdy}  & \quad\cpp{MinBranching}(\cpp{double})
  \\ & & \cpp{DarkMatter_ID} & \cpp{dsIBhhdxdy}   &
  \\ & & & \cpp{dsIBwhdxdy} &
  \\ & & & \cpp{dsIBwwdxdy} &
  \\ & & & \cpp{IBintvars} &
  \\ \cmidrule{2-5}
      & \multirow{4}{*}{\parbox{\descwidth}{\cpp{TH_ProcessCatalog_SingletDM}\\
              (\cpp{DarkBit::TH_ProcessCatalog}):
              \\  Generates process catalogue for scalar singlet dark matter. }}
   & \cpp{SingletDM_spectrum}  & &  \cpp{ProcessCatalog_}
  \\ & & \cpp{decay_rates} & & \quad\cpp{MinBranching} (\cpp{double})
  \\ & & &
  \\ & & &
   \\ \hline
 \cpp{DD_couplings}
      & \multirow{3}{*}{\parbox{\descwidth}{\cpp{DD_couplings_DarkSUSY}
              (\cpp{DM_nucleon_couplings}):
              \\  Determine the WIMP mass and couplings using \ds.  }}
 & \cpp{DarkSUSY_PointInit} & \cpp{dsddgpgn}  & \cpp{rescale_}
 \\ & & & \cpp{mspctm} & \quad\cpp{couplings} (\cpp{double})
 \\ & & & \cpp{ddcom} & \cpp{loop}* (\cpp{bool})
 \\ & & & & \cpp{pole}* (\cpp{bool})
 \\ \cmidrule{2-5}
   & \multirow{3}{*}{\parbox{\descwidth}{\cpp{DD_couplings_MicrOmegas}
              (\cpp{DM_nucleon_couplings}):
              \\  Determine the WIMP mass and couplings using \micromegas.  }}
 & & \cpp{nucleonAmplitudes} & \cpp{box}* (\cpp{bool})
 \\ & & & \cpp{FeScLoop} & 
 \\ & & & \cpp{MOcommon} &
 \\ & & & &
 \\ \cmidrule{2-5}
   & \multirow{3}{*}{\parbox{\descwidth}{\cpp{DD_couplings_SingletDM}
              (\cpp{DM_nucleon_couplings}):
              \\  Determine the WIMP mass and couplings for scalar singlet DM.  }}
 & \cpp{SingletDM_spectrum} & &
 \\ & & & &
 \\ & & & &
 \\ & & & &
 \\ \hline
\end{tabular}
\caption{Central \DB capabilities that store details about annihilation and scattering processes. The starred (*) options are only available in \GB \textsf{1.1.0}. \label{tab:darkbitcentral}
}
\end{table*}

\renewcommand\metavar\metavarf

\renewcommand\metavar\metavars
\renewcommand\descwidth{11cm}

\begin{table*}[tbp]
  \centering
  \scriptsize
\begin{tabular}{l|p{\descwidth}|l}
   \toprule
  \textbf{Capability}
      & \multirow{2}{*}{\parbox{\descwidth}{\textbf{Function} (\textbf{Return Type}):
             \\  \textbf{Brief Description}}}
          & \textbf{Dependencies}
          \\ & &
  \\ \hline
 \cpp{mwimp}
      & \multirow{2}{*}{\parbox{\descwidth}{\cpp{mwimp_generic}
              (\cpp{double}):
              \\  Retrieve the DM mass in GeV for generic models. }}
 & \cpp{TH_ProcessCatalog}
 \\ & &\cpp{DarkMatter_ID}
 \\ \hline
 \cpp{sigmav}
      & \multirow{3}{*}{\parbox{\descwidth}{\cpp{sigmav_late_universe}
              (\cpp{double}):
              \\  Retrieve the total thermally-averaged
          annihilation cross-section for indirect detection (cm$^3$\,s$^{-1}$),
      at $v=0$. }}
 & \cpp{TH_ProcessCatalog}
 \\ & &\cpp{DarkMatter_ID}
 \\ & &
  \\ \hline
 \cpp{sigma_SI_}\metavar{N}
      & \multirow{2}{*}{\parbox{\descwidth}{\cpp{sigma_SI_}\metavar{N}\cpp{_simple}
              (\cpp{double}):
              \\  Simple calculator of the spin-independent WIMP-proton or WIMP-neutron cross-section. }}
  & \cpp{DD_couplings}
 \\ & & \cpp{mwimp}
 \\ \hline
 \cpp{sigma_SD_}\metavar{N}
      & \multirow{2}{*}{\parbox{\descwidth}{\cpp{sigma_SD_}\metavar{N}\cpp{_simple}
              (\cpp{double}):
              \\  Simple calculator of the spin-dependent WIMP-proton or WIMP-neutron cross-section. }}
  & \cpp{DD_couplings}
 \\ & & \cpp{mwimp}
  \\ \hline
   \cpp{DarkMatter_ID}
      & \multirow{2}{*}{\parbox{\descwidth}{\cpp{DarkMatter_ID_SingletDM}
              (\cpp{std::string}):
              \\  Returns string ID for dark matter particle. }}
  &
  \\ & &
 \\ \cmidrule{2-3}
      & \multirow{2}{*}{\parbox{\descwidth}{\cpp{DarkMatter_ID_MSSM30atQ}
              (\cpp{std::string}):
              \\  Returns string ID for dark matter particle. }} &
  \\ & &
 \\ \hline
       \end{tabular}
       \caption{\DB capabilities for WIMP-nucleon couplings (\metavar{N} = $p, n$
       refers to the relevant nucleon), annihilation cross-section, dark matter
     mass and dark matter particle ID. \label{tab:darkbitdirect}}
\end{table*}

\renewcommand\metavar\metavarf

\renewcommand\metavar\metavars
\renewcommand\descwidth{8cm}

\begin{table*}[tbp]
  \centering
  \scriptsize
\begin{tabular}{l|p{\descwidth}|l|l}
   \toprule
  \textbf{Capability}
      & \multirow{2}{*}{\parbox{\descwidth}{\textbf{Function} (\textbf{Return Type}):
             \\  \textbf{Brief Description}}}
          & \textbf{Dependencies}
          & \multirow{2}{*}{\parbox{\optwidth}{\textbf{Options}
             \\  (\textbf{Type})}}
          \\ & &
  \\ \hline
\cpp{GalacticHalo} & \multirow{2}{*}{\parbox{\descwidth}{\cpp{GalacticHalo_gNFW}
              (\cpp{GalacticHaloProperties}):
              \\  Provides the generalised NFW density profile $\rho(r)$ and $r_{\rm sun}$.}}
 & &
 \\ & & &
  \\ \cmidrule{2-4}
 & \multirow{2}{*}{\parbox{\descwidth}{\cpp{GalacticHalo_Einasto}
              (\cpp{GalacticHaloProperties}):
              \\  Provides the Einasto density profile $\rho(r)$ and $r_{\rm sun}$. }}
 & &
 \\ & & &
  \\ \hline
\cpp{LocalHalo} & \multirow{3}{*}{\parbox{\descwidth}{\cpp{ExtractLocalMaxwellianHalo}
              (\cpp{LocalMaxwellianHalo}):
              \\  Provides the local density $\rho_0$ as well as the velocity parameters $v_0$, $v_{\rm rot}$ and $v_{\rm esc}$.}}
 & &
 \\ & & &
  \\ & & &
  \\ \hline
\cpp{lnL_rho0} & \multirow{2}{*}{\parbox{\descwidth}{\cpp{lnL_rho0_lognormal}
              (\cpp{double}):
              \\  Log of the log-normal likelihood for the local DM density. }}
 &  \cpp{LocalHalo}
 &  \cpp{rho0_obs} (\cpp{double})
 \\ & & & \cpp{rho0_obserr} (\cpp{double})
  \\ \hline
\cpp{lnL_vrot} & \multirow{2}{*}{\parbox{\descwidth}{\cpp{lnL_vrot_gaussian}
              (\cpp{double}):
              \\  Log of the Gaussian likelihood for the local disk rotation speed.  }}
  &  \cpp{LocalHalo}
 &  \cpp{vrot_obs} (\cpp{double})
 \\ & & & \cpp{vrot_obserr} (\cpp{double})
  \\ \hline
\cpp{lnL_v0} & \multirow{2}{*}{\parbox{\descwidth}{\cpp{lnL_v0_gaussian}
              (\cpp{double}):
              \\  Log of the Gaussian likelihood for the most-probable DM speed. }}
 &  \cpp{LocalHalo}
 &  \cpp{v0_obs} (\cpp{double})
 \\ & & & \cpp{v0_obserr} (\cpp{double})
  \\ \hline
\cpp{lnL_vesc} & \multirow{2}{*}{\parbox{\descwidth}{\cpp{lnL_vesc_gaussian}
              (\cpp{double}):
              \\  Log of the Gaussian likelihood for the escape velocity.  }}
 &  \cpp{LocalHalo}
 & \cpp{vesc_obs} (\cpp{double})
 \\ & & & \cpp{vesc_obserr} (\cpp{double})
  \\ \hline
  \end{tabular}
\caption{Capabilities connected to the Milky Way halo parameters. \label{tab:halocap}}
\end{table*}

\renewcommand\metavar\metavarf

\renewcommand\metavar\metavars
\renewcommand\descwidth{5.2cm}

\begin{table*}[p]
  \scriptsize
  \centering
\begin{tabular}{l|p{\descwidth}|l|l|l}
   \toprule
  \textbf{Capability}
      & \multirow{2}{*}{\parbox{\descwidth}{\textbf{Function} (\textbf{Return Type}):
             \\  \textbf{Brief Description}}}
          & \textbf{Dependencies}
          & \multirow{2}{*}{\parbox{\bewidth}{\textbf{Backend}
             \\  \textbf{requirements}}}
          & \multirow{2}{*}{\parbox{\optwidth}{\textbf{Options}
             \\  (\textbf{Type})}}
          \\ & & &
  \\ \hline
 \cpp{RD_spectrum}
      & \multirow{2}{*}{\parbox{\descwidth}{\cpp{RD_spectrum_SUSY}
              (\cpp{DarkBit::RD_spectrum_type}):
          \\   Returns  masses and d.o.f.~for all \mbox{(co-)}annihilating particles, plus
          location of thresholds and resonances. Information retrieved from \ds.}}
  & \cpp{DarkSUSY_PointInit} & \cpp{mspctm} & \cpp{CoannCharginos}
  \\ & & &  \cpp{widths}  & \quad\cpp{Neutralinos} (\cpp{bool})
  \\ & & & \cpp{intdof}  & \cpp{CoannSfermions} (\cpp{bool})
  \\ & & & \cpp{pacodes} & \cpp{CoannMaxMass} (\cpp{double})
  \\ & & & \cpp{particle_code} &
    \\ & & & &
  \\ \cmidrule{2-5}
   & \multirow{2}{*}{\parbox{\descwidth}{\cpp{RD_spectrum_from_ProcessCatalog}\\
              (\cpp{DarkBit::RD_spectrum_type}):
              \\   Returns mass and d.o.f.~of DM particles, plus location of
              thresholds and resonances. Information retrieved from Process catalogue.}}
  & \cpp{TH_ProcessCatalog} &  &
  \\ & & \cpp{DarkMatter_ID} & &
  \\ & & & &
  \\ & & & &
  \\ & & & &
  \\ & & & &
   \\ \hline
  \cpp{RD_spectrum_}
      & \multirow{2}{*}{\parbox{\descwidth}{\cpp{RD_spectrum_ordered_func}
              (\cpp{DarkBit::RD_spectrum_type}):
              \\  Adds co-annihilation thresholds to the  output from
               \cpp{RD_spectrum}, and orders all thresholds and resonances by energy. }}
  & \cpp{RD_spectrum} & &
 \\ \quad\cpp{ordered} & & & &
  \\ & & & &
  \\ & & & &
  \\ & & & &
 \\ \hline
 \cpp{RD_eff_annrate_} & \multirow{2}{*}{\parbox{\descwidth}{\cpp{RD_annrate_DSprep_func}
              (\cpp{int}):
              \\  Initializes \ds to be able to provide effective invariant rate $W_\mathrm{eff}$.}}
  & \cpp{RD_spectrum} & \cpp{rdmgev} &
  \\ \quad\cpp{DSprep} & & & &
  \\ & & & &
  \\ \hline
  \cpp{RD_eff_annrate}
      & \multirow{2}{*}{\parbox{\descwidth}{\cpp{RD_eff_annrate_SUSY}
              (\cpp{fptr_dd}):
              \\  Returns the effective invariant rate $W_\mathrm{eff}$ as provided by \ds. }}
  & \cpp{RD_eff_annrate_} & \cpp{dsanwx}  &
  \\ & & \quad\cpp{DSprep} & &
  \\ & & & &
  \\ \cmidrule{2-5}
   & \multirow{2}{*}{\parbox{\descwidth}{\cpp{RD_eff_annrate_from_ProcessCatalog}
              (\cpp{fptr_dd}):
              \\  Returns the effective invariant rate $W_\mathrm{eff}$ as calculated from the information contained in the process catalogue. }}
  & \cpp{TH_ProcessCatalog} & &
  \\ & & \cpp{DarkMatter_ID} &  &
  \\ & & & &
  \\ & & & &
  \\ & & & &
   \\ \hline
  \end{tabular}
\caption{General relic density capabilities provided by \darkbit.  \label{tab:darkbitrelcap}}
\end{table*}

\renewcommand\metavar\metavarf

\renewcommand\metavar\metavars
\renewcommand\descwidth{4.5cm}

\begin{table*}[p]
  \scriptsize
  \centering
\begin{tabular}{l|p{\descwidth}|l|l|l}
   \toprule
  \textbf{Capability}
      & \multirow{2}{*}{\parbox{\descwidth}{\textbf{Function} (\textbf{Return Type}):
             \\  \textbf{Brief Description}}}
          & \textbf{Dependencies}
          & \multirow{2}{*}{\parbox{\bewidth}{\textbf{Backend}
             \\  \textbf{requirements}}}
          & \multirow{2}{*}{\parbox{\optwidth}{\textbf{Options}
             \\  (\textbf{Type})}}
          \\ & & &
  \\ \hline
 \cpp{RD_oh2}
      & \multirow{2}{*}{\parbox{\descwidth}{\cpp{RD_oh2_general}
              (\cpp{double}):
              \\  The general dark matter relic density. }}
 & \cpp{RD_spectrum_ordered} & \cpp{dsrdthlim} & \cpp{fast}
              (\cpp{int})
 \\ &  & \cpp{RD_eff_annrate} & \cpp{dsrdtab}   &
  \\ & & & \cpp{dsrdeqn}   &
  \\ & & & \cpp{dsrdwintp}   &
  \\ & & & \cpp{particle_code}   &
  \\ & & & \cpp{widths}   &
  \\ & & & \cpp{rdmgev}   &
  \\ & & & \cpp{rdpth}   &
  \\ & & & \cpp{rdpars}   &
  \\ & & & \cpp{rdswitch}   &
  \\ & & & \cpp{rdlun}   &
  \\ & & & \cpp{rdpadd}   &
  \\ & & & \cpp{rddof}   &
  \\ & & & \cpp{rderrors}   &
  \\ \cmidrule{2-5}
 & \multirow{3}{*}{\parbox{\descwidth}{\cpp{RD_oh2_DarkSUSY}
              (\cpp{double}):
              \\  Routine for directly obtaining results from \ds. }}
  & \cpp{DarkSUSY_PointInit} & \cpp{dsrdomega} & \cpp{omtype} (\cpp{int})
  \\ & & &  &  \cpp{fast} (\cpp{int})
  \\ & & & &
  \\ \cmidrule{2-5}
 & \multirow{3}{*}{\parbox{\descwidth}{\cpp{RD_oh2_MicrOmegas}
              (\cpp{double}):
              \\  Routine for directly obtaining results from \micromegas. }}
  & & \cpp{oh2}   &  \cpp{fast} (\cpp{int})
  \\ & & & & \cpp{Beps} (\cpp{double})
  \\ & & & &
  \\ \hline
 \cpp{RD_fraction}
      & \multirow{3}{*}{\parbox{\descwidth}{\cpp{RD_fraction_from_oh2}
              (\cpp{double}):
              \\  The relic density expressed as a fraction of the critical density. }}
 & \cpp{RD_oh2} & & \cpp{oh2_obs} (\cpp{double})
 \\ & & & & \cpp{mode} (\cpp{std::string})
 \\ & & & &
 \\ \hline
 \cpp{lnL_oh2}
      & \multirow{3}{*}{\parbox{\descwidth}{\cpp{lnL_oh2_Simple}
              (\cpp{double}):
              \\  Gaussian log-likelihood (see Secs.\ 8.3.1 and 8.3.2 of \cite{gambit}) for the relic density.}}
 & \cpp{RD_oh2} & & \cpp{oh2_obs} (\cpp{double})
 \\ & & & & \cpp{oh2_obserr} (\cpp{double})
 \\ & & & & \cpp{profile_systematics} (\cpp{bool})
 \\ & & & & \cpp{oh2_fractional_}
 \\ & & & & \quad\cpp{theory_err} (\cpp{double})
 \\ \cmidrule{2-5}
      & \multirow{3}{*}{\parbox{\descwidth}{\cpp{lnL_oh2_upperlimit}
              (\cpp{double}):
              \\  A half-Gaussian log-likelihood (see Secs.\ 8.3.3 and 8.3.4 of \cite{gambit}) for the relic density, treating the measured value as a smeared upper bound.}}
 & \cpp{RD_oh2} & & \cpp{oh2_cental} (\cpp{double})
 \\ & & & & \cpp{oh2_obserr} (\cpp{double})
 \\ & & & & \cpp{profile_systematics} (\cpp{bool})
 \\ & & & & \cpp{oh2_fractional_}
 \\ & & & & \quad\cpp{theory_err} (\cpp{double})
  \\ \hline
  \end{tabular}
\caption{The main relic density capabilities provided by \darkbit. \label{tab:darkbitmainrelcap}}
\end{table*}

\renewcommand\metavar\metavarf

\renewcommand\metavar\metavars
\renewcommand\descwidth{5cm}

\begin{table*}[tbp]
  \centering
  \scriptsize
\begin{tabular}{l|p{\descwidth}|l|l|l}
   \toprule
  \textbf{Capability}
      & \multirow{2}{*}{\parbox{\descwidth}{\textbf{Function} (\textbf{Return Type}):
             \\  \textbf{Brief Description}}}
          & \textbf{Dependencies}
          & \multirow{2}{*}{\parbox{\bewidth}{\textbf{Backend}
             \\  \textbf{requirements}}}
          & \multirow{2}{*}{\parbox{\optwidth}{\textbf{Options}
             \\  (\textbf{Type})}}
          \\ & & &
  \\ \hline
\metavar{X}\cpp{_Calculate}
      & \multirow{3}{*}{\parbox{\descwidth}{\metavar{X}\cpp{_Calculate}
              (\cpp{bool}):
              \\  Perform rate calculations for direct detection analysis \metavar{X}.  }}
 & & \cpp{DD_CalcRates} &
 \\ & & & \cpp{DD_Experiment} &
 \\ & & & &
 \\ \hline
\metavar{X}\cpp{_Events}
      & \multirow{3}{*}{\parbox{\descwidth}{\metavar{X}\cpp{_Events_DDCalc}
              (\cpp{int}):
              \\  The number of observed events for direct detection analysis \metavar{X}.  }}
 & \metavar{X}\cpp{_Calculate} & \cpp{DD_Events} &
 \\ & & & \cpp{DD_Experiment} &
 \\ & & & &
 \\ \hline
\metavar{X}\cpp{_Background}
      & \multirow{3}{*}{\parbox{\descwidth}{\metavar{X}\cpp{_Background_DDCalc}
              (\cpp{double}):
              \\  The number of background events for direct detection analysis \metavar{X}.  }}
 & \metavar{X}\cpp{_Calculate} & \cpp{DD_Background} &
 \\ & & & \cpp{DD_Experiment} &
 \\ & & & &
 \\ \hline
\metavar{X}\cpp{_Signal}
      & \multirow{3}{*}{\parbox{\descwidth}{\metavar{X}\cpp{_Signal_DDCalc}
              (\cpp{double}):
              \\  The number of signal events for direct detection analysis \metavar{X}.  }}
 & \metavar{X}\cpp{_Calculate} & \cpp{DD_Signal} &
 \\ & & & \cpp{DD_Experiment} &
 \\ & & & &
 \\ \hline
\metavar{X}\cpp{_SignalSI}
      & \multirow{3}{*}{\parbox{\descwidth}{\metavar{X}\cpp{_SignalSI_DDCalc}
              (\cpp{double}):
              \\  The number of spin-independent signal events for direct detection analysis \metavar{X}.  }}
 & \metavar{X}\cpp{_Calculate} & \cpp{DD_SignalSI} &
 \\ & & & \cpp{DD_Experiment} &
 \\ & & & &
 \\ \hline
\metavar{X}\cpp{_SignalSD}
      & \multirow{3}{*}{\parbox{\descwidth}{\metavar{X}\cpp{_SignalSD_DDCalc}
              (\cpp{double}):
              \\  The number of spin-dependent signal events for direct detection analysis \metavar{X}.  }}
 & \metavar{X}\cpp{_Calculate} & \cpp{DD_SignalSD} &
 \\ & & & \cpp{DD_Experiment} &
 \\ & & & &
 \\ \hline
\metavar{X}\cpp{_LogLikelihood}
      & \multirow{3}{*}{\parbox{\descwidth}{\metavar{X}\cpp{_LogLikelihood_DDCalc}
              (\cpp{double}):
              \\  Calculate the log-likelihood for direct detection analysis \metavar{X}.  }}
 & \metavar{X}\cpp{_Calculate} &  \cpp{DD_LogLikelihood} &
 \\ & & & \cpp{DD_Experiment} &
 \\ & & & &
 \\ \hline
\cpp{lnL_SI_nuclear_} & \multirow{2}{*}{\parbox{\descwidth}{\cpp{lnL_sigmas_sigmal}
              (\cpp{double}):
              \\  The log-likehood for the nuclear parameters relevant for spin-independent scattering.  }}
 &  & & \cpp{sigmas_obs} (\cpp{double})
 \\ \quad\cpp{parameters} & & & & \cpp{sigmas_obserr} (\cpp{double})
 \\ & & & & \cpp{sigmal_obs} (\cpp{double})
  \\ & & & & \cpp{sigmal_obserr} (\cpp{double})
 \\ \hline
\cpp{lnL_SD_nuclear_} & \multirow{2}{*}{\parbox{\descwidth}{\cpp{lnL_deltaq}
              (\cpp{double}):
              \\  The log-likehood for the nuclear parameters relevant for spin-dependent scattering.  }}
 &  & & \cpp{a3_obs} (\cpp{double})
 \\ \quad\cpp{parameters} & & & & \cpp{a3_obserr} (\cpp{double})
 \\ & & & & \cpp{a8_obs} (\cpp{double})
 \\ & & & & \cpp{a8_obserr} (\cpp{double})
 \\ & & & & \cpp{deltas_obs} (\cpp{double})
 \\ & & & & \cpp{deltas_obserr} (\cpp{double})
 \\ \hline
\end{tabular}
\caption{\darkbit capabilities for direct detection log-likelihoods and related observables. Possible values for \metavar{X} are \cpp{XENON100_2012}, \cpp{LUX_2013}, \cpp{LUX_2015}, \cpp{LUX_2016}, \cpp{PandaX_2016}, \cpp{SuperCDMS_2014}, \cpp{SIMPLE_2014}, \cpp{PICO_2L}, \cpp{PICO_60_F} and \cpp{PICO_60_I} in version 1.0.0 and in addition \cpp{XENON1T_2017} and \cpp{PICO_60_2017} in version 1.1.0. \label{tab:ddcap}}
\end{table*}

\renewcommand\metavar\metavarf

\renewcommand\metavar\metavars
\renewcommand\descwidth{4.5cm}

\begin{table*}[tbp]
  \centering
  \scriptsize
\begin{tabular}{l|p{\descwidth}|l|l|l}
   \toprule
  \textbf{Capability}
      & \multirow{2}{*}{\parbox{\descwidth}{\textbf{Function} (\textbf{Return Type}):
             \\  \textbf{Brief Description}}}
          & \textbf{Dependencies}
          & \multirow{2}{*}{\parbox{\bewidth}{\textbf{Backend}
             \\  \textbf{requirements}}}
          & \multirow{2}{*}{\parbox{\optwidth}{\textbf{Options}
             \\  (\textbf{Type})}}
          \\ & & &
  \\ \hline
  \cpp{SimYieldTable}
      & \multirow{3}{*}{\parbox{\descwidth}{\cpp{SimYieldTable_DarkSUSY}
              (\cpp{DarkBit::SimYieldTable}):
              \\  Provides access to tabulated yields in \ds }}
  & & \cpp{dshayield}  &  \cpp{allow_yield_extrapolation} (\cpp{bool})
  \\ & & & &
  \\ & & & &
  \\ & & & &
  \\ \cmidrule{2-5}
      & \multirow{3}{*}{\parbox{\descwidth}{\cpp{SimYieldTable_MicrOmegas}
              (\cpp{DarkBit::SimYieldTable}):
              \\  Provides access to tabulated yields in
          MicrOmegas. }}
  & & \cpp{dNdE}  & \cpp{allow_yield_extrapolation} (\cpp{bool})
  \\ & & & &
  \\ & & & &
  \\ & & & &
%  \\ \cmidrule{2-5}
%
%        & \multirow{3}{*}{\parbox{\descwidth}{\cpp{SimYieldTable_PPPC}
%              (\cpp{DarkBit::SimYieldTable}):
%              \\  Provides access to tabulated yields in PPPC. }}
%  &  & &
%  \\ & & & &
%  \\ & & & &
%  \\ & & & &
 \\ \hline
 \cpp{GA_missing}
      & \multirow{2}{*}{\parbox{\descwidth}{\cpp{GA_missingFinalStates}
              (\cpp{std::vector<std::string>}):
              \\  Determines final states that are not available
          with tabulated spectra. }}
 & \cpp{TH_ProcessCatalog} & &     \cpp{ignore_all} (\cpp{bool})
  \\ \quad\cpp{FinalStates} &  & \cpp{SimYieldTable}  & &  \cpp{ignore_two_body} (\cpp{bool})
  \\ & & \cpp{DarkMatter_ID} & &   \cpp{ignore_three_body} (\cpp{bool})
  \\ & & & &
  \\ \hline
 \cpp{GA_AnnYield}
      & \multirow{2}{*}{\parbox{\descwidth}{\cpp{GA_AnnYield_General}
              (\cpp{Funk::Funk}):
              \\  General function for calculating gamma-ray yields
          from the process catalogue.}}
 & \cpp{TH_ProcessCatalog} & & \cpp{line_width} (\cpp{double})
  \\ &  & \cpp{SimYieldTable}  & &
  \\ & & \cpp{DarkMatter_ID} & &
  \\ & & \cpp{cascadeMC_} & &
  \\ & & \quad\cpp{gammaSpectra} & &
    \\ \hline
   \end{tabular}
\caption{General gamma-ray capabilities provided by \darkbit.  \label{tab:darkbitgammacap}}
\end{table*}

\renewcommand\metavar\metavarf

\renewcommand\metavar\metavars
\renewcommand\descwidth{6cm}

\begin{table*}[tbp]
  \scriptsize
  \centering
\begin{tabular}{l|p{\descwidth}|l|l|l}
   \toprule
  \textbf{Capability}
      & \multirow{2}{*}{\parbox{\descwidth}{\textbf{Function} (\textbf{Return Type}):
             \\  \textbf{Brief Description}}}
          & \textbf{Dependencies}
          & \multirow{2}{*}{\parbox{\bewidth}{\textbf{Backend}
             \\  \textbf{requirements}}}
          & \multirow{2}{*}{\parbox{\optwidth}{\textbf{Options}
             \\  (\textbf{Type})}}
          \\ & & &
 \\ \hline
 \cpp{lnL_FermiLATdwarfs}
 & \multirow{3}{*}{\parbox{\descwidth}{\cpp{lnL_FermiLATdwarfs_gamLike}
              (\cpp{double}):
              \\  Log-likelihood for the \textit{Fermi}-LAT dwarf galaxy search, using \gamlike.  }}
 & \cpp{GA_AnnYield} & \gamlike & \cpp{version (str)}
 \\ & & \cpp{RD_fraction}  & &
 \\ & & & &
 \\ \hline
 \cpp{lnL_FermiGC}
 & \multirow{3}{*}{\parbox{\descwidth}{\cpp{lnL_FermiGC_gamLike}
              (\cpp{double}):
              \\  Log-likelihood for the \textit{Fermi}-LAT GeV excess, using \gamlike.  }}
 & \cpp{GA_AnnYield} & \gamlike & \cpp{version (str)}
 \\ & & \cpp{RD_fraction} & &
 \\ & & \cpp{set_gamLike_GC_halo} & &
 \\ \hline
 \cpp{lnL_HESSGC}
 & \multirow{3}{*}{\parbox{\descwidth}{\cpp{lnL_HESSGC_gamLike}
              (\cpp{double}):
              \\  Log-likelihood for the HESS Galactic halo searches, using \gamlike.  }}
 & \cpp{GA_AnnYield} & \gamlike & \cpp{version (str)}
 \\ & & \cpp{RD_fraction}  & &
 \\ & & \cpp{set_gamLike_GC_halo} & &
 \\ \hline
 \cpp{lnL_CTAGC}
 & \multirow{3}{*}{\parbox{\descwidth}{\cpp{lnL_CTAGC_gamLike}
              (\cpp{double}):
              \\  Log-likelihood for projected dark matter searches with CTA, using \gamlike.  }}
 & \cpp{GA_AnnYield} & \gamlike & \cpp{version (str)}
 \\ & & \cpp{RD_fraction}  & &
 \\ & & \cpp{set_gamLike_GC_halo} & &
 \\ \hline
  \cpp{set_gamLike_GC_halo}
 & \multirow{3}{*}{\parbox{\descwidth}{\cpp{set_gamLike_GC_halo}
              (\cpp{bool}):
              \\  Initialises the Galactic dark matter distribution in \gamlike, based on the halo model used in the corresponding scan.  }}
 & \cpp{GalacticHalo} & \gamlike &
 \\ & & & &
 \\ & & & &
 \\ & & & &
 \\ \hline
\cpp{GalacticHalo} & \multirow{4}{*}{\parbox{\descwidth}{\cpp{GalacticHalo_gNFW}\\
              (\cpp{GalacticHaloProperties}):
              \\  Provides the generalised NFW density profile\\$\rho(r)$ and $r_{\rm sun}$.}}
 & & &
 \\ & & & &
  \\ & & & &
   \\ & & & &
  \\ \cmidrule{2-5}
 & \multirow{4}{*}{\parbox{\descwidth}{\cpp{GalacticHalo_Einasto}\\
              (\cpp{GalacticHaloProperties}):
              \\  Provides the Einasto density profile\\$\rho(r)$ and $r_{\rm sun}$. }}
 & & &
 \\ & & & &
  \\ & & & &
   \\ & & & &
  \\ \hline
     \end{tabular}
\caption{\darkbit capabilities for gamma-ray indirect detection likelihoods.  \label{tab:darkbitIDlike}}
\end{table*}

\renewcommand\metavar\metavarf

\renewcommand\metavar\metavars
\renewcommand\descwidth{5cm}

\begin{table*}[tbp]
  \centering
  \scriptsize
\begin{tabular}{l|p{\descwidth}|l|l}
   \toprule
  \textbf{Capability}
      & \multirow{2}{*}{\parbox{\descwidth}{\textbf{Function} (\textbf{Return Type}):
             \\  \textbf{Brief Description}}}
          & \textbf{Dependencies}
          & \multirow{2}{*}{\parbox{\bewidth}{\textbf{Backend}
             \\  \textbf{requirements}}}
          \\ & & &
  \\ \hline
\cpp{capture_rate_Sun}
      & \multirow{3}{*}{\parbox{\descwidth}{\cpp{capture_rate_Sun_const_xsec}
              (\cpp{double}):
              \\  Capture rate of regular dark matter in the Sun (no $v$-dependent or $q$-dependent cross-sections) ($s^{-1}$).}}
& \cpp{mwimp} & \cpp{cap_Sun_v0q0_isoscalar}
\\ & & \cpp{sigma_SI_p} &
\\ & & \cpp{sigma_SD_p} &
\\ & & &
\\ & & &
\\ \hline
\cpp{equilibration_time_Sun}
      & \multirow{3}{*}{\parbox{\descwidth}{\cpp{equilibration_time_Sun}
              (\cpp{double}):
              \\  Equilibration time for capture and annihilation of dark matter in the Sun ($s$). }}
& \cpp{mwimp} &
\\ & & \cpp{sigmav} &
\\ & & \cpp{capture_rate_Sun} &
  \\ \hline
\cpp{annihilation_rate_Sun}
      & \multirow{3}{*}{\parbox{\descwidth}{\cpp{annihilation_rate_Sun}
              (\cpp{double}):
              \\  Annihilation rate of dark matter in the Sun ($s^{-1}$). }}
& \cpp{equilibration_time_Sun} &
\\ & & \cpp{capture_rate_Sun} &
\\ & &
\\ \hline
\cpp{nuyield_ptr}
      & \multirow{2}{*}{\parbox{\descwidth}{\cpp{nuyield_from_DS}
              (\cpp{double}):
              \\  Neutrino yield function pointer and setup. }}
& \cpp{TH_ProcessCatalog} & \cpp{nuyield_setup}
\\ & & \cpp{mwimp} & \cpp{nuyield}
\\ & & \cpp{sigmav} & \cpp{get_DS_neutral_h_}
\\ & & \cpp{sigma_SI_p} & \quad\cpp{decay_channels}
\\ & & \cpp{sigma_SD_p} & \cpp{get_DS_charged_h_}
\\ & & \cpp{DarkMatter_ID} & \quad\cpp{decay_channels}
\\ \hline
  \end{tabular}
\caption{General \darkbit capabilities for neutrino indirect detection processes. \label{tab:darkbitnucap}}
\end{table*}

\renewcommand\metavar\metavarf

\renewcommand\metavar\metavars
\renewcommand\descwidth{5.5cm}

\begin{table*}[tbp]
  \centering
  \scriptsize
\begin{tabular}{l|p{\descwidth}|l|l|l}
   \toprule
  \textbf{Capability}
      & \multirow{2}{*}{\parbox{\descwidth}{\textbf{Function} (\textbf{Return Type}):
             \\  \textbf{Brief Description}}}
          & \textbf{Dependencies}
          & \multirow{2}{*}{\parbox{\bewidth}{\textbf{Backend}
             \\  \textbf{requirements}}}
          & \multirow{2}{*}{\parbox{\optwidth}{\textbf{Options}
             \\  (\textbf{Type})}}
          \\ & & &
  \\ \hline
\metavar{X}\cpp{_data}
      & \multirow{3}{*}{\parbox{\descwidth}{\metavar{X}\cpp{_full}
              (\cpp{nudata}):
              \\  Do signal, likelihood and related calculations for neutrino indirect detection analysis \metavar{X}. }}
 & \cpp{mwimp} & \cpp{nubounds} & \cpp{nulike_speed}
\\ & & \cpp{annihilation_rate_Sun} & & \quad(\cpp{int})
\\ & & \cpp{nuyield_ptr} &
\\ \hline
\metavar{X}\cpp{_signal}
      & \multirow{3}{*}{\parbox{\descwidth}{\metavar{X}\cpp{_signal}
              (\cpp{double}):
              \\  Number of signal events for neutrino indirect detection analysis \metavar{X}. }}
 & \metavar{X}\cpp{_data} & &
\\ & &  & &
\\ & &  & &
\\ \hline
\metavar{X}\cpp{_bg}
      & \multirow{3}{*}{\parbox{\descwidth}{\metavar{X}\cpp{_bg}
              (\cpp{double}):
              \\  Number of background events for neutrino indirect detection analysis \metavar{X}. }}
 & \metavar{X}\cpp{_data} & &
\\ & &  & &
\\ & &  & &
\\ \hline
\metavar{X}\cpp{_loglike}
      & \multirow{3}{*}{\parbox{\descwidth}{\metavar{X}\cpp{_loglike}
              (\cpp{double}):
              \\  Log-likelihood for neutrino indirect detection analysis \metavar{X}. }}
 & \metavar{X}\cpp{_data} & &
\\ & &  & &
\\ & &  & &
\\ \hline
\metavar{X}\cpp{_bgloglike}
      & \multirow{3}{*}{\parbox{\descwidth}{\metavar{X}\cpp{_bgloglike}
              (\cpp{double}):
              \\  Background-only log-likelihood for neutrino indirect detection analysis \metavar{X}. }}
 & \metavar{X}\cpp{_data} & &
\\ & &  & &
\\ & &  & &
\\ \hline
\metavar{X}\cpp{_pvalue}
      & \multirow{3}{*}{\parbox{\descwidth}{\metavar{X}\cpp{_pvalue}
              (\cpp{double}):
              \\  $p$-value for neutrino indirect detection analysis \metavar{X}. }}
 & \metavar{X}\cpp{_data} &
\\ & &  & &
\\ & &  & &
 \\ \hline
\metavar{X}\cpp{_nobs}
      & \multirow{3}{*}{\parbox{\descwidth}{\metavar{X}\cpp{_nobs}
              (\cpp{int}):
              \\  Number of observed events for neutrino indirect detection analysis \metavar{X}. }}
 & \metavar{X}\cpp{_data} &
\\ & &  & &
\\ & &  & &
\\ \hline
\cpp{IC79_loglike}
      & \multirow{2}{*}{\parbox{\descwidth}{\cpp{IC79_loglike}
              (\cpp{double}):
              \\  The full 79-string IceCube log-likelihood. }}
 & \metavar{Y}\cpp{_loglike}  & &
\\ & & \metavar{Y}\cpp{_bgloglike}  & &
\\ & & for all $Y\in\{$\cpp{IC79WH}, & &
\\ & & \quad\cpp{IC79WL},\cpp{IC79SL}$\}$ & &
\\\hline
\cpp{IceCube_}
      & \multirow{2}{*}{\parbox{\descwidth}{\cpp{IC_loglike}
              (\cpp{double}):
              \\  The complete IceCube log-likelihood. }}
 & \metavar{Y}\cpp{_loglike}  & &
\\ \quad\cpp{likelihood} & & \metavar{Y}\cpp{_bgloglike}  & &
\\ & & for all \metavar{Y}$\in\{$\cpp{IC22},\cpp{IC79WH}, & &
\\ & & \quad\cpp{IC79WL},\cpp{IC79SL}$\}$ & &
\\ \hline
  \end{tabular}
\caption{\darkbit capabilities for neutrino indirect detection likelihoods. Possible values for \metavar{X} are \cpp{IC22}, \cpp{IC79WH}, \cpp{IC79WL}, and \cpp{IC79SL}. \label{tab:darkbitnulikecap}}
\end{table*}

\renewcommand\metavar\metavarf

\renewcommand\metavar\metavars
\renewcommand\descwidth{5cm}

\begin{table*}[tbp]
  \centering
  \scriptsize
\begin{tabular}{l|p{\descwidth}|l|l}
   \toprule
  \textbf{Capability}
      & \multirow{2}{*}{\parbox{\descwidth}{\textbf{Function} (\textbf{Return Type}):
             \\  \textbf{Brief Description}}}
          & \textbf{Dependencies}
          & \multirow{2}{*}{\parbox{\optwidth}{\textbf{Options}
             \\  (\textbf{Type})}}
          \\ & & &
  \\ \hline
 \cpp{cascadeMC_FinalStates}
      & \multirow{2}{*}{\parbox{\descwidth}{\cpp{cascadeMC_FinalStates}
              (\cpp{std::vector<std::string>}):
              \\  Function for retrieving list of final states for cascade decays. }}
 &  & \cpp{cMC_finalStates}
 \\ &  &  & ~~ (\cpp{std::vector<std::string>})
 \\ & &  &
 \\ & &  &
 \\ \hline
 \cpp{cascadeMC_DecayTable}
      & \multirow{3}{*}{\parbox{\descwidth}{\cpp{cascadeMC_DecayTable}
              (\cpp{DarkBit::DecayChain::DecayTable}):
              \\  Function setting up the decay table used in decay chains. }}
 & \cpp{TH_ProcessCatalog} &
 \\ &  & \cpp{SimYieldTable}  &
 \\ & &  &
 \\ & &  &
  \\ \hline
 \cpp{cascadeMC_gammaSpectra}
      & \multirow{3}{*}{\parbox{\descwidth}{\cpp{cascadeMC_gammaSpectra}
              (\cpp{DarkBit::stringFunkMap}):
              \\  Function requesting and returning gamma ray spectra from cascade decays. }}
 & \cpp{GA_missingFinalStates} &
 \\ &  & \cpp{cascadeMC_FinalStates}  &
 \\ & &\cpp{cascadeMC_Histograms}  &
 \\ & &\cpp{cascadeMC_EventCount}  &
   \\ \hline
  \end{tabular}
\caption{Cascade decay capabilities provided by \darkbit that do not run inside the cascade decay Monte Carlo loop. \label{tab:darkbitcascadecap}}
\end{table*}

\renewcommand\metavar\metavarf

\renewcommand\metavar\metavars
\renewcommand\descwidth{5.6cm}

\begin{table*}[tbp]
  \centering
  \scriptsize
\begin{tabular}{l|p{\descwidth}|l|l}
   \toprule
  \textbf{Capability}
      & \multirow{2}{*}{\parbox{\descwidth}{\textbf{Function} (\textbf{Return Type}):
             \\  \textbf{Brief Description}}}
          & \textbf{Dependencies}
          & \multirow{2}{*}{\parbox{\optwidth}{\textbf{Options}
             \\  (\textbf{Type})}}
          \\ & & &
  \\ \hline
 \cpp{cascadeMC_LoopManager}
      & \multirow{3}{*}{\parbox{\descwidth}{\cpp{cascadeMC_LoopManager}
              (\cpp{void}):
              \\  Controls the loop for the cascade decay Monte Carlo simulation. }}
 & \cpp{GA_missingFinalStates} & \cpp{cMC_maxEvents} (\cpp{int})
              \\ &  & &
              \\ &  & &
  \\ \hline
 \cpp{cascadeMC_InitialState}
      & \multirow{3}{*}{\parbox{\descwidth}{\cpp{cascadeMC_InitialState}
              (\cpp{std::string}):
              \\  Function selecting the initial state for the cascade decay chain. }}
 & \cpp{GA_missingFinalStates} &
 \\ &  &   &
 \\ &  &   &
  \\ \hline
 \cpp{cascadeMC_EventCount}
      & \multirow{2}{*}{\parbox{\descwidth}{\cpp{cascadeMC_EventCount}
              (\cpp{DarkBit::stringIntMap}):
              \\  The event counter for cascade decays. }}
 & \cpp{cascadeMC_InitialState} &
 \\ &  &   &
 \\ &  &   &
 \\ \hline
 \cpp{cascadeMC_ChainEvent}
      & \multirow{3}{*}{\parbox{\descwidth}{\cpp{cascadeMC_GenerateChain}
              (\cpp{DarkBit::DecayChain::ChainContainer}):
              \\  Function for generating decay chains. }}
 & \cpp{cascadeMC_InitialState} & \cpp{cMC_maxChainLength} (\cpp{int})
 \\ &  & \cpp{cascadeMC_DecayTable}  & \cpp{cMC_Emin} (\cpp{double})
 \\ &  &   &
  \\ \hline
  \cpp{cascadeMC_Histograms}
      & \multirow{2}{*}{\parbox{\descwidth}{\cpp{cascadeMC_Histograms}
              (\cpp{DarkBit::simpleHistContainter}):
              \\  Function responsible for histogramming and evaluating the end conditions for the event loop in the cascade decay Monte Carlo simulation. }}
 & \cpp{cascadeMC_InitialState} & \cpp{cMC_numSpecSamples} (\cpp{int})
 \\ &  & \cpp{cascadeMC_ChainEvent}  &  \cpp{cMC_NhistBins} (\cpp{int})
 \\ &  & \cpp{TH_ProcessCatalog}  & \cpp{cMC_binLow} (\cpp{double})
 \\ &  & \cpp{SimYieldTable}  & \cpp{cMC_binHigh} (\cpp{double})
 \\ &  & \cpp{cascadeMC_FinalStates}  & \cpp{cMC_gammaBGPower} (\cpp{double})
 \\ & & & \cpp{cMC_gammaRelError} (\cpp{double})
 \\ & & & \cpp{cMC_endCheckFrequency} (\cpp{int})
  \\ \hline
   \end{tabular}
\caption{The loop manager capability for the \darkbit cascade decay Monte Carlo, and the capabilities that are filled within the loop. Each of these depends on the \cpp{cascadeMC_LoopManagement} capability. \label{tab:darkbitcascadeloopcap}}
\end{table*}

\renewcommand\metavar\metavarf

\renewcommand\metavar\metavars
\renewcommand\descwidth{4.5cm}

\begin{table*}[tbp]
  \scriptsize
  \centering
 \begin{tabular}{l|p{\descwidth}|l|l|l}
   \toprule
  \textbf{Capability}
      & \multirow{2}{*}{\parbox{\descwidth}{\textbf{Function} (\textbf{Return Type}):
             \\  \textbf{Brief Description}}}
          & \textbf{Dependencies}
          & \multirow{2}{*}{\parbox{\bewidth}{\textbf{Backend}
             \\  \textbf{requirements}}}
          & \multirow{2}{*}{\parbox{\optwidth}{\textbf{Options}
             \\  (\textbf{Type})}}
          \\ & & &
  \\ \hline
  \cpp{DarkSUSY_PointInit}
      & \multirow{3}{*}{\parbox{\descwidth}{\cpp{DarkSUSY_PointInit_MSSM}
              (\cpp{bool}):
              \\  Function to initialise \ds to a specific
          model point. The generic \ds initialisation is done in the
      backend initialisation; this here is only necessary for other capabilities that make use of model-specific \ds routines. }}
  &  \cpp{MSSM_spectrum} & \cpp{dswwidth}  &  \cpp{use_DS_isasugra} (\cpp{bool})
  \\ & & \cpp{decay_rates}  & \cpp{dsprep} & \cpp{use_dsSLHAread} (\cpp{bool})
  \\ & & & \cpp{mssmpar}  & \cpp{debug_SLHA_filenames}
  \\ & & & \cpp{dssusy}   & \quad(\cpp{std::vector<str>})
  \\ & & & \cpp{dsSLHAread}  &
  \\ & & & \cpp{dssusy_isasugra} &
  \\ & & & \cpp{dsgive_model_}   &
  \\ & & & \quad\cpp{isasugra} &
  \\ & & &\cpp{initFromSLHAea}  &
  \\ & & & \quad\cpp{AndDecayTable}  &
  \\ \hline
  \multirow{2}{*}{\parbox{2.7cm}{\cpp{DarkSUSY_PointInit_}\\ \mbox{ \cpp{LocalHalo}}}}
      & \multirow{8}{*}{\parbox{\descwidth}{\cpp{DarkSUSY_PointInit_} \mbox{\ \cpp{LocalHalo_func}}
              (\cpp{bool}):
              \\  Function to initialise Milky Way halo model parameters in \ds. Any \gambit function that uses a \ds function that depends on the structure of the Milky Way halo should have this as a dependency.}}
   &  \cpp{RD_fraction} & \cpp{dshmcom}  & \cpp{v_earth} (\cpp{double})
  \\ & & \cpp{LocalHalo}& \cpp{dshmisodf} &
  \\ & & & \cpp{dshmframevelcom}  &
  \\ & & & \cpp{dshmnoclue}   &
  \\ & & & &
  \\ & & & &
  \\ & & & &
  \\ & & & &
  \\ \hline
 \cpp{dump_GammaSpectrum}
      & \multirow{2}{*}{\parbox{\descwidth}{\cpp{dump_GammaSpectrum}
              (\cpp{double}):
              \\  Dumps gamma-ray yield into ASCII table.}}
 & \cpp{GA_AnnYield} & & \cpp{filename} (\cpp{std::string})
 \\ & & & &
 \\ & & & &
  \\ \hline
  \cpp{UnitTest_DarkBit}
      & \multirow{2}{*}{\parbox{\descwidth}{\cpp{UnitTest_DarkBit}
              (\cpp{int}):
              \\  Prints various \DB results into \YAML file.}}
  & \cpp{DD_couplings} &  & \cpp{fileroot} (\cpp{std::string})
  \\ & & \cpp{RD_oh2}  & & \cpp{GA_AnnYield:Emin}
  \\ & & \cpp{GA_AnnYield} & & \quad(\cpp{double})
  \\ & & \cpp{TH_ProcessCatalog} & & \cpp{GA_AnnYield:Emax}
  \\ & & \cpp{DarkMatter_ID} & & \quad(\cpp{double})
  \\ & & & & \cpp{GA_AnnYield:nbins}
  \\ & & & & \quad(\cpp{double})
  \\ \hline
\end{tabular}
\caption{Miscellaneous capabilities for \ds initialisation and debugging.  \label{tab:darkbitmisc}}
\end{table*}

\bibliography{R1}

\end{document}